# Microfluidic Isotachophoresis: Theory and Applications

Ashwin Ramachandran[1] and Juan G. Santiago[2,*]

[1] Department of Aeronautics & Astronautics, Stanford University, California, USA 94305.
[2] Department of Mechanical Engineering, Stanford University, California, USA 94305.
Email: juan.santiago@stanford.edu
* Corresponding author

**Abstract**

Isotachophoresis (ITP) is a versatile electrophoretic technique which can be used for sample preconcentration, separation, purification, mixing, and control and acceleration of chemical reactions. Although the basic technique is nearly a century old and widely used, there has been a persistent need for an easily approachable, succinct, and rigorous review of ITP theory and analysis. This is important as interest and adoption of the technique has grown over the last two decades, especially because of its implementation into microfluidics and integration with on-chip chemical and biochemical assays. We here provide a review of ITP theory with a strong emphasis on steady and unsteady transport starting from physicochemical first principles including conservation of species, conservation of current, the approximation of charge neutrality, pH equilibrium of weak electrolytes, and so-called regulating functions governing transport dynamics. We combine these generally applicable (to all types of ITP) theoretical discussions with applications of ITP in the field of microfluidic systems, particularly on-chip biochemical analyses. Our discussion includes principles governing ITP focusing of weak and strong electrolytes, ITP dynamics in peak and plateau modes, review of simulation tools, experimental tools and detection methods, applications of ITP for on-chip separations and trace analyte manipulation, and design considerations and challenges for microfluidic ITP systems. We conclude with remarks on possible future research directions. The intent of this review is to help make ITP analysis and design principles more accessible to the scientific and engineering communities, and to provide a rigorous basis for increased adoption of ITP into microfluidics.

# Contents

# 1. Introduction

## 1.1 Qualitative introduction and definition of isotachophoresis

We begin with a qualitative description of isotachophoresis (ITP) to frame our summary of its background and development. ITP is an electrophoretic technique useful for purification, preconcentration, and separation of analytes.[1,2] **Figure 1** shows schematically a simple ITP process including an arrangement of ions of various types and these ions are migrating due to an applied electric field. This initial, basic example of ITP is of so-called anionic ITP where two different anions (and a single, common cation) are used to focus anionic samples, but, as we shall see, ITP can also be performed as cationic ITP (where two cationic species are used to focus cationic samples).

As shown in the schematic, the system includes an interface between two buffer mixtures. There is a leading electrolyte (LE) mixture and a trailing electrolyte (TE) mixture. These mixtures contain a leading ion and a trailing co-ion, respectively. As we shall see, the ion displacement nature of ITP is such that co-ions in purified zones travel with the same ion migration velocity. Plotted in the figure are ion concentrations $c$ as a function of axial dimension along the channel, $x$. $c_T^{T'}$ and $c_L^L$ refer to the concentrations of the trailing and leading anions. The concentration of the counterion (here, a cation) is denoted by $c_C$ and this counterion is assumed to be the same species for the leading and trailing electrolytes. The superscript on the trailing ion refers to the concentration of that ion in regions formerly occupied by the LE ion. Here, $E$ is applied from right to left and the LE-to-TE interface propagates to the right. Such a sharp moving interface between a leading high mobility ion and a trailing low mobility ion is sometimes referred to as a moving boundary in electrophoresis.[3–5]

In this anionic ITP example, these species are anions with relatively high and low ion mobilities, respectively. Upon application of an electric field, the interface(s) between the LE and TE moves due to electromigration (in the direction opposed to that of the electric field). Importantly, we apply the electric field such that the TE to LE co-ions move in the TE-to-LE direction. The latter electric field orientation creates a self-steepening TE-to-LE interface that moves at precisely the velocity of the anion in the LE zone. As we shall also see, the relative mobility magnitudes of the LE and TE lead to LE and TE regions which are respectively of high and low ionic conductivity. The gradient in conductivity in turn establishes a gradient in the electric field so the electric field magnitude is high in the TE and low in the LE. Sample species whose effective mobilities are greater than the TE ion will, if placed in the TE, migrate faster than the TE co-ion, and catch up to and focus at the TE-to-LE interface. Similarly, sample species placed into the LE zone will migrate slower than the LE co-ion and be caught by and focused into the TE-to-LE interface. In this way, sample ions of a specific mobility range are eventually focused at this interface irrespective of where they are introduced. As we shall show, the cause of the self-sharpening nature of the interface is due to the arrangement of mobilities and the resulting electric field gradient This self-steepening interface is effectively an ion concentration shock wave whose minimum width is limited by balance between non-uniform electromigration (established by the electric field gradient) and molecular diffusion.

The observable concentration and shape of the focused sample ions are a function of their initial concentration and time. As shown in **Figure 1b**, multiple analyte species initially strongly overlap and focus in "peak mode" within a small zone. This initial "peak mode" has a local concentration distribution that is at least roughly Gaussian in shape and of a magnitude much lower than that of the LE or TE. Given sufficient initial concentration and/or sufficient time, analyte species accumulate and increase concentration to the point where the analyte contributes significantly to local conductivity. Thereafter, as shown in **Figure 1c**, an analyte can be purified relative to other co-ions and form a plateau which is purified from and displaces other co-ions. The ion displacement nature of ITP is such that co-ions in purified zones travel with the same ion migration velocity. That is, the co-ions of the LE ion (including the TE ion) and any fully formed plateau travel at the same velocity as the LE ion. Lastly, the self-focusing nature of the various interfaces (e.g., the moving LE-to-TE interface as well as any LE-to-plateau, TE-to-plateau, and plateau-to-plateau interfaces) is robust to disturbances including channel roughness, channel turns, pressure-driven flow disturbances, and variation in channel geometry.

Lastly, we briefly note here that an electric field oriented in the "wrong" direction, (such that LE ions move toward the TE zone), leads to a phenomenon called electromigration dispersion (EMD). This phenomenon is characterized by an ion migration rarefaction wave.[6–9] In EMD, the ion drift velocities rapidly broaden the LE-TE interface and electromigration drives rapid dispersion of the interface and mixing of the co-ions on either side of the interface. We note that, in performing experiments with various buffer mixtures and applied electric fields, EMD is easily confused with rapid diffusion or Taylor dispersion. However, EMD is distinct as it is a deterministic, electromigration-driven mixing phenomenon where, for constant applied current, interface lengths increase linearly in time (compared to the square root of time expected from diffusion).

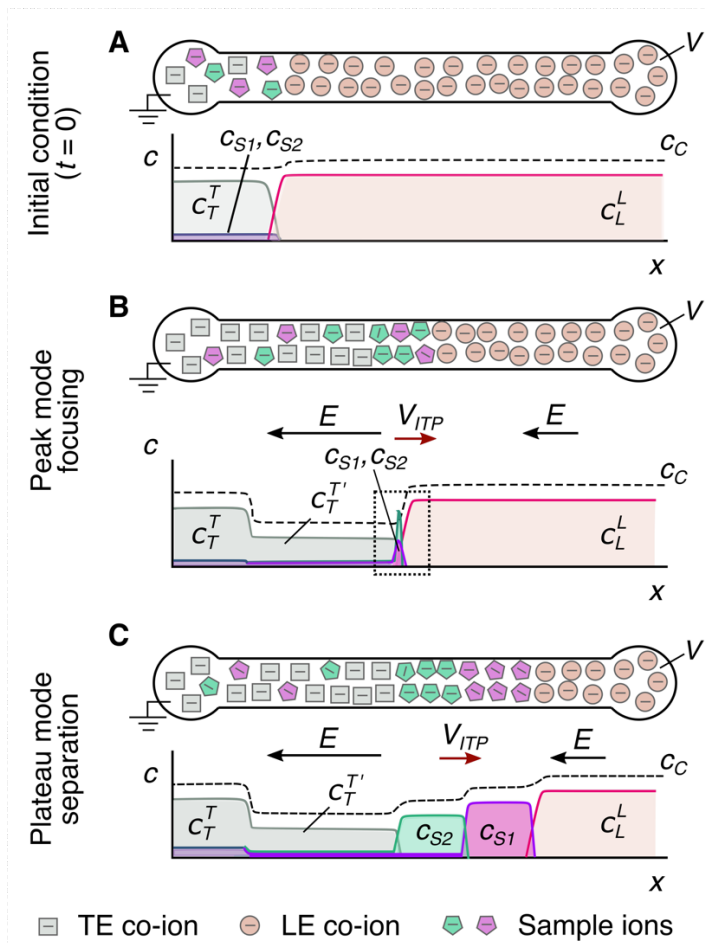

**Figure 1**: Schematics for qualitative understanding of microfluidic ITP. (A) Initial conditions at $t = 0$, common to both peak and plateau modes ITP. (B) Peak mode ITP focusing, which corresponds to either trace analyte focusing or the early stages of any ITP focusing (c.f. **Sections 2.9** and **5**). (C) Plateau mode ITP for the case where the analytes are present in high concentration and the experiment is run for a sufficiently long duration (c.f. **Sections 2.9** and **6**). Each subfigure shows the location of ions within the channel (top) and the concentration profiles of the electrolytes (bottom) in anionic ITP. For illustration, only two sample species $S_1$ and $S_2$ are considered. Peak-mode sample concentration fields are exaggerated for depiction, while sample mode concentration fields are drawn to relative scale. The loading configuration depicted in (A) and the schematics in (B) and (C) correspond to a semi-infinite sample loading configuration (c.f. **Section 2.8**).

## 1.2 Brief history and development of ITP

Although the term "isotachophoresis" was introduced only in the 1970s,[10,11] similar techniques based on the principles of ITP have existed for nearly a century. For example, Kendall and Crittenden[12] in 1923 described a technique fundamentally identical to ITP to separate rare earth metals and some acids, which they called the "ion migration method". Likewise, studies on "moving boundary electrophoresis",[13] steady-state stacking in disc electrophoresis,[14] and "displacement electrophoresis"[15] describe processes that are nearly identical (and in some cases identical) to ITP. "Displacement electrophoresis" and "transphoresis" have also been used synonymously with ITP.[16] The term "isotachophoresis" refers to the fact that all zones migrate at the same velocity. (The Greek roots "isos" and "tachos" mean equal and velocity, respectively.) ITP gained significant popularity in the 1970s as an analytical separation tool which, unlike capillary electrophoresis (CE), could be performed in capillaries with large inner diameters (typically several hundred micrometers) in a stable manner.[17–19] The 1980s saw a decline in ITP's popularity due to the wide availability of high quality capillaries with small inner diameters (order tens of micrometers), the easier design of CE buffers, and the high separation performance of CE.[20,21] The early 1990s saw a revival of ITP, but primarily as a tool for preconcentration of analytes and hence as a method of improving sensitivity of other separation methods, most notably the sensitivity of various CE separation modalities. The earliest form of this type of implementation involved two-stage, on-line coupling of ITP and CE.[22] Later, in 1993, this approach was more conveniently implemented via column-coupling of transient ITP and electrophoresis (mostly zone electrophoresis).[23] Refer to Bahga and Santiago[24] for a review of methods for coupling CE and ITP.

The late 1990s saw the first implementations of ITP in a microfluidic format, beginning with the work of Walker et al. in 1998.[25] As with many other bioassays, microfluidics offers ITP smaller channel volumes and lower reagent use. Microfluidics also provides ease of optical access, and the ability to create on-chip networks which can be accessed and controlled via end-channel reservoirs. Moreover, microfluidics offers ITP on-chip electric field control useful in initiating and terminating electrokinetic processes, switching electric field direction among intersections and bifurcations within on-chip networks, and, in the case of glass or fused silica, a relatively efficient heat sink to mitigate the effects of Joule heating resulting from ITP. Consistent with and following the successes of capillary ITP-CE systems, initial applications of microfluidic ITP in late 1990s and early 2000s combined on-chip transient ITP preconcentration with zone electrophoresis.[26–30] Later in the 2000s, other on-chip applications for ITP were developed including sample purification and preconcentration,[31,32] focusing and separation,[33–37] and the control and acceleration of biochemical reactions.[38,39] Over the last two decades, ITP has enjoyed increased adoption and growing interest in microfluidic formats due to its ease of adaptability and compatibility with miniaturized devices. In **Figure 2**, we summarize the growth of microfluidic ITP studies. Currently, there are around 200 papers published per year which describe developments or applications of microfluidic isotachophoresis across various disciplines including chemistry, engineering, molecular biology, materials and environmental sciences.

The theory and physicochemical principles of ITP have been addressed in a classic book and several book chapters. Perhaps most influential of these reviews of theory is the book by Everaerts et al.[1] titled "Isotachophoresis". Everaerts et al. presented models of the ITP ion mobility and migration dynamics. Their descriptions are confined mostly to simple algebraic relations for coupling of pH equilibrium, electroneutrality, mass balances, and current conservation (at steady state), and they discussed only qualitatively unsteady dynamics including plateau zone development and growth. Coxon and Binder[16] presented 1D conservation (partial differential) equations, leading to analytical solutions (including steady interface distributions) and a confirmation of the Kohlrausch regulating function, but unfortunately their work is limited to strong electrolytes. In the chapter "Analytical Isotachophoresis", Boček[2] presents formulations for dynamics of strong electrolytes and only summarizes some classic results of weak electrolyte plateau concentrations by Svensson[40] and Dismukes and Alberty[41]. Boček covers the analytical solution for the steady-state distribution at the interface of two neighboring plateau co-ions.[2] Boček also presents formulations for separation capacity and expert discussions of practical aspects including detection of plateaus with temperature, the mitigation of Joule heating, counterflow ITP, and separation assays using plateau mode ITP. Subsequently, Krivankova et al.[42] published a book chapter covering ITP which offers very good and practical advice for buffer and experimental setup designs (including column coupling) and interesting application examples, but very little by the way of theoretical development (aside from stating the so-called ITP condition). Despite all of these texts, we know of no summary of ITP physics which both summarize the derivations of the Alberty and Jovin functions and formulates the dynamics of ITP processes. The current review also uniquely includes a summary of unsteady ITP ion concentration fields starting from first principles (including species and current conservations, the electroneutrality approximation, and chemical equilibrium). The current review also uniquely covers details of peak mode ITP, including accumulation rates and peak distributions.

In accordance with the very strong interest in ITP, a good number of review articles addressing various aspects of ITP have been published over the last 20 years.[43–58] The majority of these are articles on ITP were published periodically and incrementally covering developments in instrumentation, experimental techniques, modeling, and simulation for both capillary and microfluidic ITP systems. The latter reviews typically cover progress over a few years at a time (e.g., most commonly, periods of two years). To our knowledge, the first article which exclusively reviewed microfluidic ITP was the work of Chen et al.[26] in 2006. The latter work largely focused on the developments in microchip-based technologies between 1998-2006 for ITP-based analysis and pretreatment of biomolecules and ionic compounds. We know of only one other such broad review article on microfluidic ITP, which was published by Smejkal et al.[19] in 2013. The latter article primarily discussed applications of microfluidic ITP and placed scant emphasis on discussions of theoretical principles, analysis, or assay design. The field of microfluidic ITP has significantly evolved since the article of Smejkal et al., with many new developments both in the theoretical aspects and fundamental understanding of ITP (for e.g., sample zone dynamics, dispersion in peak-mode ITP, and ITP-aided reactions),[38,39,59–62] as well as new applications (e.g., single-cell analyses, ITP-aided reactions).[38,63,64] Several topical reviews have focused on various sub-fields of microfluidic ITP over the last decade. For example, in 2013, Bahga and Santiago[24] reviewed in detail the coupling of microfluidic ITP with zone electrophoresis. The application of ITP to nucleic acid sample preparation including purification and preconcentration was reviewed by Rogacs et al.[31] in 2014, and later in 2017 by Datinská et al.[65]. In 2018, Eid and Santiago[38] reviewed applications of ITP to biomolecular reactions. Recently, in 2020, Khnouf and Han[66] reviewed challenges and opportunities for ITP-based immunoassays. Despite several such examples, no review article broadly covers all aspects of microfluidic ITP systems including theory, applications, experimental tools, detection methods, practical considerations for system design, and limitations. For example, we know of no succinct and rigorous (in any format including review papers or textbooks) which attempts to present a review of the physicochemical fundamentals of ITP dynamics starting from first principles. In fact, we believe the lack of any such quantitative review of the physics and chemistry has inhibited adoption and spread of ITP as a technique in any format (traditional or microfluidic). We therefore believe that such a review is timely and can have potential to strongly influence the rapidly growing research field of microfluidic ITP.

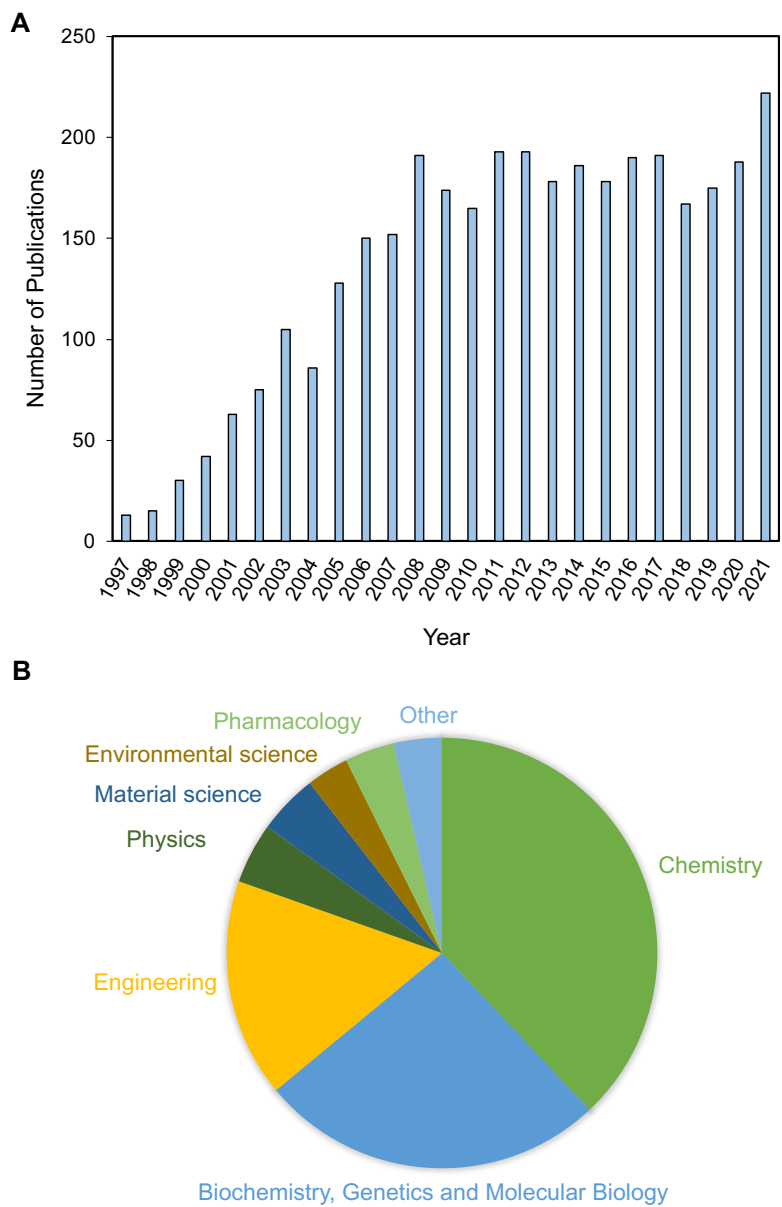

**Figure 2**. (A) Number of publications related to microfluidic isotachophoresis published per year between 1997 and 2021. Data was obtained from Scopus using the query: ALL(("microfluidic" OR "microchip" OR "microfluidics") AND ("isotachophoresis" OR "isotachophoretic")). (B) Distribution of subject areas of microfluidic ITP-related publications. Data was obtained from Scopus using the same query as in (A).

### 1.3 Overview of this review

We here provide an overall review of microfluidic ITP systems with a strong focus on the description of fundamental physicochemistry of ITP starting from first principles. We combine this fairly general introduction to the theory of ITP with a review of emerging applications of, specifically, microfluidic ITP systems. Accordingly, most applications reviewed in this work are from studies published within the last two decades, and we particularly emphasize microfluidic

ITP applications involving biological analyses. In **Section 2**, we begin by reviewing key conservation principles and terminology useful in analyzing ITP systems. We also provide a brief review of chemical buffers and general electrophoresis principles for weak electrolyte systems including the concept of effective mobility, and then discuss configurations of sample loading strategies and ITP modes (peak vs. plateau). Next, in **Section 3**, we review the theory for ITP processes for strong electrolytes, including derivations of the Kohlrausch regulating function and analytical expressions for the concentration of focused sample and LE-to-TE interface width. Later in **Section 4**, we review the theory for ITP of weak electrolytes, including derivations of Jovin and Alberty regulating functions. Later, we discuss the theory of the identification of trace analytes in peak mode ITP in **Section 5**, and for separation process in plateau mode ITP in **Section 6**. As part of **Sections 5** and **6**, we review the theory for estimation of analyte accumulation rates, zone lengths for plateau mode, and sample zone dynamics for peak mode. In **Section 7**, we review the theory and various models for systems in which ITP is used to initiate, control, and accelerate both homogeneous and heterogeneous chemical reactions. We subsequently review several practical considerations and limitations for microfluidic ITP systems in **Section 8** including dispersion, Joule heating effects, buffering, separation capacity, operation method, sample volume vs sensitivity tradeoff, and channel materials. In **Section 9**, we summarize publicly available simulation tools useful for modeling and studying ITP systems. In **Section 10**, we review various experimental tools and methods for analyte detection compatible with microfluidic ITP and provide scaling arguments around the sensitivity and resolution for detection of plateau zones. In **Section 11**, we provide an overview of several types of systems that leverage ITP for trace analyte detection and separations, including bioassay systems which involve nucleic acids, proteins, and single cell analyses. Then, in **Section 12**, we provide a brief overview of miscellaneous techniques used in microfluidic ITP which include coupling of ITP preconcentration with zone electrophoresis, cascade ITP, counterflow and gradient elution ITP, and free-flow ITP. Lastly, in **Section 13**, we conclude with remarks on possible future research directions for microfluidic ITP.

## 2. Basic concepts and terminology

In this section, we first define the mobility of an ion under an applied electric field and briefly discuss basic terminology and notation we use for descriptions of such ions and ion families. Refer to Supplementary Information Table S1 for a detailed nomenclature list including variable names, brief descriptions, and units. Our discussions in this section involve both strong and weak electrolytes, which respectively refer to fully ionized and partially ionized species. We review steady and unsteady transport phenomena starting from physicochemical first principles including conservation of species, conservation of current, and the approximation of charge neutrality. These concepts are generally applicable to all electrophoresis systems, and more specifically, are useful in describing systems over the relevant length and time scales involved in ITP. To aid our understanding, we also provide here a qualitative introduction to weak electrolyte ITP. We then review briefly pH equilibrium and electrophoresis of weak electrolytes, including a discussion around total concentration and effective mobility of a weak electrolyte. We then describe two useful approximations of safe and moderate pH conditions, which we will use in subsequent sections to simplify ITP analyses. We conclude this section with qualitative descriptions of peak vs. plateau mode ITP, and finite vs. semi-infinite sample injection.

**2.1 Ion mobility and ion families**

Our discussions of ITP will require descriptions of multi-species systems which change in space and time, and so we begin by describing our notation for ion mobility and concentration and present some basic physicochemical relations. Our notation is designed to simplify descriptions of ITP systems, including weak electrolytes (which require additional specification of mobilities and local ionization state; see below). Firstly, we describe our definition and the associated dimensions of ion mobility (a.k.a. electrophoretic mobility). Our basic definition of ion mobility shall be simply the ratio of the ion drift velocity divided by the local electric field. The drift velocity will, of course, be measured as the ion velocity relative to that of the local continuum fluid velocity of the aqueous solvent. The dimensions of this mobility are then the square of a length per unit electric potential and per time. In the SI system, the units of our mobility definition will be $[m^2 V^{-1} s^{-1}]$. We shall use the symbol $\mu$ for mobility and use subscripts to distinguish among ion types and properties. For example, the relation among mobility of ion $i$ with (integer-valued) valence $z$, its drift velocity vector $\bar{u}_{i,z}$ and local applied electric field $\bar{E}$ is $\bar{u}_{i,z} = \mu_{i,z} \bar{E}$.

In ITP systems, there is often some finite region in space where the chemistry is locally uniform. The most common of example of this is the LE zone but locally uniform properties could also exist within trailing and plateau zones.

First, the quantities $\mu_{A,z}^X$ and $c_{A,z}^X$ refer respectively to the fully dissociated electrophoretic mobility (a signed quantity) and the concentration of the species with a specific valence $z$ and which is a member of some ion family $A$. Here, the ion family is defined as all species within a particular chemical group with the ability to donate or accept protons. For example, $A$ can refer to all ionic forms of the aqueous phosphoric acid "family": $PO_4^{3-}$, $HPO_4^{2-}$, $H_2PO_4^-$, or $H_3PO_4$, which have respective ionization states $z$ equal to -3, -2, -1, and 0 (c.f. **Section 2.5**). The superscript $X$ refers to the zone of interest (i.e. a location in space, at a specific time). Examples of zones include the leading electrolyte zone, trailing electrolyte zone, or some generic plateau zone. We next describe our notation for strong and weak electrolytes. Recall that a strong electrolyte refers to a solution where all the solute species are fully ionized (e.g., NaCl), while a weak electrolyte refers to a solution where the solute species are only partially ionized (c.f., **Section 2.5**). For strong electrolytes, we omit the superscript in mobility since the mobility of a fully ionized (electrolyte is independent of the zone. The presence of a superscript in mobility refers to the description of a weak electrolyte, whose mobility can vary depending on the degree of dissociation and zone of interest. Note that the superscript in mobility is redundant while describing fully ionized electrolytes. Superscripts can always be used for concentration since concentrations in ITP vary in space and time (c.f. **Section 2.2**). Next, since the mobility of weak electrolyte ions are functions of local pH and the various ionization states, we need to describe an "effective" mobility for such a species family which accounts for all the relevant ionization states of $A$ (see formal discussion in **Section 2.6**). We will denote the effective mobility of $A$ with an overbar as $\bar{\mu}_A^X$. Note this overbar can never be confused with vector notation as mobility is always a scalar. Likewise, we use the notation $c_A^X$ (i.e., lack of second subscript) to denote the total concentration (a.k.a. analytical concentration) of $A$ across all ionization states. For example, the total concentration of the phosphoric acid family described earlier is $c_A = c_{A,-3} + c_{A,-1} + c_{A,-2} + c_{A,0}$. Concentrations of strong electrolytes, of course, require only one specified or implied (e.g., unwritten) subscript (since there is only one relevant ionization state).

**2.2 Basic conservation and transport relations**

We here introduce basic conservation principles including ion transport and current. For simplicity, we will assume ionic solutions sufficiently dilute to apply the well-known Nernst-Planck[67] for the vector flux $\bar{J}$ (units of mol m$^{-2}$ s$^{-1}$) of ion $i$ with valence $z$ :

$$\bar{J} = -D_{i,z}\nabla c_{i,z} + c_{i,z}\mu_{i,z}\bar{E} + c_{i,z}\bar{u}_b. \tag{1}$$

Here, $c_{i,z}$ is the ion's concentration (in molar density units), $D_{i,z}$ is a molecular diffusivity, and $\bar{u}_b$ is the bulk fluid (i.e., solvent) velocity. Note further that the mobility $\mu_{i,z}$ can be related to diffusivity according to the Nernst-Einstein relation[68] given by $\mu_{i,z} = zD_{i,z}F/RT$, where $R$ is the universal gas constant, $F$ is the Faraday's constant, and $T$ is the absolute temperature. From the relation in Eq. (1), we can derive a general relation for the conservation of species over a differential control volume of the fluid. This derivation can be found in several classic references, and, for interested readers, we here recommend the textbooks of Probstein[67] and Deen[69]. The resulting conservation equation is

$$\frac{\partial c_{i,z}}{\partial t} + \bar{u}_b \cdot \nabla c_{i,z} = \nabla \cdot \left(D_{i,z}\nabla c_{i,z} - c_{i,z}\mu_{i,z}\bar{E}\right) + R_{i,z} . \tag{2}$$

$R_{i,z}$ is the production rate and area-averaged concentration of species $i$ (valence $z$) (i.e. due to chemical reactions). For this conservation of a dilute ion, we have assumed that the solvent is an incompressible liquid such that $\nabla \cdot \bar{u}_b = 0$. The latter equation is useful in the study of a large number of electrokinetic systems. For simplicity, and a simple introduction to ITP dynamics, in this section we consider the very simple case of one-dimensional transport and negligible bulk fluid velocity. Under these assumptions, we derive the following

$$\frac{\partial c_{i,z}}{\partial t} = \frac{\partial}{\partial x}\left(D_{i,z}\frac{\partial c_{i,z}}{\partial x} - c_{i,z}\mu_{i,z}E\right) + R_{i,z} . \tag{3}$$

Here, $x$ is the streamwise direction along a channel. We will consider all aspects and terms of this equation in our description of ITP, and so it is worth reviewing this expression and building intuitions for their coupling. For example, we shall typically consider uniform values of diffusivity and mobility for any specific ion. Importantly, the equation nevertheless includes the product $c_{i,z}E$ which makes it non-linear. We shall see that it is this term which can result in shock and rarefaction wave behavior of ion transport. We shall also consider cases where the reaction term is very significant, including cases of chemical equilibrium (c.f. **Section 4**) and unsteady chemical kinetics (c.f. **Section 7**).

We next describe the important concept of current density in multi-ion systems which we will then use (in **Section 2.4**) in our discussion of current conservation. Again, assuming one-dimensional transport and negligible bulk current, the Nernst-Planck flux reduces to $J = -D_{i,z}\frac{\partial c_{i,z}}{\partial x} + c_{i,z}\mu_{i,z}E$. If we multiply this relation by the valence $z$ of each ion type and Faraday's constant $F$ and sum over all ionic species, we naturally derive an expression for current density $j$ as follows

$$j = \sigma E + F \sum_{i=1}^{N}\sum_{z=n_i}^{p_i} zD_{i,z}\frac{\partial c_{i,z}}{\partial x} . \tag{4}$$

From this summation, we see that we inherently formulate the Ohmic conductivity, $\sigma$, of these mixture as follows:

$$\sigma = F \sum_{i=1}^{N}\sum_{z=n_i}^{p_i} z\mu_{i,z}c_{i,z} . \tag{5}$$

Here, the double summation implies summations across the various ionization states within each family and across all species families ($i = 1$ to $N$). The bounds $n_i$ and $p_i$ are respectively the most negative and most positive valence state within each species family. For example, for an ampholyte which can, in some relevant pH range, acquire valence states -1, 0, and 1, $n_i$ and $p_i$ are respectively -1 and 1. Importantly, the expression for current density demonstrates an important consequence of the Nernst-Planck relations above. Namely, we see that current can be transported not just by ion mobility (Ohmic-type transport) but by diffusion.[67] Note that neglecting of the bulk velocity in formulating current is typically a good approximation for typical electrokinetic systems where ionic strength is sufficiently such that charge relaxation times are much shorter than characteristic advection times.[70,71]

In subsequent sections, we will discuss in more detail applications of conservation of species and current. For now, we note simply that, away from regions of sharp interfaces (i.e., high diffusion gradients), such as the interface between ITP zones, we will be able to assume that the current density is given simply by the Ohmic current, $j \approx \sigma E$. An important consequence of this is that regions of low conductivity necessarily imply a high local electric field and vice versa. This approximation provides additional insight to the qualitative discussion of **Section 1.1** above.

We conclude this section by listing the results for species conservation and current for the simpler case of strong electrolytes (i.e., species with uniform and constant valences and no chemical reactions). For this case, our species conservation reduces to

$$\frac{\partial c_i}{\partial t} = \frac{\partial}{\partial x}\left(D_i \frac{\partial c_i}{\partial x} - c_i \mu_i E\right). \tag{6}$$

Despite the severe simplifications, we see that the aforementioned non-linearity remains, and this non-linearity is an essential feature of ITP. Lastly, the expression for current density $j$ summed over of $i = 1$ to $N$ strong electrolytes is then

$$j = \sigma E + F \sum_{i=1}^{N} z_i D_i \frac{\partial c_i}{\partial x}, \tag{7}$$

where ionic conductivity $\sigma$ (from Eq. (5)) is simply $\sigma = F \sum_{i=1}^{N} z_i \mu_i c_i$.

## 2.3 Qualitative description of weak electrolyte ITP

Unlike most electrophoretic methods, the conditions leading to ITP are created using a discontinuous electrolyte system consisting of a minimum of two electrolyte solutions.[1,15,72,73] Likely the first choice in any ITP process is whether the user wishes to focus anions or cations. Hence, the simplest ITP system is a system with three species: a single LE co-ion, a single TE co-ion, and a single (counter-migrating) counterion. In anionic (cationic) ITP, the TE and LE co-ions are anions (cations) and the counterion is a cation (anion). In the case of strong electrolytes, the LE, sample, and TE ions have the highest, intermediate, and lowest magnitudes of mobility. (Note that we are careful to specify "mobility magnitude" since, for anionic ITP, the LE co-ion is negative and hence a lower mobility).

As we shall discuss, the most interesting applications of ITP are weak electrolyte systems for the simple reason that such mixtures enable strong pH buffering. Consider that protein mobility, function, and solubility are each a strong function of pH. For example, proteins are often focused and/or separated using cationic ITP since a large number of proteins have a positive charge in electrolytes buffered near pH 6-8 (i.e. relatively high isoelectric points, p*I*'s).[74–77] Anionic ITP is typically the most useful mode for assays involving focusing and separation of nucleic acids

(NAs as in DNA and/or RNA) and negatively charged proteins (i.e. proteins with relatively high p$I$ values).[32,78–80] NAs typically remain negatively charged over a broad range of pH (with a p$K_a$ of ~1.5 due to the phosphate backbone) although nucleic acid solutions should be pH-buffered for stability and function.

For weak electrolyte ITP, a general statement about mobilities is more complex since ion mobilities are a function of local ion makeup and stoichiometries (and therefore space and time). However, for weak electrolytes, we can say that ITP occurs if the TE-to-LE interface is stable and self-correcting. That is, if a TE co-ion diffuses into the LE zone, the TE ion should have a mobility less than that of the LE co-ion and will then fall back to the TE. Conversely, if an LE co-ion diffuses into the TE zone, then it should have a mobility greater than the TE zone and should therefore migrate back to the LE. If there is a sample ion, then it will focus somewhere between the TE and LE zones if sample ions have higher and lower mobility magnitude than the respective co-ions in the TE and LE zones.

Once the initial TE-to-LE interface is established, an electric field is applied to initiate electromigration and ITP. This electric field is applied typically using either a constant current or constant voltage source (c.f. **Section 3.1**). The electric field direction is necessarily directed through this interface from the LE to the TE for anionic ITP (and from TE to LE for cationic ITP). This orientation will result in a self-steepening ion concentration shock wave which for anionic (cationic) ITP propagates toward the positive (negative) electrode. Note that the dynamics of weak electrolyte ITP are such that the LE counter-ion continuously migrates into the TE zone. As it does so, the concentration at which it enters, and the acid dissociation constants of this counter-ion and the TE co-ion necessarily drive the pH equilibrium in the TE zone. We shall discuss these dynamics and the idea of a well-buffered ITP process in **Section 8.4**.

**2.4 Charge neutrality approximation and divergence of current**

We here introduce the powerful approximation of electroneutrality (a.k.a. charge neutrality) for electrokinetic systems. In ITP, we shall use electroneutrality to formulate the dynamics which lead to equal electromigration velocities of the LE and TE zones (c.f. **Section 3.1**). Persat and Santiago[81] describe scaling arguments originating in Gauss law for diffuse ion systems which lead to the electroneutrality assumption as well as the so-called Ohmic model for electrokinetics (see also Refs.[82,83]). We here present a similar scaling argument but in the context of ITP. We consider a region subject to some electric field that is imposed by electrodes placed far from this region (e.g., at end-channel reservoirs). These electrodes have sufficient potential difference to sustain Faraday reactions and hence force a net ionic current through the system including our region of interest. The region has approximately uniform permittivity, $\varepsilon$ (e.g., dominated by the polarizability of water under approximately isothermal conditions), but has significant gradients in species concentration (and therefore ionic conductivity as per Eq. (5)). The general result is a spatially varying electric field $\bar{E}$ within this region. The differential form of Gauss law then relates the gradients of this electric field to any net charge density as follows:

$$\varepsilon \nabla \cdot \bar{E} = \sum_{i=1}^{N} \sum_{z=n_i}^{p_i} F z c_{i,z} . \qquad (8)$$

The right-hand side of this equation is the net charge density, $\rho_E$, here expressed in terms of ion concentrations for some arbitrary mixture of weak electrolytes (summed over $i = 1$ to $N$ species families, each of which valence states between $n_i$ and $p_i$). $F$ is the Faraday's constant. Without loss of generality, we decompose the local electric field into external and internal components as follows: $\bar{E} = \bar{E}_{ext} + \bar{E}_{int}$. We define $\bar{E}_{ext}$ as the nominal, applied electric field which would result from the electrodes if the species concentrations and properties were uniform, so that $\nabla \cdot \bar{E}_{ext} = 0$. Hence, $\bar{E}_{int}$ is the electric field component strictly resulting from the net charge and associated species gradients. **Figure 3** depicts the situation in the context of a one-dimensional treatment of an ITP interface. We analyze species gradients over some diffuse interface of length scale $\delta$. Sufficiently far from the interface, regions L and T have locally uniform species concentrations and ionic conductivities $\sigma^L$ and $\sigma^T$, and hence have locally uniform current densities $\sigma^L E^L$ and $\sigma^T E^T$ (c.f. Eq. (4)). Here, $E^L$ and $E^T$ are the local electric fields in the L and T regions, respectively. We will discuss current conservation later in this section and show that, to a high degree of approximation, $\sigma^L E^L \cong \sigma^T E^T$, even for typical unsteady ITP processes. Hence, under quasi-steady conditions, conservation of current demands that $E^L$ and $E^T$ be different and hence there must be (from Gauss' law) net charge within the interface. As we discuss below, the net charge within the interface and the sharp gradient in electric field are associated with an internally generated electric field $\bar{E}_{int}$, which is directed away from the interface.

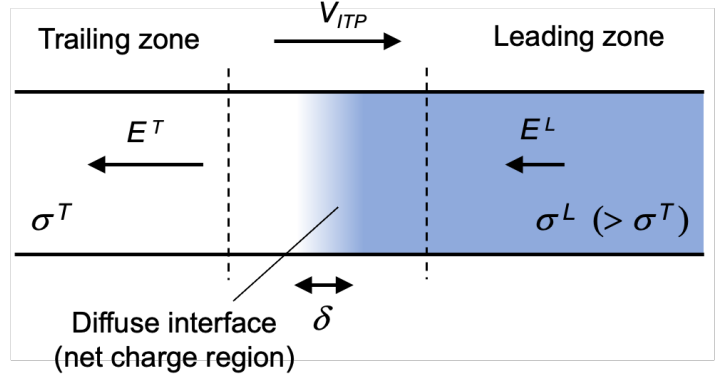

**Figure 3**: Simple, one-dimensional treatment of an ITP interface. The diffuse interface between the two region has a length scale $\delta$. The regions away from the interface have locally uniform conductivities $\sigma^L$ and $\sigma^T$. Conservation of current requires a sharp gradient in electric field and therefore local net charge.

Given this situation of a one-dimensional, monotonic gradient in conductivity, we can bound the scale of $E_{int}$. In ITP, typical values of the high-to-low conductivity ratio $\sigma^L/\sigma^T$ are roughly a factor 2 or less. Hence, we expect that, at most, the electric field between the regions changes by a factor of 2 or less, so that $E_{int}$ is on the order of $E_{ext}$. Even for some extreme case where $\sigma^L/\sigma^T$ is order 20, $E_{int}$ will remain order $E_{ext}$ so long as the region occupying the lower conductivity has an axial length that is on order of the total length between electrodes. We can now proceed with our scaling analysis. We scale the left-hand side of Eq. (8) as $\varepsilon E_{int}/\delta$ (where $E_{int} \approx E_{ext}$). For the purpose of scaling, we also simplify the right-hand side for the case of strong monovalent electrolytes and obtain

$$\frac{\varepsilon E_{int}}{\delta F} = \sum_{i=1}^{N_{cat}} c_i - \sum_{i=1}^{N_{ani}} c_i, \tag{9}$$

where $N_{cat}$ and $N_{ani}$ are respectively the local number of (monovalent) cationic and anionic species (and $N = N_{cat} + N_{ani}$). Next, define the dimensional parameters $c_+$ and $c_-$ as the (local) characteristic sum of cations and anions, respectively. Also, define $c_o$ as the characteristic sum of concentration of all (monovalent) ionic species, $c_o = \sum_{i=1}^{N} c_i$ and write

$$\frac{\varepsilon E_{int}}{\delta F c_o} = \frac{(c_+ - c_-)}{c_o} \equiv \alpha . \tag{10}$$

We thereby define the parameter $\alpha$ as a measure of the characteristic difference in concentration between positive and negative charges relative to the characteristic background concentration of all ionic species. We can estimate $\alpha$ for typical ITP systems as follows. Empirically we observe that ITP systems exhibit interfaces with minimum diffusion-limited lengths of roughly 10 µm. Externally applied fields are at most order $10^4$ V/m (and hence $E_{int}$ is order $10^4$ V/m). Due to pH buffering considerations (c.f. **Section 8.4**), minimum practical values of $c_o$ are order 1 mM. Substituting these dimensional scales into Eq. (10), we see for aqueous solutions that $\alpha$ is a maximum of roughly $10^{-5}$ (see also[81]). That is, a very small mismatch between the concentration of cations and anions is enough to account for internally generated electric fields in ITP. This tiny mismatch is very important in conserving electric flux but, for the purposes of conserving species and applying Eq. (3), we assume the system is approximately net neutral. Hence, at any position and time, the concentrations of negative and positive charges are assumed to be equal for the purpose of computing the transport of local ion concentrations (via conservation of species relations). This is the essence of the electroneutrality approximation.

We shall leverage the net neutrality approximation in our formulations of both ion transport and chemical equilibrium. Basically, we will assume (to a high degree of accuracy), that, any position and time, the concentrations of negative and positive charges are equal when computing local ion concentrations via transport principles and expressions of chemical equilibria. For example, for a univalent, three-ion case (TE co-ion, LE co-ion, and a common counterion) described in **Section 2.3**, charge neutrality in the adjusted TE region implies that $c_C^{T'} + c_H = c_T^{T'} + c_{OH}$.

A second concept arising from the electroneutrality is the so-called Ohmic model of electrokinetics[81] and how we compute electric fields in the system. A rigorous and exact computation of electric fields in general requires a careful computation of the unsteady net charge fields in the system (particularly at conductivity interfaces) via careful balances of Gauss law in addition to all species conservations. However, for most electrokinetic systems, we can finesse the electric field calculation with a different approach. Namely, we can assume quasi-steady charge accumulation[81,84] (a.k.a. relaxed charge approximation) and assume electroneutrality whenever we compute species conservation. We then compute electric fields from the conservation of current. In ITP, this approach is especially useful in obtaining closed-form solutions for ion concentrations at points within plateau regions and/or in control volume analyses of ITP ion fluxes. A fully three-dimensional version of Eq. (7) yields a more general description of current density flux for multispecies, weak electrolyte electrokinetic systems as follows:

$$\bar{j} = \sigma \bar{E} + F \sum_{i=1}^{N} \sum_{z=z_1}^{z_n} z D_{i,z} \nabla c_{i,z} . \tag{11}$$

Using this expression of current flux, the conservation of net charge density $\rho_E$ over a differential element yields[67]

$$\frac{\partial \rho_E}{\partial t} + \nabla \cdot \bar{j} = 0 . \tag{12}$$

We next apply the approximation of relaxed charge which assumes the time-scale for accumulation of net charge (as in the net charge region in the diffuse interface of the example of **Figure 3** above) is much smaller than the time-scale of interest.[81,84] In electrokinetic systems, the latter time scales are typically milliseconds or less. For this regime, we have simply
$$\nabla \cdot \bar{j} = 0 . \tag{13}$$
This equation can be interpreted as a form of Kirchoff's law but is in fact a more general expression since it includes contributions of diffusive fluxes to current transport (c.f. Eq. (4)). It is valid for unsteady processes whose characteristic time-scales are significantly larger than the charge relaxation time scale of the system. A useful form of this equation can be derived by simply integrating over a finite volume and applying the Divergence Theorem to obtain
$$\int_{CS} \bar{j} \cdot d\bar{A} = 0 , \tag{14}$$
where CS refers to a (closed) integral over a control surface. This three-dimensional form is useful as it can be applied over complex volumes spanning intersections among an arbitrary number of channels (e.g., as in microfluidic systems). For control volumes whose control surfaces are drawn far from sharp concentration gradients (e.g., away from an LE-TE interface) and/or at the wall/liquid interface, this expression of the conservation of current reduces to a traditional Kirchoff's law expressed in terms of Ohmic current as follows:
$$\int \sigma \bar{E} \cdot d\bar{A} \cong 0 . \tag{15}$$

We saw an example of this concept in the scaling argument presented at the beginning of this section. In that example, our control volume yielded $\sigma^L E^L \cong \sigma^T E^T$.

The net neutrality approximation will also be useful in analyses of both chemical equilibrium and transport phenomena associated with ITP. We conclude this section with the following example formulation of the net neutrality approximation for a weak electrolyte mixture:
$$\sum_{i=1}^{N} \sum_{z=n_i}^{p_i} z c_{i,z} + c_H - c_{OH} = 0 . \tag{16}$$
Here $c_H$ and $c_{OH}$ are the concentration of the hydronium and hydroxyl ions, respectively. In this way, we will treat these ions (associated with the autoprotolysis of water) separately from the species families of interest. Note that protons (or more exactly hydronium ions) and hydroxyl ions are always of critical importance in weak electrolyte systems due to the strong pH-dependence of species mobility.

## 2.5 Brief review of pH buffers

Before we continue with our discussion of ITP of weak electrolyte systems, it is instructive to briefly review the concept of a chemical buffer. A detailed review of buffers and the electrophoretic transport of weak electrolytes can be found in Refs.[85,86]. We shall assume here a working knowledge of these topics and adapt the Lowry-Bronsted definition of acids and bases (as, respectively, proton donors and acceptors). We will emphasize the case of anionic ITP systems wherein the TE and LE co-ions are (strong or weak acid) anions and the common counter ion is a weak base used to provide pH buffering for the ITP system.

Consider a buffer created using a singly ionized weak base and a strong acid, such as HCl. The (proton exchange) chemical equilibrium reactions are:
$$\begin{aligned} BH^+ &\rightleftharpoons B + H^+ \\ HCl &\rightarrow H^+ + Cl^- \end{aligned} \tag{17}$$

$$H_2O \rightleftharpoons H^+ + OH^-$$

Here we use "B" to denote a generic Bronsted-Lowry weak base, that is, a proton acceptor. To achieve pH buffering in this mixture, the weak base is obviously the "buffering species" and the chloride ion (here, from the strong acid) is the "titrant". Note we present reactions with the proton species on the right-hand side as this facilitates bookkeeping among significant number of species and coupling of simulations with large data bases of weak acid species properties.[85–87] The latter arrangement of species in the reactions in Eq. (17) also lets us specify all dissociation constants, $K$, as the appropriate acid dissociation constants, $K_a$. Henceforth, we will drop the "$a$" in the subscript of $K_a$ where appropriate with the understanding that $K$ refers to the acid dissociation constants. We instead use the subscripts of $K$ to indicate the species family and the ionization state, respectively. The two equilibrium reactions and mass conservation relations for Eq. (17) are then

$$K_{B,0} = \frac{c_H c_{B,0}}{c_{B,1}}, \quad K_w = c_H c_{OH} \tag{18}$$

$$c_{B,tot} = c_{B,0} + c_{B,1}, \quad c_{HCl,tot} = c_{Cl}$$

where $c_{B,tot}$ and $c_{B,tot}$ are respectively the total amounts of (generic) base B and strong acid HCl initially added to the mixture. The definition of the other concentration variables follows from our discussions of **Section 2** (i.e., $c_{B,0}$ refers to the concentration of the weak base $B$ and $c_{B,1}$ is for the corresponding conjugate acid, $BH^+$ ). From charge neutrality, we have

$$c_{Cl} + c_{OH} = c_{B,1} + c_H \tag{19}$$

Though it is possible to solve these equations as is (by solving a parabolic equation with the concentration of hydronium as the root), assuming moderate pH (i.e. anticipating a resulting pH between about 4 and 10, see **Section 2.7**) is a useful simplification wherein we rewrite Eq. (19) as $c_{B,1} \approx c_{Cl} = c_{HCl,tot}$. Combining these equations, we derive

$$c_H = \frac{K_{B,0} c_{HCl,tot}}{c_{B,tot} - c_{HCl,tot}}, \tag{20}$$

whence

$$\text{pH} - \text{p}K_{B,0} = \log_{10}\left(\frac{c_{B,tot} - c_{Cl,tot}}{c_{Cl,tot}}\right). \tag{21}$$

Here, $\text{pH} = -\log_{10} H^+$, $\text{p}K_{B,0} = -\log_{10} K_{B,0}$, and $c_{HCl,tot} = c_{Cl,tot}$. For a well-designed, classic buffer the pH is near the p$K_a$ of the weak electrolyte. The strongest buffering capacity condition is achieved when the $\text{pH} = \text{p}K_{B,0}$, or equivalently, $c_H = K_{B,0}$. Further, $\text{pH} = \text{p}K_{B,0}$ implies $c_{BH^+,tot} = 2\, c_{HCl,tot}$. We see the strongest buffering capacity where buffering species half dissociated by the ionized concentration of the (typically fully ionized) titrant.

Similar analyses can be performed for a variety of cases including fairly arbitrary mixtures of weak and strong acids and weak and strong bases and/or salts. Refer to Persat et al.[85] for details.

## 2.6 Electrophoresis of weak electrolytes

We review concepts around the electromigration of mixtures comprising weak electrolytes. In such systems, typical acid/base chemical equilibrium reaction kinetics occur over a much smaller time scale than characteristic advection and diffusion time scales.[88] Hence, we will here assume each species is in chemical equilibrium at all times. See Refs.[89,90] for examples where finite reaction kinetics may be important in ITP.

### *2.6.1 Total concentration, species conservation, and effective mobility*

For weak electrolyte solutions, the conservation of species requires that the sum total of members of each chemical "family" be conserved. To show this, we define this sum across members of a chemical family as the total (a.k.a. analytical) concentration of "family" $i$

$$c_i = \sum_{z=n_i}^{p_i} c_{i,z}. \tag{22}$$

The total concentration of buffer species is typically a known quantity. For example, this quantity may be known by weighing some amount of a weak base stock-supply powder. Alternately, the quantity is known from a dilution of stock solution of the electrolyte. Further, since the various members of a single family can only accept or donate protons, the sum of various production rates in the species conservation equation (Eq. (3)) across the family, $\sum_{z=n_i}^{p_i} R_{i,z}$, is identically zero. Summing the weak electrolyte conservation equation across the members of a single family, we can eliminate the production term and obtain a simplified conservation equation for the total concentration $c_i$ as

$$\frac{\partial c_i}{\partial t} = \frac{\partial}{\partial x}\left(\sum_{z=n_i}^{p_i} D_{i,z}\frac{\partial c_{i,z}}{\partial x} - E\sum_{z=n_i}^{p_i} \mu_{i,z} c_{i,z}\right). \tag{23}$$

Further, for simplicity, assuming that all the members of family $i$ have similar diffusivity $D_i$, we have can write

$$\frac{\partial c_i}{\partial t} = \frac{\partial}{\partial x}\left(D_i \frac{\partial c_i}{\partial x} - E c_i \bar{\mu}_i\right). \tag{24}$$

The resulting relation for the net transport of each species family has a strong similarity to the conservation equation for fully ionized species. The important difference here (compared to Eq. (6)) is that we have formulated in terms of a total concentration and defined a new quantity $\bar{\mu}_i$ as the *effective mobility* for family $i$. This effective mobility describes the net rate of migration of all members of the family in terms of the absolute mobilities $\mu_{i,z}$ of the individual species of the family weighted by the molar fraction of the species within the family. $\bar{\mu}_i$ is given by[88,91]

$$\bar{\mu}_i = \frac{\sum_{z=n_i}^{p_i} \mu_{i,z} c_{i,z}}{c_i}. \tag{25}$$

Interestingly, the modified conservation equation for the total concentration (Eqs. (23) and (24)) describes the net transport of a group of species which are being created and destroyed by acquiring and donating protons, but the equation contains no explicit reaction term. The physicochemistry of the acid-base reactions that is embedded into the chemical equilibrium determines the various values of species concentrations $c_{i,z}$ and, of course, the concerted interaction of all families determines the local and instantaneous pH.

### *2.6.2 Effective mobility: an example calculation*

To illustrate the procedure to calculate the effective mobility of a species family, we here consider the example of a singly ionized weak acid electrolyte $A$ (e.g., an analyte in anionic ITP). The equilibrium reaction for $A$ is given by $A \rightleftharpoons A^- + H^+$. The associated equilibrium reaction and species conservation relations can be written as

$$K_{A,-1} = \frac{c_H c_{A,-1}}{c_{A,0}}, \quad c_{A,tot} = c_{A,0} + c_{A,-1} \tag{26}$$

where $c_{A,tot}$ is the total (or analytical) concentration.

In the ideal limit when the conjugate base ion $A^-$ is fully dissociated at $z = -1$ (and neglecting the effects of ionic strength on ion mobility and dissociation constants[86]), the electromigrative drift velocity $u_{A^-}$ of the ion $A^-$ can be written in terms of its absolute mobility $\mu_{A,-1}$ as $u_{A^-} = \mu_{A,-1} E$. However, more generally, the observed drift velocity of the species family is described by the effective mobility which accounts for the time-averaged velocity of the species $A$ as it accepts and donates protons. This effective mobility is defined by the relation

$$u_{A^-} = \bar{\mu}_A E, \tag{27}$$

where

$$\bar{\mu}_A = \mu_{A,-1} \frac{c_{A,-1}}{c_{A,-1} + c_{A,0}} = \mu_{A,-1} \frac{1}{1 + c_{A,0}/c_{A,-1}} = \mu_{A,-1} \frac{1}{1 + c_H/K_{A,-1}}, \tag{28}$$

where $K_{A,-1}$ is the acid dissociation constant. The overbar here denotes the average mobility observed for this species family. The species is quickly acquiring and donating a proton, but this process is so fast that we observe only the species' time-averaged mobility.

Consider two example buffer conditions for Eq. (28). First, for a well-designed buffer such that $\text{pH} = \text{p}K_a$, i.e., when $c_H = K_{A,-1}$, the effective mobility $\bar{\mu}_A = 0.5\,\mu_{A,-1}$ which is equal to half of the fully dissociated value. Second, consider a buffer such that $\text{pH} < \text{p}K_{A,-1} + 2$. In this regime, $c_H \gg K_{A,-1}$, so the effective-to-fully-ionized mobility ratio is nearly zero, i.e., $\bar{\mu}_A/\mu_{A,-1} \approx 0$. Thus, the spatiotemporal development of pH plays a crucial role in determining the dynamics of weak electrolyte species in ITP.

Equation (25), of course, applies to any weak electrolyte including multivalent ionizations, weak bases, and ampholytes.[86] For example, an ion family that "hops" (transitions) among a few states (e.g. $z = +1, 0, -1$, and $-2$) would have observable mobilities calculated using the general expression Eq. (25) above.

For a simple weak base electrolyte $B$ described by $BH^+ \rightleftharpoons B + H^+$, the effective mobility is

$$\bar{\mu}_B = \mu_{B,1} \frac{1}{1 + K_{B,0}/c_H}. \tag{29}$$

where $K_{B,0} = c_H c_{B,0}/c_{B,1}$, $c_{B,tot} = c_{B,0} + c_{B,1}$, and $u_{BH^+} = \mu_{B,1} E$.

It is worth noting that the magnitude of free solution mobilities $\mu$ of a large variety of ions (both buffer and analyte ions) applicable in ITP typically vary by roughly a factor of 3 to 4, at most. At the same time, the acid dissociation constants of interest $K_a$ vary by roughly 10 orders of magnitude. Therefore, effective mobility quantity $\bar{\mu}$ (as per from Eq. (25)) varies from 0 to a maximum equal to the fully ionized value of the largest magnitude valence state. The quantity $\bar{\mu}$ is of most importance in designing and analyzing ITP systems since it governs conductivity, electrophoretic mobility, and contribution to local charge. This quantitative is also the most intuitive as it is directly observable in experiments.

## 2.7 The concepts of moderate and safe pH

We here introduce two very useful approximations in descriptions of buffers and electrophoresis of weak electrolytes which we shall use in the analysis of ITP.[85,86] First, is the concept of *moderate pH*.[86] By moderate pH, we will refer to an approximation valid for buffer concentrations of about 10 mM or greater and for a solution pH range between about 4 and 10. In this regime, the concentration of hydronium and hydroxyl ions *can be neglected in determining charge neutrality* in writing the charge neutrality approximation (Eq. (7)). For example, for a buffer composed of some generic weak base $BH^+$ and the strong acid HCl, the charge neutrality relation, under the moderate pH assumption, is

$$0 \cong \sum_{i=1}^{N} \sum_{z=z_1}^{z_n} zc_{i,z} = c_{B,1} + c_{H,1} - c_{Cl,-1} - c_{OH,-1} \cong c_{B,1} - c_{Cl,-1}. \quad (30)$$

Moderate pH will be very important in developing intuitions and closed-form solutions for the ion concentrations expected within plateau ITP zones.

Second, we introduce the concept of *safe* pH[86] Unlike "moderate pH", the term safe pH has appeared explicitly (by that name) in the electronics literature for decades. Safe pH refers to the approximation that hydronium and hydroxyl ions *carry negligible current*. Again, considering the example above of a buffer composed of some generic weak base $BH^+$ and the strong acid HCl, we have

$$\sigma = \sum_{i=1}^{N} \sum_{z=n_i}^{p_i} zF\mu_{i,z}c_{i,z} = F(\mu_{B,1}c_{B,1} + \mu_{H,1}c_{H,1} - \mu_{Cl,-1}c_{Cl,-1} - \mu_{OH,-1}c_{OH,-1}) \quad (31)$$
$$\cong F(\mu_{B,1}c_{B,1} - \mu_{Cl,-1}c_{Cl,-1}).$$

We note that the assumptions of safe pH and moderate pH are not exactly equivalent (see Persat et al.[86]). This is because the mobility of protons and hydroxyls are fairly high relative to many other ions,[92] and so safe pH is in practice more restrictive than moderate pH. However, we can fairly accurately assume both safe and moderate pH for pH values between 4 and 10 and buffer concentrations of order 10 mM and greater (which are typical of most microfluidic ITP applications). We will significantly expand on the relations presented here in **Sections 3** to **7** to derive fundamental principles governing ITP of both strong and weak electrolytes.

## 2.8 Finite versus semi-infinite injection

How does one initiate a simple ITP experiment? In microfluidic devices, likely the simplest way is to fill a single, straight channel between two reservoirs with LE and then replace the contents of one reservoir with the TE buffer (it is a good idea to rinse the reservoir once or twice with deionized water prior to filling with TE). This simple configuration is depicted in **Fig. 1A**. Sample ions included in the TE and/or LE will focus at the TE-to-LE interface as it moves away from the TE. A second way to achieve the initial LE-TE interface is to establish bulk flow (e.g. using pressure-driven flow) of LE and TE streams to and from an intersection within a microchannel network (see Refs.[93,94] for examples).

Given this basic requirement of an initial TE-to-LE interface, there are important differences in the manner in which sample ions are initially introduced (a.k.a. injected) into the system. In this section, we classify two basic forms of sample injection: finite-injection versus semi-infinite injection modes. In finite injection mode, the sample (either raw sample or diluted with either TE or LE) is initially loaded in a finite region within the main channel between two regions containing pure TE and LE mixtures prior to the application of the electric field, as shown in **Fig. 4A**. This configuration be achieved using an intermediate reservoir along the main channel between the TE and LE reservoirs, or by designing branched channels to aid in loading particular sections of the channel using pressure driven flow. In finite injection, the sample can be dissolved in TE (typical case) or LE, or the sample can contain purely its own inherent ions different than the LE or TE. On application of the electric field, TE ions from the TE reservoir electromigrate into the initial sample region, displacing higher mobility co-ions, while the sample ions of interest (with mobilities bracketed by the LE and TE) electromigrate into the former LE region (displacing LE co-ions), as shown in **Figs. 1** and **4**. This scheme is often preferred over plateau mode ITP as it can yield purified analyte zones of constant (in time) concentrations during ITP focusing, and these can be identified based on the analyte physicochemical properties. Moreover, finite injection is more compatible with undiluted complex samples (e.g., blood, serum, urine) where inherent ion densities (including contributions from analytes, impurities, and background ions) are typically on the same order of the LE and TE concentrations. Unlike semi-infinite injection, finite injection results in steady spatial distribution and concentration of the focused analytes, and these distributions are independent of their initial concentrations. Thus, a major advantage of finite injection is that the downstream ion fields are insensitive to sample composition or ionic strength. A drawback of finite injection is the requirement for more complex microfluidic network design and flow control schemes. Also, creation of plateau modes starting from trace ions (e.g., of micromolar concentrations or less) may require processing of large volumes of samples using large channel volumes.

A second approach of introducing sample in ITP is called semi-infinite injection. In semi-infinite injection, the sample is initially mixed with TE buffer (but can also be mixed with LE or both TE and LE) prior to loading on chip and this mixture serves as the effective trailing electrolyte. This is depicted in **Fig. 1A**. For the particular case when the sample is mixed with the TE buffer in the TE reservoir, this injection scheme has been called electrokinetic supercharging or electrokinetic injection.[95,96] Unlike finite injection, upon application of electric field, analyte peaks/zones never reach a steady state distribution. Instead, the concentration of analytes focused in peak-mode increase directly proportionally in time (until plateau mode is reached). After plateau mode is achieved, the length of zones continuously increases linearly in time. This continuous accumulation can be used to improve detection sensitivity for both peak and plateau modes. However, in semi-infinite injection, sample ions from the TE reservoir are never typically fully processed by the ITP process. A larger portion of the analytes in the TE reservoir can be processed by decreasing the conductivity (equivalently, ionic strength) of the TE in the reservoir to achieve higher local electric fields.[39,97] A key advantage of the semi-infinite injection approach is that it is easy to implement and it is compatible with a simple, straight channel geometry (i.e., no branched channels or other complex designs) and one can use a simple pipette to set up the initial TE-LE interface (at the interface between the reservoir and a channel) . A drawback of the semi-infinite injection scheme is that variability in sample conductivities (as is typical of several biological samples) can affect the rate of focusing and quantification of the amount of target analytes. This is because the rate of accumulation depends on the (possibly unknown) ratio of sample concentration to TE ion concentration. Hence, semi-infinite injection is most easily applied if the raw sample can be substantially diluted by the TE mixture. As an example, for anionic ITP of physiological samples and semi-infinite injection, background sodium and chloride ions from the raw sample (which are typically present in high concentrations) can significantly modify TE buffer properties if the sample is not diluted sufficiently.

Lastly, we note that electrode shape and configuration determine electric field lines within the reservoir, and consequently, affect focusing dynamics and amount of sample focused in ITP. For example, Rosenfeld and Bercovici[98] demonstrated that a C-shaped electrode encircling half the circumference of the reservoir resulted in nearly twice the amount of sample focused in ITP compared to a straight electrode. Refer to Refs.[99–101] for further discussions around the importance of electrode configuration for efficient sample injection in ITP.

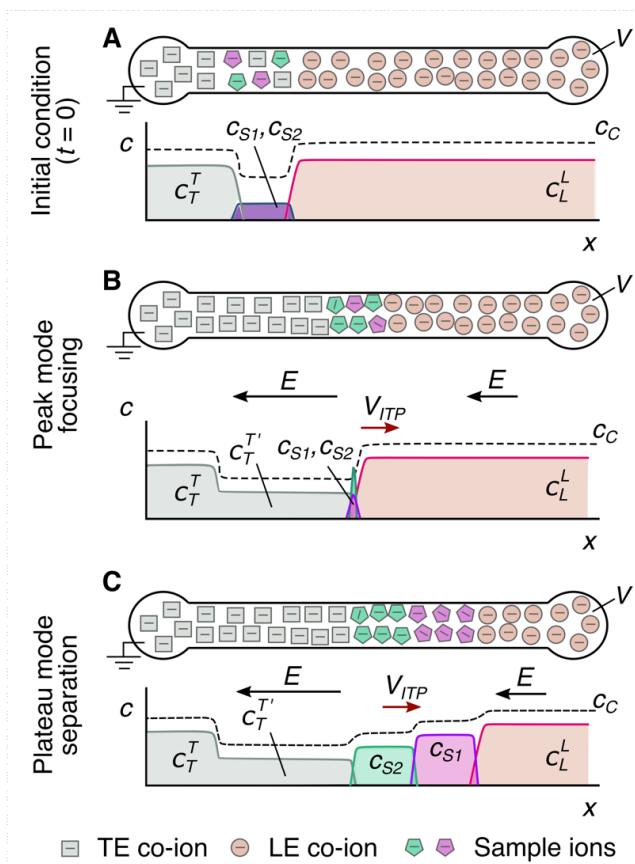

**Figure 4**. Schematic for finite sample injection configuration for microfluidic ITP. (A) Initial placement of LE, TE, and sample, common to both peak and plateau modes ITP. (B) and (C) depict Peak mode and Plateau mode ITP focusing, respectively. Each subfigure shows the location of ions within the channel (top) and the concentration profiles of the electrolytes (bottom) in anionic ITP. For finite injection, when the experiment is sufficiently long, all sample ions focus in ITP, leading to steady state concentration fields of focused ions. This is unlike the case of semi-infinite injection (**Figure 1**) which is associated with continuous sample focusing in ITP (i.e., focused sample amount increases with time).

## 2.9 Qualitative descriptions of peak versus plateau modes of ITP

Peak mode ITP is characterized with relatively low initial sample concentrations (relative to the LE and TE concentrations) and by brief focusing times. Peak mode (which has been referred to as spike mode[102]) ITP between and LE and TE results from a transient process wherein ions continuously focus into approximately overlapping peaks whose concentration are still much lower than that of the neighboring leading and trailing ions. This is shown schematically in **Figs. 1B** and **4B**, where we exaggerate the concentration of peak-mode ions for clarity. If the applied current is constant (in time), the peak (and leading and trailing ions) will all travel at an equal and constant velocity, $V_{ITP}$. Often, the focused ions are several orders of magnitude lower concentration than ions from the neighboring zone and contribute negligibly to the local current. Ions in peak mode are therefore typically only detected directly, as in the case of fluorescent sample ions.[34]

If sample ions are present at higher concentrations and/or given sufficient focusing time (and assuming sample ions are sufficiently soluble in the solvent), peak mode ITP eventually transitions into a state where sample species contribute significantly to local conductivity, and therefore to the local electric field. Such sample species begin to displace neighboring ions and segregate themselves into contiguous, adjoining "plateau" regions/zones of locally uniform concentrations, as shown in **Figs. 1C** and **4C**. We refer to the latter configuration as plateau mode ITP. When the sample is injected in a finite zone between LE and TE zones (e.g., finite injection), species in plateau mode reach a maximum concentration established by the properties of the LE zone. As we shall discuss, this is due to constraints imposed by the system's regulating functions (c.f. **Sections 3** and **4**). If the applied current is constant, the plateaus (and leading and trailing ions) travel at an equal, constant velocity, $V_{ITP}$. For fully ionized co-ionic species, the zone order in plateau mode is typically the decreasing order of effective mobility (and exactly in decreasing order of mobility for the case of fully ionized co-ionic species). Also, the width of interfaces between adjoining plateaus are usually small relative to length of the plateaus. This condition of relatively large plateau-to-interface-width ratio can be used to distinguish and identify one ion from the next in the train of plateaus using a variety of detection techniques such as electrical conductivity,[103] fluorescence intensity,[104,105] UV absorption,[106] and temperature.[1,107] Plateau mode also lends itself to indirect detection of sample species, for example, using non-focusing directly-detectable tracers such as labeled counterions, and overspeeder or underspeeder co-ions.[33] Lastly, in addition to the plateau-forming sample ions, there may be other sample ions with significantly lower concentration. In that case, the trace sample ions focus in peak mode amid other plateau zones (i.e., mixed peak and plateau mode) at a location(s) which depends on the trace ion mobility value(s).

Peak mode and plateau mode ITP are each useful in a variety of settings (see detailed discussion in **Sections 3-7** and **11**). Briefly, peak mode ITP provides a method of highly effective preconcentration and simultaneous mixing.[30,108–110] Peak mode can also be used to control and accelerate chemical reactions of species.[39,111] Plateau mode allows for preconcentration, separation (including identification), and segregation of multiple sample ions.[33,103–106]

## 3. ITP using strong electrolytes

The earliest demonstration of ITP involved separations of strong electrolytes including salts and strong acids and strong bases.[2,12,112] Most applications involving ITP of strong electrolytes were demonstrated prior to the 1980s. In this section, we review the theory associated with strong electrolyte ITP and provide illustrations and important limitations of this approach. We derive the so-called ITP condition using a control volume approach and then estimate the ITP interface width. We then consider a simple strong electrolyte system consisting of three ions in total and use it to illustrate plateau mode ITP of strong electrolytes. Next, we derive the so-called Kohlrausch regulating function from first principles and use it to provide analytical expressions for the "adjusted" ion concentrations in plateau mode ITP. We conclude this section by mentioning a few demonstrations and validation studies involving ITP of strong electrolytes. We stress that, although the strong electrolyte theory is not very practical for modern microfluidic applications which largely use weak electrolytes (including pH buffers), strong electrolyte ITP is an excellent starting point to introduce researchers to the theory and practice of ITP processes. Strong electrolyte conservation analyses are often also more intuitive for researchers with experience in classic binary electrolyte electrokinetics.[67,81]

## 3.1 Single-interface ITP and the ITP condition

In this section we use an analysis of the simplest possible form of ITP to show the connection between the net neutrality approximation and the so-called ITP condition. The simplest ITP system consists of three fully dissociated, strong electrolyte species, namely two co-ions and a counter ion (and no sample ion). The situation was described qualitatively in **Section 2.3.** We consider the case of anionic ITP where the leading ($L$) and trailing ($T$) ions are anions and the common counterion ($C$) is a cation. The LE and TE contain an anion of high and low magnitude (negative) mobility, respectively, and the electric field is applied from LE to TE. This arrangement leads to a self-sharpening interface with the TE and LE zones electromigrating with equal velocities $V_{ITP}$. This result of co-ions of differing mobility but traveling at the same velocity is known as the ITP condition.

Next, we derive an expression for $V_{ITP}$ in terms of the physicochemical properties of the LE and TE, and the electric field. The relations obtained here apply to the case of cationic ITP with minor modifications. We first develop relations relating the jump conditions across a sharp ITP interface (e.g., the electric fields and ion concentrations) as a function of the ITP interface velocity. For simplicity, we consider strong electrolytes (c.f. Eq. (6)). Unlike classic ITP texts,[1,2] we will formulate these jump conditions using an approach similar to the formulation of the Rankine-Hugoniot jump conditions for supersonic shock waves in compressible fluid mechanics.[113–116] In ITP, we deal with ion concentration shock waves between co-ions of varying mobilities and concentrations. In general, these waves may have velocities which vary over time and space (e.g., for constant voltage sources). We here follow an approach similar to LeVeque,[116] where we consider a finite distance $\Delta x$ in a stationary reference frame over which the wave propagates over a finite time $\Delta t$. The distance $\Delta x$ is taken as significantly larger than the instantaneous ITP interface shock width but sufficiently small (e.g., compared to the total length of propagation) such that the shock velocity is approximately constant. We consider then an integral formulation of the one-dimensional species conservation equation (Eq. (6)) where we integrate over distance $\Delta x$ and time $\Delta t$ as follows:

$$\int\limits_{x_0,t_0}^{x_0+\Delta x,t_0+\Delta t} \frac{\partial c_i}{\partial t} dt\, dx + \int\limits_{x_0,t_0}^{x_0+\Delta x,t_0+\Delta t} \frac{\partial}{\partial x}\left[\left(-D_i \frac{\partial c_i}{\partial x}\right) + (\mu_i c_i E)\right] dt\, dx = 0, \qquad (32)$$

where $x_0$ and $t_0$ are respectively the initial position where and time that the ITP interface enters the stationary region of width $\Delta x$. During most of the time the thin shock traverses the distance region $\Delta x$, the diffusive fluxes at the boundaries of the region are within locally uniform concentrations and so we can well approximate these integrals as

$$\int\limits_{x_0,t_0}^{x_0+\Delta x,t_0+\Delta t} \frac{\partial c_i}{\partial t} dt\, dx + \int\limits_{x_0,t_0}^{x_0+\Delta x,t_0+\Delta t} \frac{\partial}{\partial x}(\mu_i c_i E) dt\, dx = 0, \qquad (33)$$

The first term on the left-hand side of this equation describes the change in species $i$ concentration (from just ahead of the shock in the leading zone to just behind of the shock in the trailing zone) in the Eulerian reference frame. The second term describes the change in electromigration species fluxes on either side of the wave. Using the approach of LeVeque,[116] the result of the integrations is well approximated as

$$(c_i^L - c_i^T)\Delta x \cong j\left(\frac{\mu_i^L c_i^L}{\sigma^L} - \frac{\mu_i^T c_i^T}{\sigma^T}\right)\Delta t. \tag{34}$$

Here $c_i^L$ and $c_i^T$ are the concentration of species $i$ in the leading and trailing zones, respectively; $\mu_i^L$ and $\mu_i^T$ are the electrophoretic mobilities of species $i$ in the leading and trailing zones, respectively; $\sigma^L$ and $\sigma^T$ are the local conductivity of the leading and trailing zones, respectively; is the velocity of the interface; and $j$ is the applied current density. Note $j/\sigma^k$ is merely the local electric field in region $k$. Dividing both sides by $\Delta t$ and taking the differential limit, we obtain

$$(c_i^L - c_i^T) V_{ITP} \cong j\left(\frac{\mu_i^L c_i^L}{\sigma^L} - \frac{\mu_i^T c_i^T}{\sigma^T}\right), \tag{35}$$

where $V_{ITP}$ is the wave speed of the ITP interface. The equation becomes exact for shock waves with uniform and constant wave speeds (as in many constant current ITP experiments). Note that the ionic current density, $j$, is the same on either side of the shock wave. Further, we can obtain the relation between ITP wave velocity and mobilities of the co-ions by evaluating Eq. (35) independently for the leading and trailing co-ions. Thus, we have

$$\begin{aligned} V_{ITP} &= j\frac{\mu_L^L}{\sigma^L} = j\frac{\mu_T^T}{\sigma^T} \\ &= \mu_L^L E^L = \mu_T^T E^T. \end{aligned} \tag{36}$$

That is, the leading and trailing co-ions travel at the same velocity equal to $V_{ITP}$. This follows from the conservation of current; i.e., $j = \sigma^L E^L = \sigma^T E^T$ (c.f. **Section 2.4**) and noting that electromigration velocity of the LE and TE co-ions are $\mu_L^L E^L$ and $\mu_T^T E^T$, respectively. This result (Eq. (36)) has historically been called the ITP Condition.[117–120] Basically, the ITP Condition states that the LE co-ion will electromigrate in the same direction and at the same velocity as the TE co-ion. Any trailing plateau co-ion will also travel at this velocity. Hence, this jump condition in Eq. (35) can be interpreted as describing the required wave velocity such that leading ions do not "run away" from the trailing ions leaving behind counter ions, which would otherwise violate net neutrality. Researchers new to ITP may find this result counter-intuitive as it describes a condition where ions of like charge but different mobilities travel through the system with the precise same velocity. We next provide a qualitative explanation for this result.

The result that even ions of even widely different mobility can travel in the same direction at the same velocity is in fact a requirement of the conservation of current and the net neutrality approximation. Consider that this equal velocity condition is consistent with precluding the possibility of forming a gap between LE and TE co-ions. That is, the demands of net neutrality preclude LE co-ions from speeding away from the TE co-ions. The gap would constitute a region of locally unbalanced counterions. As described in **Section 2.4**, the typical ion densities and electric field magnitudes in ITP preclude the possibility of such grossly unbalanced charge regions.

How then does the electric fields in the TE and LE respectively increase and decrease to achieve this equal velocity condition? We discussed the mechanism in **Section 2.4**. For a very short time (on the order of the charge relaxation time scale[70]), the LE ions slightly move away from TE ions and form a diffuse region which contains a tiny amount of unbalanced charge. The sign of this net charge (negative for anionic ITP) is such that it will raise the field magnitude in the TE zone while decreasing it in the LE. This tiny but important imbalance in charge builds quickly and self-limits as soon as the TE and LE co-ion velocities are equal. The latter condition ensures the conservation of current in the system, i.e., $j = \sigma^L E^L = \sigma^T E^T$. The result is a region of relative high field in the TE and low field in the LE and the two regions are interfaced by an electric field gradient. Away from the interface, diffusive current is negligible, and we can assume conductivity is inversely proportional to local field (c.f. **Section 2.4**).

As we shall see below, the established electric field gradient tends to self-sharpen and preserve a sharp interface between the ions of high and low mobility. The latter self-steeping conditions can be described qualitatively as follows. Consider that a TE co-ion that diffuses out of the LE-TE interface into the LE zone. Such an ion experiences the relatively low electric field of the LE but it is "outraced" by its neighboring LE ions, and so it falls back to the interface. Similarly, an LE co-ion diffusing into the TE zone experiences a higher local electric field and this tends to drive it back to the LE-TE interface. Even as the electric field gradient sharpens the interface, diffusion tends to try to broaden it–and its width is determined by a balance between these effects (see **Section 3**).[88,97]

We next present a more detailed derivation of the current densities, electric fields, and the ITP velocity in terms of the specific ion valences and mobilities in the system. To this end, we will assume that hydronium $H^+$ and hydroxyl $OH^-$ ions do not contribute significantly to current (see "safe pH" approximation; c.f. **Section 2.7**), so from Eq. (7) current density is expressed in terms of the mobilities and concentrations of the LE and TE co-ions and counterion as

$$j = (z_L^L c_L^L \mu_L^L + z_C^L c_C^L \mu_C^L) F E^L = (z_T^T c_T^T \mu_T^T + z_C^T c_C^T \mu_C^T) F E^T. \tag{37}$$

where $F$ is again Faraday's constant, and $E^L$ and $E^T$ are the local electric field of the leading and trailing zones, respectively. We note Eq. (37) is valid for both weak and strong electrolytes. For strong electrolytes, we can avoid the subscript (location) values in valences and mobilities, and for weak electrolytes, the mobilities are effective, local mobilities. For example, the term $z_C^L c_C^L \mu_C^L$ ensures the correct contribution of current by the partially ionized counterion from the multiplication of the valence, effective mobility, and the total ion concentration in the LE region.

Combining Eqs. (36) and (37), we have

$$\begin{aligned} V_{ITP} = V_L^L = V_T^T = \mu_L^L E^L = \mu_T^T E^T &= \mu_L^L \frac{j}{(z_L^L c_L^L \mu_L^L + z_C^L c_C^L \mu_C^L)F} \\ &= \mu_T^T \frac{j}{(z_T^T c_T^T \mu_T^T + z_C^T c_C^T \mu_C^T)}. \end{aligned} \tag{38}$$

where $V_L^L$ and $V_T^T$ are the velocity of the leading and trailing ions in the leading and trailing zones, respectively. Lastly, the position of the ITP interface $x(t)$ and the voltage difference $\Delta V$ across the microchannel can be related as

$$\Delta V = E^T x(t) + E^L(L - x(t)) = j \left( \frac{x(t)}{\sigma^T} + \frac{L - x(t)}{\sigma^L} \right). \tag{39}$$

where $L$ is the length of the microchannel. This equation is valid for uniform cross-section channels and for either constant current or constant voltage operation. Moreover, the resistance in the channel increases with time as the TE replaces the LE zone. For constant current operation, the ITP velocity is constant and uniform, and the voltage (and resistance) increases linearly with time. For constant voltage operation, the current and ITP velocity decreases over time in accordance with the increase in the resistance in the channel. Refer to Bahga et al.[121] for a detailed derivation of ITP velocity for the case of constant voltage operation.

## 3.2 The width of the interface between two ITP plateau zones

For constant applied current, the ITP interface achieves a steady state width independent of the initial conditions. This results from a self-focusing of ion concentration gradients driven by non-uniform electromigration and limited by molecular diffusion. We here first present simple scaling arguments for ITP interface width, and then derive the analytical solution based on the classic work of MacInnes and Longsworth.[122] We focus our discussion on the example of the interface between a TE and an LE zone, but the current discussion applies to the interface between any two adjoining ITP plateau zones.

We begin with an instructive scaling argument for the interface that provides intuition into the self-focusing nature of ITP. The electromigration velocity magnitude of a TE ion in the LE region of the interface (or an LE ion in the TE region) can be expressed as the quantity $(|\mu_L - \mu_T|)E$, where $E$ is some characteristic local electric field in the interface region (e.g., average of the LE and TE electric fields). A characteristic velocity due to diffusion for the same ion scales as $D/\delta$, where $D$ is the diffusivity of the ion and $\delta$ is the ITP interface width. At the ITP interface, we then balance the fluxes associated with diffusion and (non-uniform) electromigration to obtain a scaling for $\delta$ as

$$\delta \sim \frac{D}{(|\mu_L - \mu_T|)E}. \tag{40}$$

This scaling suggests that the ITP interface width is inversely proportional to the electric field and the mobility difference between the LE and TE ions, and directly proportional to the diffusivity of the species.

An exact solution for the interface width between two plateau zones, $\delta$, was first given by MacInnes and Longsworth[122] in 1932. First, the strong electrolyte species conservation Eq. (6) for the LE and TE co-ions in the frame of the moving ITP interface (i.e., after performing a Galilean transformation[123]) at steady state can be expressed as

$$(V_{ITP} - \mu_L E)c_L + D_L \frac{\partial c_L}{\partial \xi} = 0 \tag{41}$$

$$(V_{ITP} - \mu_T E)c_T + D_T \frac{\partial c_T}{\partial \xi} = 0 \tag{42}$$

where $\xi = x - V_{ITP}t$ is the moving coordinate of the interface, $V_{ITP}$ is the ITP velocity. We next invoke Nernst-Einstein relation between diffusivity and ion mobility (see Ref.[68]) and write as $D_i = RT\mu_i/z_iF$. Here, $R$ is the universal gas constant and $T$ is the absolute temperature. Combining these two equations to eliminate $E$ explicitly, we derive

$$\frac{RT}{F}\left(\frac{1}{z_T c_T}\frac{\partial c_T}{\partial \xi} - \frac{1}{z_L c_L}\frac{\partial c_L}{\partial \xi}\right) = V_{ITP}\left(\frac{1}{\mu_L} - \frac{1}{\mu_T}\right) \tag{43}$$

Integrating with respect to $\xi$ and defining $\xi = 0$ as the location in the interface where $c_L = c_T$, we obtain for the case of $z_L = z_T = -1$,

$$\frac{c_L}{c_T} = \exp\left[-\frac{FV_{ITP}}{RT}\left(\frac{\mu_L - \mu_T}{\mu_L\mu_T}\right)\xi\right] \equiv \exp\left(-\frac{\xi}{\delta}\right), \tag{44}$$

where $\delta$ is the characteristic length scale for the ITP interface width given by

$$\delta = \frac{RT}{F\mu_l E^L}\left(\frac{\mu_L\mu_T}{\mu_L - \mu_T}\right) = \frac{RT}{F\mu_L \frac{j}{\sigma^L}}\left(\frac{\mu_L\mu_T}{\mu_L - \mu_T}\right). \tag{45}$$

where we have substituted $V_{ITP} = \mu_L E^L$ (see Eq. (36)).

Eq. (45) shows that $\delta$ scales in proportion to diffusivity (via the absolute value of the characteristic mobility of the Nernst-Einstein ion) and inversely with the applied current density (or equivalently, the electric field). Such proportionality is observable in experiment up to some maximum applied electric field, beyond which electrokinetic instabilities can disrupt a stable ITP interface.[97,124] It is also very useful in obtaining an intuition for the ideal conditions of (one-dimensional) ITP wherein interface width is limited strictly the by the competing effects of non-uniform electromigration and diffusion. However, in practice, sufficiently high currents can lead to electrohydrodynamic instabilities which limits the minimum ITP interface width (as will be discussed in **Section 8**). Further, note that the expression for $\delta$ in Eq. (45) was derived exclusively for strong electrolytes in ITP, and this relation does not hold exactly for weakly ionized species. Nevertheless, the length scale $\delta$ (i.e., LE-to-TE interface width) is a useful estimate of the order of magnitude of length scale relevant for ITP applications involving peak-mode ITP. In peak-mode ITP, the focused peak-mode sample minimum achievable width of the sample is governed by the finite length scale $\delta$. Hence, this interface width affects the degree to which sample can be preconcentrated and reactions accelerated (c.f. **Sections 5** and **7**), including for applications involving weak electrolytes. For typical ITP experiments in order 10-100 μm channels with (typical) applied field magnitudes of order 20 kV/m, the interface width is order 10 μm.[97]

### 3.3 Condition for focusing of a strong analyte

For strong electrolyte ITP, the condition which results in an analyte ion (with fully dissociated mobility equal to $\mu_S$) focusing between the LE and TE zones is simply[1,125]
$$|\mu_T| < |\mu_S| < |\mu_L|, \qquad (46)$$
where the fully dissociated mobilities of the TE and LE co-ions are, respectively, equal to $\mu_T$ and $\mu_L$. We use absolute values in Eq. (46) to account for both anionic and cationic ITP.

### 3.4 Closed form calculation for plateau-mode ITP of strong electrolytes

The electromigration of multiple analyte species into and out of zones may at first be counter intuitive. Consider again **Figure 1** which shows a simple single-interface anionic ITP system including an LE and a TE zone. We assume that all ion motion is due to electromigration alone (i.e., no bulk flow). The LE ions move to the right and are replaced by the influx of TE ions. Interestingly, ion displacement dynamics in this system are such that the concentration of the TE ions as they replace LE ions is completely independent of the initial TE ion concentration. Instead, the concentration at which TE ions enter the region formerly occupied by LE are determined by the properties of the LE and the TE ion mobility. Hence, we find a situation where the local concentration of TE ions (in the former LE region) is independent of the initial concentration of that ion.

To further describe this effect, it is useful to define two TE-related concentrations: an initial concentration of the trailing ion (e.g., in the reservoir well), $c_T^T$, and the concentration of this ion after it enters the region $T'$ formerly occupied by the leading electrolyte, $c_T^{T'}$. We refer to the latter as adjusted TE (ATE) concentration. As depicted in **Figs. 1** and **4**, the interface between the TE reservoir and ATE zones ($c_T^T$ and $c_T^{T'}$) is stationary while the interface between the ATE and LE ($c_T^{T'}$ and $c_L^L$) moves at $V_{ITP}$. We next derive the concentration $c_T^{T'}$ of the trailing co-ion in the ATE region using the concepts of current conservation, charge neutrality, and the ITP condition.

Assuming safe pH (c.f. **Section 2.7**), current conservation yields
$$j = (z_L c_L^L \mu_L + z_C c_C^L \mu_C) F E^L = (z_T c_T^{T'} \mu_T + z_C c_C^{T'} \mu_C) F E^{T'}. \qquad (47)$$
We do not use superscripts for mobility or valence in Eq. (47) since these quantities are constant and uniform for fully ionized species. (e.g., $\mu_C^L = \mu_C^{T'} = \mu_C$). We also assume a moderate pH (c.f. **Section 2.7**), so that charge neutrality in the LE and ATE regions yield
$$z_L c_L^L = -z_C c_C^L \text{ (in LE) , and } z_T c_T^{T'} = -z_C c_C^{T'} \text{ (in ATE) .} \qquad (48)$$

Lastly, the ITP condition yields

$$\mu_L E^L = \mu_T E^{T'}. \tag{49}$$

Combining Eqs. (47)-(49), we obtain

$$\frac{z_L c_L^L}{\mu_L} + \frac{z_C c_C^L}{\mu_C} = \frac{z_T c_T^{T'}}{\mu_T} + \frac{z_C c_C^{T'}}{\mu_C}. \tag{50}$$

Eq. (50) can be interpreted as the condition that the sum $\Sigma(z_i c_i/\mu_i)$ in the relevant zone (here the region of ATE) is preserved whether it is occupied by LE or ATE ions. We shall show in the next section that the preservation of this sum, called the Kohlrausch regulating function, must hold in all channel locations for an arbitrary number and initial distribution of strongly ionized species. From Eqs. (50) and (48), we obtain the concentration of the trailing ion in the ATE zone for the univalent case of $z_L = z_T = -z_C = -1$ as

$$c_T^{T'} = c_L^L \left(\frac{\mu_T}{\mu_L}\right) \left(\frac{\mu_C - \mu_{LL}}{\mu_C - \mu_T}\right). \tag{51}$$

Notice that the concentration of the TE ion in the ATE zone $c_T^{T'}$ is independent of the initial TE concentration $c_T^W$, and is solely dependent on the LE co-ion concentration and the mobilities of the LE and TE ions. The concentration of the counterion in the ATE can simply be obtained from charge neutrality $c_C^{T'} = -(z_T/z_C) c_T^{T'} = c_T^{T'}$, where the second equality is for monovalent ions, $z_T = -z_C = -1$. We will consider multivalent species in the next section.

### 3.5 Kohlrausch regulating function and the concept of adjusted concentrations

We can generalize plateau mode ITP for an arbitrary number of strong electrolytes and also derive the "adjusted" concentration of a focused ion in any arbitrary plateau zone between the LE and TE using the Kohlrausch regulating function (KRF). KRF is a conservation principle which is applicable to the arbitrary electrophoresis of mixtures of fully ionized species. This was originally derived by F. Kohlrausch in 1897 and called by Kohlrausch the "Beharrliche funktion".[126] Below, we present our derivation of KRF, summarize its assumptions, and apply it to a generic case of plateau mode ITP of strong electrolytes.

We begin with the species conservation equation (Eq. (6)) for the case of no bulk flow and for a strong electrolyte $i$. We multiply and divide Eq. (6) by the valence and (fully ionized) mobility of the species to obtain

$$\frac{\partial}{\partial t}\left(\frac{z_i c_i}{\mu_i}\right) = \frac{\partial}{\partial x}\left(-z_i c_i E + \frac{z_i D_i}{\mu_i} \frac{\partial c_i}{\partial x}\right). \tag{52}$$

Summing Eq. (52) across all the species $i = 1$ to $N$, we have

$$\frac{\partial}{\partial t}\left(\sum_{i=1}^{N} \frac{z_i c_i}{\mu_i}\right) = \frac{\partial}{\partial x}\left(-E\left\{\sum_{i=1}^{N} z_i c_i\right\} + \frac{\partial}{\partial x} \sum_{i=1}^{N} \frac{z_i D_i c_i}{\mu_i}\right). \tag{53}$$

Next, we assume moderate pH and use the charge neutrality approximation for the entire system, i.e., $\sum_{i=1}^{N} z_i c_i = 0$, and neglect diffusive terms (since $\partial c_i/\partial x$ vanishes away from sharp ITP interfaces). Thus, Eq. (53) reduces to

$$\frac{\partial}{\partial t}\left(\sum_{i=1}^{N} \frac{z_i c_i}{\mu_i}\right) \equiv \frac{\partial w}{\partial t} = 0. \tag{54}$$

where $w(x,t)$ is now only a function of $x$, i.e., $w = w(x)$. The function $w(x)$ is defined as the Kohlrausch regulating function (KRF), and is given by

$$w(x) \equiv \sum_{i=1}^{N} \frac{z_i c_i}{\mu_i}. \tag{55}$$

We see from Eq. (55) that the KRF is simply the summation of the ratios $w_i = z_i c_i / \mu_i$ over all ionic species at each point in space. Since the partial derivative of $w$ with time is zero, the regulating function establishes a constraint on ion concentrations in space. This constraint is established by the initial condition. In practice, this initial condition is determined by, for example, the result of some preliminary process where pressure-driven flow was used to inject various electrolyte mixtures into various contiguous zones along some channel. The analysis shows that the value of the KRF along the channel is thereafter invariant with time (but can vary with $x$, as determined by the initial condition). In ITP, since $i$ can refer to any ion in the system, the KRF relation applies to (i.e., governs) all zones including any trailing zones. Importantly note that Eq. (54) is not accurate at the interface between zones where diffusion effects due to strong concentration gradients are important.[127,128] Consistent with this, and importantly, the KRF cannot be used to compute the concentration of ions in peak mode. Peak-mode ions do not contribute significantly to local net neutrality (since these ions have concentrations which are several orders of magnitude lower than the background LE and TE ions) and exist solely within a region of strong diffusive gradients. For example, consider a trace sample ion initially at 100 pM focused in peak-mode by 1,000-fold to a new maximum peak concentration of 100 nM using a 100 mM LE zone. Such a species would have a concentration predicted by the KRF (and established by the LE zone) of ~10-100 mM, which is about 5-6 orders of magnitude higher than the actual peak concentration. The latter discrepancy arises from the incorrect application of KRF to predict peak mode ion concentrations.

The KRF can be mathematically interpreted as the Riemann invariant corresponding to the zeroth eigenvalue of the governing hyperbolic equations for species transport.[114,115] The levels of concentration of ions which replace an original mixture of ions via electromigration are constrained by the concentrations and mobilities of all ions which initially occupied that space. For fully ionized species, the constraint of the KRF requires that the quantity $\Sigma_i z_i c_i / \mu_i$ of "newcomer" ions to match the scalar value of $\Sigma_i z_i c_i / \mu_i$ as governed by the "former occupant" ions.

An intuition for the KRF is as follows. For each region in space, ions are not "allowed" to enter with arbitrary concentrations. Instead, ions enter each region with species-specific fluxes and species-specific contributions to current density. These species-specific flux quantities are determined by a (global) electric field and by the respective valence and mobility of each ion. The KRF reflects the precise balance by which these species fluxes and current contributions at all times satisfy both species conservation and the local net neutrality of charge.

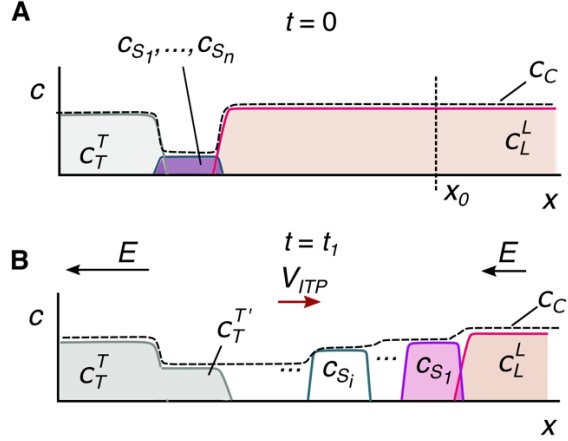

**Figure 5**: Schematic of concentration profiles in plateau-mode ITP for strong electrolytes. (A) Finite injection sample loading configuration. Sample contains a mixture of ions $S_1, \ldots S_n$ which focus in ITP. (B) Sample ions are focused and stack in decreasing order of (magnitude of) mobility values between the LE and TE. Indicated in (A) by a vertical dashed line is an example location along the channel where we choose to illustrate the application of Kohlrausch regulating function in Section 3.5. $S_1$ ion and counterion at $x = x_0$ must meet the KRF set by the LE zone which originally occupied this region.

To show its utility, we now apply KRF to obtain the concentration of plateaus in ITP. For simplicity, we consider a single counterion $C$ common to all plateau zones. We consider the plateau region of a sample ion, $X$, which moves into a region formerly occupied by the LE zone. Note $X$ represents some co-ion which can be any one of the (several) sample ion zones, or which can be the trailing ion $T$ in the adjusted TE zone, see **Figure 5**. As an illustration, consider the location marked by the dashed vertical line at $x = x_0$ in **Figure 5**, where $X = S_1$. The initial condition at $t = 0$ is set by the ions originally occupying that region. The initial KRF value at $x_0$, is then

$$w(x_0)|_{t=0} = \frac{z_C c_C^L}{\mu_C} + \frac{z_L c_L^L}{\mu_L} = z_L c_L^L \left( \frac{1}{\mu_L} - \frac{1}{\mu_C} \right). \tag{56}$$

Here, we have used charge neutrality in the second equality. At some later time $t_1$, consider that the location at $x = x_0$ is occupied by the "newcomer" ion $X$ as well as the original (same) counter ion. The KRF at this location at time $t_1$ is

$$w(x_0)|_{t=t_1} = \frac{z_C c_C^X}{\mu_C} + \frac{z_X c_X^X}{\mu_X} = z_X c_X^X \left( \frac{1}{\mu_X} - \frac{1}{\mu_C} \right). \tag{57}$$

Equating the KRF at times $t = 0$ and $t_1$, we obtain

$$z_L c_L^L \left( \frac{1}{\mu_L} - \frac{1}{\mu_C} \right) = z_X c_X^X \left( \frac{1}{\mu_X} - \frac{1}{\mu_C} \right). \tag{58}$$

Eq. (58) can be rearranged to obtain the concentration of the arbitrary sample $X$ at $x = x_0$ as

$$c_X^X = c_L^L \left( \frac{z_L \mu_X}{z_X \mu_L} \right) \left( \frac{\mu_C - \mu_L}{\mu_C - \mu_X} \right). \tag{59}$$

Hence, $c_L^L$ determines $c_X^X$ by establishing a local, invariant KRF value. Note also that Eq. (51) is a special case of Eq. (59) where $X$ represents the TE ion in the ATE zone. Typical values of a fairly wide range of mobilities yield $c_X^X$ of roughly a factor 0.5 to 0.9 of the value of $c_L^L$. Eq. (59) shows how plateau concentrations are proportional to leading ion concentrations and have values roughly equivalent to (and typically a bit lower than) $c_L^L$. This also shows high initial-to-final sample concentration ratios can be achieved using a high $c_L^L$ and low value of concentration in the sample reservoir, $c_X^W$ (here $W$ is used to denote "well" value, i.e., the concentration in the sample reservoir). As before, the concentration of the counterion in zone $X$ can simply be obtained from charge neutrality as

$$c_C^X = -\frac{z_X}{z_C} c_X^X = -c_L^L \left(\frac{z_L \mu_X}{z_C \mu_L}\right)\left(\frac{\mu_C - \mu_L}{\mu_C - \mu_X}\right). \tag{60}$$

Note that $z_L$ and $z_C$ have opposite signs, so the second equality in Eq. (60) is overall a positive quantity. Later in **Sections 6** and **10**, we extend the current analysis to include sample accumulation rates in plateau zones, and the sensitivity and resolution associated with detection of the separated zones, respectively.

### 3.6 Illustrations and limitations of strong electrolyte ITP theory

The earliest demonstration of ITP of strong electrolytes was in 1932 by Kendall and Crittenden[12] who reported separation of rare earth metals and simple acids using ITP. They called their process the "ion migration method". Importantly, this study qualitatively described that the zone concentrations automatically "adjusted" themselves to ratios governed by KRF. Much later in 1980, the study of Hjertén et al.[129] was the first to quantitatively and experimentally validate the KRF conservation principle in ITP. They validated KRF in both cationic ($K^+$, $Co^{2+}$, $Cu^{2+}$) and anionic ($Cl^-$, 5-sulphosalicylic acid) ITP systems involving strong electrolytes. To obtain accurate data for validation, Hjertén et al.[129] carefully performed ITP experiments in free solution, suppressed EOF, used a photoelectric scanner to determine attainment of steady state conditions, and collected ITP-separated zone fractions for downstream measurements of conductivity and concentrations.

ITP applications involving strong electrolytes were significantly developed and widely employed throughout the 1950s and 1980s. We do not review these in detail but refer the reader to a few studies involving the theory and applications of strong electrolyte ITP, including separation of trace elements, heavy metals, and isotopes, among many others.[13,112,122,127–137] Typically, ITP systems involving strong electrolytes alone are not buffered, and such systems can exhibit highly variable (and difficult to control) pH across zones.[138] For this reason, modern microfluidic ITP applied to biological systems seldom involve strong electrolytes exclusively. Extreme pH values can result in additional zones and dispersed interfaces (e.g. due to EOF[61,139]), and lead to complex zone structures.[132,140] We refer to studies of Gas et al.[141] and Ermakov et al.[140] for a detailed theoretical discussion and associated experimental validation of ITP at extreme pH conditions.

As mentioned earlier, we here cover strong electrolyte ITP as a starting point of instruction. This theory is also useful in providing closed-form analytical solutions which can be used to benchmark new computational approaches and simulation results.[87,127,142–144] As an example illustration and benchmarking of strong electrolyte ITP simulation, see **Fig. 6**.[132] **Fig. 6** shows simulation results of an ITP system involving exclusively strong acids and bases. The study validated steady state predicted concentrations with numerical results shown in **Fig. 6**. **Fig. 6** also shows KRF invariance in time at every location between the initial and final states. Notably, **Fig. 6** also highlights a small, inverted peak (indicated by an arrow in **Fig. 6**) in the computed KRF value at the zone boundary. Note this is a small region of high diffusive gradients in the initial condition.

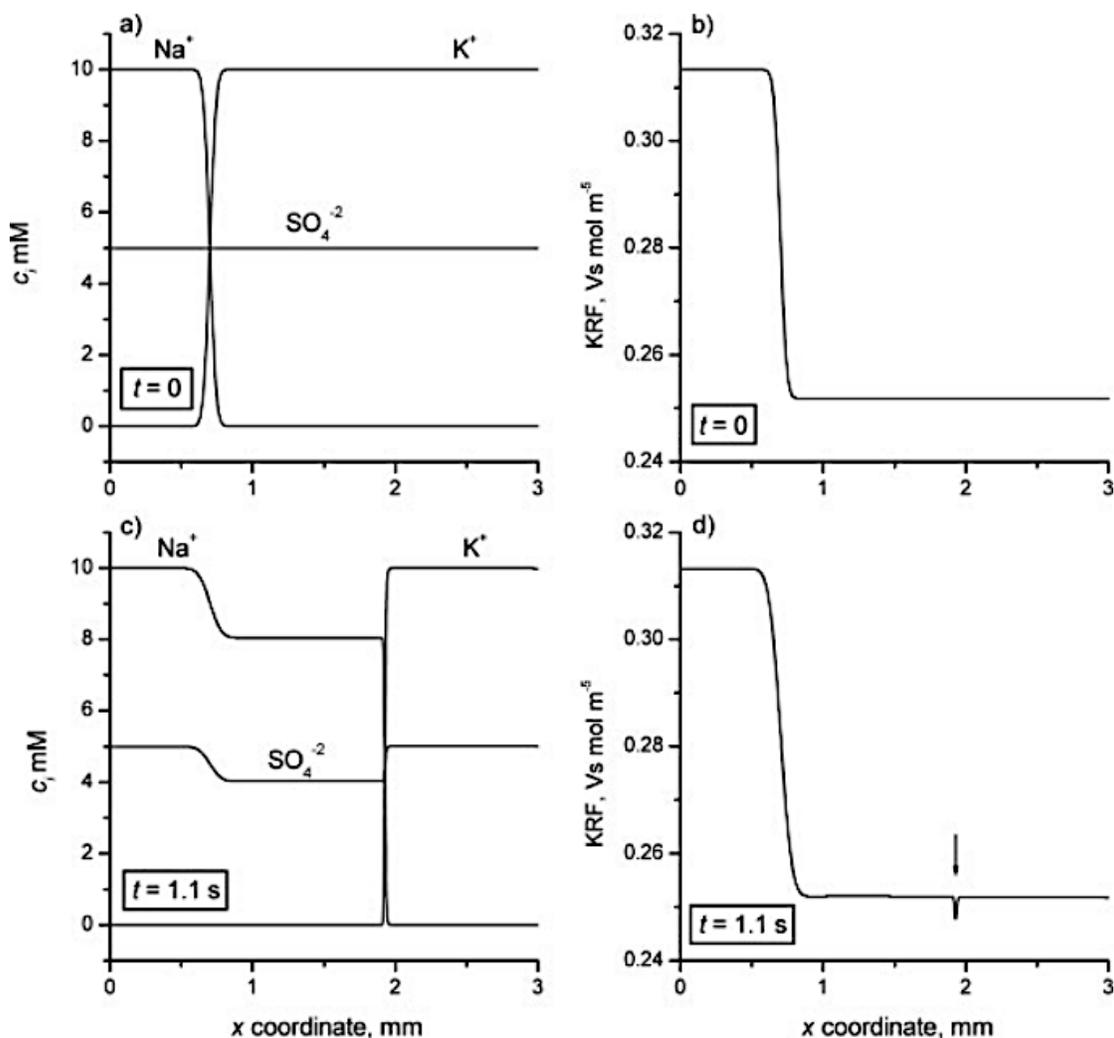

**Figure 6**. Simulated cationic ITP run for LE mixture consisting of 10 mM potassium and 5 mM sulfate ions, and TE mixture consisting of 10 mM sodium and 5 mM sulfate ions. (A) and (C) show the concentration fields at $t = 0$ and $t = 1.1$ s, respectively. (B) and (D) show the value of the Kohlrausch regulating function (KRF) versus the channel location $x$ at $t = 0$ and $t = 1.1$ s, respectively. Simulations were performed using the Simul software,[144] with parameters: capillary length of 3 mm, driving voltage of 50 V, $x$-coordinate mesh consisting of 1000 nodes. Figure is reproduced with permission from Ref. [132]. Copyright 2007 John Wiley and Sons.

## 4. ITP using weak electrolytes

In this section, we review the theory for ITP processes involving weak electrolytes. Weak electrolytes are important as they enable pH buffers and these are crucial to systems processing or analyzing biological species. Weak electrolytes are also used extensively for several applications including basic biochemistry, cell biology, food and plant sciences, and biological assays at the molecular, cellular, or tissue level.[31,145–153] pH buffering is also important for well controlled processes and assays involving chemical assays and synthesis.[31,86,154]

An important complexity of weak electrolytes is that, unlike Eq. (6) above, conservation equations necessarily involve multi-species source terms and where chemical conditions have a strong impact on the spatiotemporal development of ionization state (including effective species mobility). To illustrate this complexity (and the failure of Kohlrausch regulating function), consider a special case of weak electrolyte transport Eq. (3) for the following special case. We neglect diffusive fluxes, multiply by valence, divide by mobility, and sum over all species families and ionization states to derive

$$\frac{\partial}{\partial t}\sum_{i=1}^{N}\sum_{z=n_i}^{p_i}\left(\frac{zc_{i,z}}{\mu_{i,z}}\right) = \frac{\partial}{\partial x}\left(-E\left\{\sum_{i=1}^{N}\sum_{z=n_i}^{p_i}zc_{i,z}\right\}\right) + \sum_{i=1}^{N}\sum_{z=n_i}^{p_i}\frac{zR_{i,z}}{\mu_{i,z}}. \quad (61)$$

Applying the net-neutrality approximation, i.e., $\Sigma_{i,z}\, zc_{i,z} \approx 0$, we derive the following expression for the temporal change of a Kohlrausch-like variable, $w$:

$$\frac{\partial}{\partial t}\sum_{i=1}^{N}\sum_{z=n_i}^{p_i}\left(\frac{zc_{i,z}}{\mu_{i,z}}\right) \equiv \frac{\partial w(x,t)}{\partial t} = \sum_{i=1}^{N}\sum_{z=n_i}^{p_i}\frac{zR_{i,z}}{\mu_{i,z}}. \quad (62)$$

The result shows that we must express $w$ as a function of time and space. Note that the production term results in spatiotemporal development of the variable $w$ and so we are unable to derive a "simple" KRF for weak electrolytes. This is an important point as the KRF is very often improperly applied to analyze ITP systems involving weak electrolyte[14,73,155–160] and such analyses can result in grossly incorrect predictions of ITP dynamics and species concentrations.

We are unaware of regulation functions applicable to generalized weak electrolyte systems involving multiple species where one or more is a multi-valent species and/or involving extreme pH values outside of the aforementioned "safe" and "moderate" ranges. Fortunately, there are known and convenient regulation functions for weak electrolytes applicable to the specialized case of univalent families (chemical families with only neutral and univalent members) and for safe and moderate pH. The latter regulating functions are called the Jovin[161] and Alberty[41] functions, which we discuss in this section. In addition, we discuss conditions for focusing of a weak analyte in ITP, derive steady state concentrations in plateau mode ITP using the regulating functions, and provide some results of the weak electrolyte theory.

### 4.1 Conditions for focusing of a weak analyte

We know of no reference which summarizes the general conditions required for focusing of sample species for the case of weak electrolytes and for both plateau and peak-mode ITP. We here propose a set of relations for such focusing between LE and (adjusted) TE co-ions. Consider that various zones in ITP will have different pH and the varying ion mixtures in these zones can lead to different degrees of ionization. Hence, for weak electrolytes, we have to describe conditions for focusing in terms of effective (not fully ionized) ion mobilities and in terms of position in space. We first consider the case of an ion $X$ focused in a plateau region X somewhere between the LE and ATE. The propensity of the ion to electromigrate out of the ATE region and focus within zone X (and for the TE ion to fall back from zone $X$ into the ATE region) is given by

$$\left|\mu_X^{T'}\right| > \left|\mu_T^{T'}\right| \quad \text{and} \quad \left|\mu_X^X\right| > \left|\mu_T^X\right|. \quad (63)$$

Likewise, we also require such stability conditions for the interface between zone X and the LE as

$$\left|\mu_X^L\right| < \left|\mu_L^L\right| \quad \text{and} \quad \left|\mu_X^X\right| < \left|\mu_L^X\right|. \quad (64)$$

For the case of a single ion $X$ focused in peak mode and sandwiched by the ATE and LE, the condition for focusing can be expressed as
$$\left|\mu_X^{T'}\right| > \left|\mu_T^{T'}\right| \text{ and } |\mu_X^L| < |\mu_L^L|. \tag{65}$$
That is, for peak mode, no formal (plateau) zone X exists and ion $X$ is subject to the electric field gradients (and relative ion velocities) established by the TE and LE co-ions.

Lastly, we note that, in all cases, ITP should exist and hence
$$\left|\mu_L^{T'}\right| > \left|\mu_T^{T'}\right| \text{ and } |\mu_T^L| < |\mu_L^L|. \tag{66}$$

## 4.2 Weak electrolyte regulating functions: Jovin and Alberty regulation

As we discussed earlier, the KRF conservation principle is not applicable for weak electrolytes. We can, however, define two regulating functions applicable to weak electrolyte systems for the special case of, at most, singly ionized species and systems which obey the safe and moderate pH conditions. These two very useful and fundamental functions are the Jovin and Alberty functions and described in the following two sections. To our knowledge, the current review is the first presentation of the derivation of both of these functions starting from the one-dimensional unsteady conservation equations. The authors find this remarkable given that Alberty and Jovin functions are fundamental to the analysis of ITP, a useful method in use for nearly a century.

### *4.2.1 Jovin function*

The Jovin function imposes fundamental restrictions on the singly ionized ion concentrations achievable at any location within a multispecies electrokinetic process. It is a very useful quantity in deriving closed-form relations and analyses of ITP processes. Assuming negligible effects from diffusive fluxes (i.e., negligibly influenced by sharp interfaces), the weak electrolyte species conservation equation (Eq. (3)) reduces to
$$\frac{\partial c_{i,z}}{\partial t} = \frac{\partial}{\partial x}\left(-\mu_{i,z}c_{i,z}E\right) + R_{i,z}. \tag{67}$$
Summing this equation over all subspecies $z$ of species family $i$ and recognizing that proton donation/acceptance results in no net generation of species family $i$ (i.e., $\Sigma_z R_{i,z} = 0$), we have
$$\frac{\partial}{\partial t}\sum_z c_{i,z_i} = \frac{\partial c_i}{\partial t} = \frac{\partial}{\partial x}\sum_z -\mu_{i,z}c_{i,z}E. \tag{68}$$
We next restrict ourselves to the simple case of a univalent weak electrolyte family (comprising only a neutral state and a single charge state of charge 1 or -1), Eq. (68) simplifies to
$$\frac{\partial c_i}{\partial t} = \frac{\partial}{\partial x}\left(-\mu_{i,z_i}c_{i,z_i}E\right), \tag{69}$$
where $z_i$ is $+1$ or $-1$. Multiplying Eq. (69) by $z_i$, and summing over all species families, and applying the safe pH assumption (i.e., that hydronium or hydroxyl ions do not carry significant current), we have
$$\frac{\partial}{\partial t}\sum_i z_i c_i = \frac{\partial}{\partial x}\sum_i -z_i\mu_{i,z_i}c_{i,z_i}E = \frac{\partial}{\partial x}\left(\frac{\sigma E}{F}\right) = 0. \tag{70}$$
Here, the last equality follows from current conservation (for negligible diffusive current). Hence, we find the so-called Jovin's relation, originally derived by T.M. Jovin in 1973)[161]

$$\sum_i z_i c_i = \text{constant} . \tag{71}$$

Interestingly, the left-hand-side of this relation looks similar to the summation in charge neutrality (e.g., compare to Eq. (30)), however the summation of Eq. (71) is in fact a very different quantity since here $c_i$ is total concentration (and not just ion concentration). Here, the summation is performed over the (monovalent) species families and includes a contribution(s) from the concentration of the neutral species family member (whose $z = 0$) multiplied by the ionization state of that species family ($z_i = 1$ or -1).

The Jovin function implies that total species concentrations impose a strong regulation on the entry (exit) of all ions into (out of) any particular zone in space. Unlike charge neutrality, the Jovin function involves balances among both charged and uncharged species. For example, consider the following system consisting of just three species families. A certain zone is originally occupied by a single weak acid (comprised of members A1$_{-1}$ and A1$_0$) and a single weak base (comprised of C$_0$ and C$_{+1}$). The Jovin function dictates that the influx of each single molecule of a new acidic species family A2 (comprised of A2$_{-1}$ and A2$_0$) must necessarily result in either removal of one molecule which is a member of the co-ionic family A1 or alternately the addition of one molecule member of the counter-ionic family C. Of course, the net neutrality assumption is in addition to (and separate from) the Jovin function constraint.

### 4.2.2 Alberty function

The Alberty function is a second fundamental constraint on singly ionized ion concentrations. Like the Jovin function, it is very useful in deriving closed-form relations and analyses of ITP processes. To derive the Alberty function, we begin with Eq. (69) and again restrict ourselves to a single ion type per family:

$$\frac{\partial c_i}{\partial t} = \frac{\partial}{\partial x}(-z_i |\mu_{i,z_i}| c_{i,z} E). \tag{72}$$

Here, we have used the relation $\mu_{i,z_i} = z_i |\mu_{i,z_i}|$ since we have considered a single ion per family which is univalent (i.e., $z_i$ is restricted to values of -1, 0, or 1). Dividing Eq. (72) by $|\mu_{i,z_i}|$ and summing over all species we derive

$$\frac{\partial}{\partial t} \sum_i \frac{c_i}{|\mu_{i,z_i}|} = -\frac{\partial}{\partial x} E \sum_i z_i c_{i,z} = 0 , \tag{73}$$

where the last equality follows from the charge neutrality approximation for moderate pH. We thus obtain the so-called Alberty's condition (originally derived by E.B. Dismukes and R.A. Alberty in 1984) as[41]

$$\sum_i \frac{c_i}{|\mu_{i,z_i}|} = \text{constant} . \tag{74}$$

At first glance, Alberty's relation is similar to the KRF as it involves a summation of ratios of concentration and mobility across species. However, the Alberty function is in fact much different since here $c_i$ is total (analytical) concentration. For example, consider that ratios involved in this summation can include ratios of a total concentration of an ion family which is primarily (and can be entirely) in an uncharged state, divided by the mobility of the (possibly locally non-existent) charged state of that species.

Like the Jovin relation, the Alberty relation imposes a constraint involving the concentration of both neutral and ionic species. Conceptually, the Alberty condition states that displacement of one ion from a zone (e.g., the LE co-ion zone being replaced by the influx of a plateau zone co-ion) must be accompanied by a decreased concentration if the mobility of the new ion is lower. One important consequence of the Alberty relation is that successively trailing plateaus in ITP tend to have progressively lower concentrations (as one moves from LE to TE).

Together, the Jovin and Alberty functions, net neutrality, and conservation of current help us design, analyze, and build intuitions for the ITP of weak electrolytes. We will therefore invoke these relations in subsequent sections of this review.

### 4.3 Plateau-mode ITP for weak electrolytes

As an important example of plateau-mode ITP of weak electrolytes, here were discuss the common situation of a well-buffered anionic ITP system. Such as system can often use a weak base counterion to buffer the LE as well as every trailing plateau zone. Consider the ITP system depicted in **Fig. 7**. Here, $C$ is the weak base (comprising the neutral species and the cation with $z = +1$), $X$ is a weak acid analyte, and a leading ion $L$. As a specific example of a leading ion in anionic ITP, we here consider chloride, $Cl^-$ (although the discussion will apply generally to any leading ion). Note that the conjugate acid ion $C^+$ implies migration of the species family $C$ from LE toward the TE and hence helps establish acid/base equilibrium in all zones. To make the LE zone (and the trailing sample zones) a robust buffer (c.f. **Section 8**), a good choice for the total concentration of counterion $c_C$ is approximately twice the concentration of HCl (equivalently, $Cl^-$ since HCl is fully dissociated) in the LE zone. We can formulate and solve for the co-ion and counter ion concentrations in all zones trailing the LE zone (including the ATE). One approach to this is to formulate current conservation, mass balances, and the ITP condition in an approach similar to what we did in **Section 3** for the case of strong electrolytes. However, the formulation of the problem is much more compact if we restrict our discussion to at most univalent weak electrolytes in the safe and moderate pH range, and so leverage the Jovin and Alberty functions. Interestingly, the resulting formulation will hold for both strong and weak electrolytes with the understanding that the mobilities for strong electrolytes are equal to the fully dissociated values (which are independent of the location).

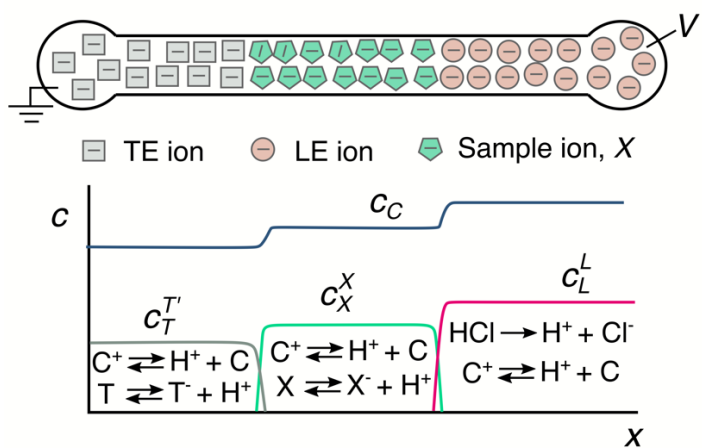

**Figure 7**: Schematic of plateau mode anionic ITP. Top panel depicts spatial location of leading $L$, trailing $T$, and sample $X$ co-ions in the channel, and the bottom panel depicts the associated concentration profiles. Shown also in the bottom panel are the counter ion $C$ concentration profile and the local (rapid) acid/base chemical equilibrium reactions within each zone.

The concentration of a trailing univalent sample species $X$ (strong or weak electrolyte) in its ITP zone can be related to the LE concentration using the Jovin and Alberty conservation laws. First, Jovin's condition is applied to the system in **Fig. 7** for the spatial location currently occupied by zone X. We evaluate the Jovin summation at the initial state (when that location was occupied by the LE zone) and the current state as

$$c_L^L - c_C^L = c_X^X - c_C^X, \qquad (75)$$

where $c_i^j$ denotes the total (analytical) concentration of species $i$ in zone $j$. Here, subscripts $i = L$, $X$, and $C$ refer to the leading ion, species $X$, and counterion, respectively, and superscripts $j = L$ and $X$ are the leading and sample zones, respectively.

Next, Alberty's condition for the leading and sample zones at the initial and final states in zone $X$ gives

$$\frac{c_L^L}{|\mu_L^L|} + \frac{c_C^L}{|\mu_C^L|} = \frac{c_X^X}{|\mu_X^X|} + \frac{c_C^X}{|\mu_C^X|}. \qquad (76)$$

Here, $|\mu_i^j|$ refers to the magnitude of fully ionized mobility associated with species family $i$ in zone $j$. Combining Jovin and Alberty relations (Eqs. (75) and (76)), we obtain the total concentrations of the species $X$ and the counterion $C$ in zone $X$ as

$$c_X^X = \left(\frac{\frac{|\mu_C^X|}{|\mu_L^L|} + 1}{\frac{|\mu_C^X|}{|\mu_X^X|} + 1}\right) c_L^L + \left(\frac{\frac{|\mu_C^X|}{|\mu_C^L|} - 1}{\frac{|\mu_C^X|}{|\mu_X^X|} + 1}\right) c_C^L. \qquad (77)$$

and

$$c_C^X = \left(\frac{\frac{|\mu_C^X|}{|\mu_L^L|} - \frac{|\mu_C^X|}{|\mu_X^X|}}{\frac{|\mu_C^X|}{|\mu_X^X|} + 1}\right) c_L^L + \left(\frac{\frac{|\mu_C^X|}{|\mu_C^L|} + \frac{|\mu_C^X|}{|\mu_X^X|}}{\frac{|\mu_C^X|}{|\mu_X^X|} + 1}\right) c_C^L. \qquad (78)$$

Note the latter expression for $c_C^X$ follows from Eq. (65) and charge neutrality. This is a closed-form solution for the total concentration of the ion $X$ in zone X. We next simplify the above expressions by neglecting the effects of ionic strength on the mobility of species and neglecting temperature differences between the leading and trailing zones (see Refs.[86,162,163] for further discussion of these effects). Under these assumptions, the fully ionized mobility of the counterion (originating in the leading zone and moving to the sample zone) remains unchanged, so that $|\mu_C^X| \approx |\mu_C^L| = |\mu_C|$. For simplicity, we drop the superscripts on the fully ionized mobilities of $X$ and $L$ and assume that these values are constant (independent of zone), i.e., $|\mu_X^X| = |\mu_X|$ and $|\mu_L^L| = |\mu_L|$. Whence, $c_X^X$ and $c_C^X$ can be simplified as

$$c_X^X = \frac{|\mu_X|}{|\mu_L|} \left(\frac{|\mu_C| + |\mu_L|}{|\mu_C| + |\mu_X|}\right) c_L^L \qquad (79)$$

and

$$c_C^X = c_C^L + \frac{|\mu_C|}{|\mu_L|} \left(\frac{|\mu_X| - |\mu_L|}{|\mu_C| + |\mu_X|}\right) c_L^L. \qquad (80)$$

Again, it is important to emphasize the mobilities here are each fully ionized mobilities (for valence magnitudes of unity) while the concentrations are total (analytical concentrations—involving charged and uncharged species). Note also that, for a strong electrolyte leading ion like chloride, the analytical concentration is equal to the ionic concentration. Species family $X$ in this equation is any species family comprising an anion which occupies a space formerly occupied by the leading ion (including any plateau sample ion or the TE ion of the ATE zone). Note further that, for the special case of strong electrolytes, we can drop all superscripts and show that, for the case of univalent species, Eqs. (79) and (80) are equivalent to Eqs. (59) and (60), respectively, described earlier.

As with strong electrolytes, a good rule of thumb for typical weak electrolyte ITP system is that a trailing ion $X$ achieves a total concentration of about $c_X^X \approx 0.5$ to $0.9$ of $c_L^L$. In other words, the total (analytical) concentration is "governed" by the LE co-ion $c_L^L$, while the ionic concentration of the trailing ion depends on $c_L^L$ and pH (which can depend on the relative $pK_a$ of the trailing and counterion). Note that the adjusted TE properties can be obtained by simply replacing species $X$ and zone X with properties of the TE co-ion $T$ and the adjusted TE zone ATE, respectively. Finally, for a trailing ion with a $pK_a$ significantly lower than that of the weak base, the concentration of H$^+$ (equivalently, the pH) in any trailing zone can be solved in terms of the analytical concentrations. In this case, the trailing anion is approximately fully ionized, and hence we obtain a simple expression for $c_H$ (from Eq. (20)) given by

$$c_H = \frac{K_{B,0} c_X^X}{c_C^X - c_X^X}. \tag{81}$$

For such anionic ITP, the pH's of trailing zones are typically near that of and slightly higher than the leading electrolyte zone.

The pH buffering dynamics presented in this section can be summarized as follows. The typical situation is to choose a common (to both TE and LE) counterion with a convenient $pK_a$. This ion will serve as the pH buffering species and the LE co-ion will serve as a titrant. The counter ion migrates back into the TE with a concentration that is roughly twice that of the TE co-ion. In this way, the LE and TE co-ions of anionic play the role of titrant to the weak base buffering counter-ion. Again, we refer the reader to Refs.[85,86] for a good introduction and/or review of the electrophoresis of weak electrolytes including pH buffers.

### 4.4 Illustration of weak electrolyte theory and limitations

**Figure 8** shows simulation results from Ref.[132] of a cationic ITP process consisting of two buffer solutions comprising 10 mM univalent strong cations Na$^+$ and K$^+$ buffered by 20 mM of a common univalent (counterion) weak acid, MES. The entire system is buffered around pH of 6.1 and is within the safe and moderate pH range, so both Jovin and Alberty conservation principles are applicable. Indeed, this is seen by the value of the Jovin $\left(W_{00}^{(2)}\right)$ and Alberty $\left(W_{00}^{(1)}\right)$ functions at the initial and final states on the right-side plots of **Fig. 8** (these variables are equivalent to the small case $w$ used in this review). Notice also that the value of MES (counterion) and Na$^+$ (trailing ion) "adjusts" in the adjusted TE zone based on the concentration of the LE and the pH of the zone (Eqs. (79) and (80)). Similar to the case depicted in **Fig. 8**, many of the species of interest for ITP, especially for biological applications, are weak electrolyte species. In **Section 8**, we will discuss the rationale behind the need for buffering in ITP using weak acids and bases, and also provide some guidelines for the choice of buffering ions. As with strong electrolytes, the weak electrolyte theory presented above for steady state concentration profiles of plateau zones can be used to validate the accuracy of numerical codes. We here also point the reader to Refs.[1,164–168] which contain extended lists of the $pK_a$, fully ionized mobility and possible ionization states (valence), and other physicochemical properties relevant to ITP of both strong and weak acids and bases.

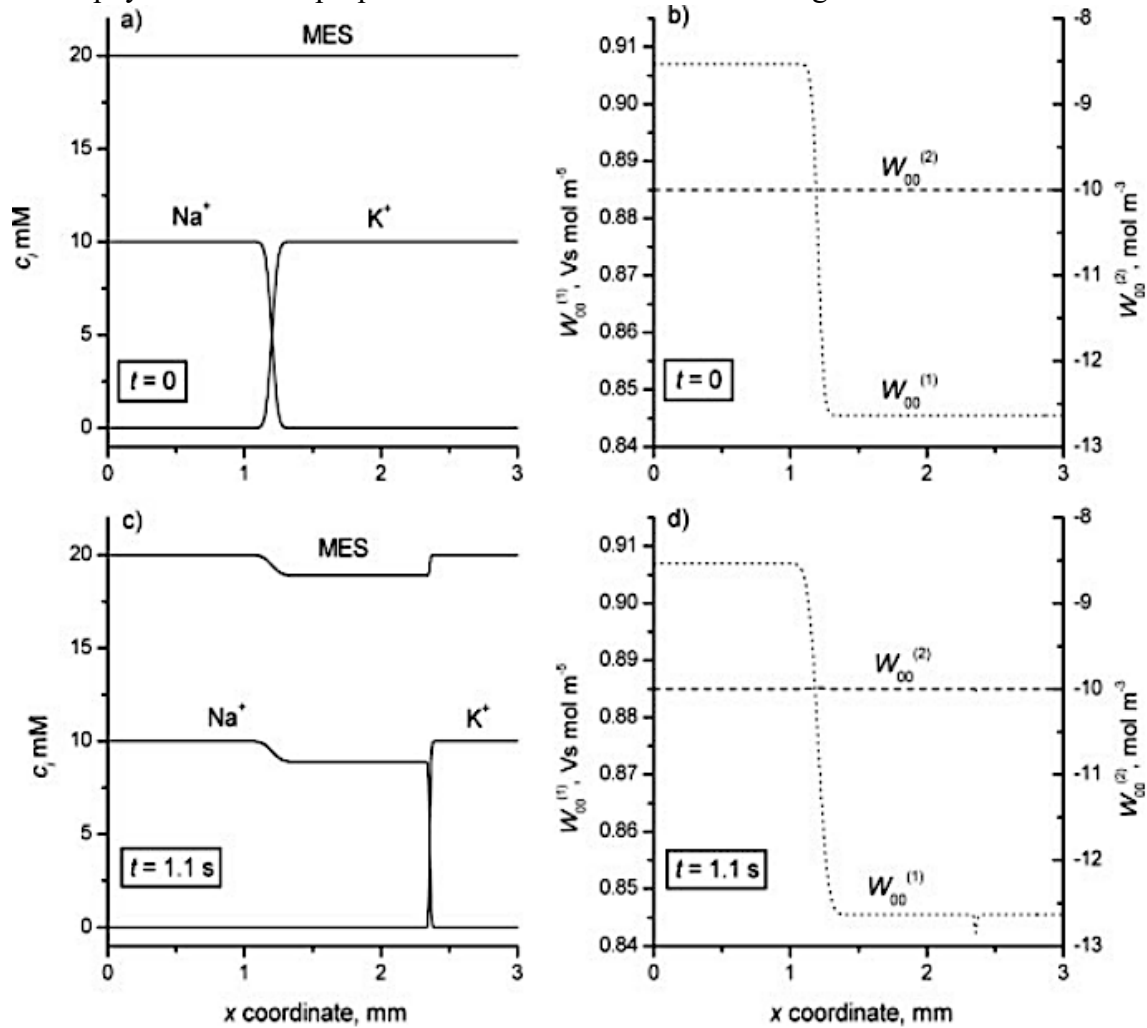

**Figure 8**. Simulated cationic ITP run for LE mixture consisting of 10 mM potassium ions and 20 mM MES, and TE mixture consisting of 10 mM sodium ions and 20 mM MES (pH 6.1). (A) and (C) show the concentration fields at $t = 0$ and $t = 1.1$ s, respectively. (B) and (D) show values of the Alberty[41] $\left(W_{00}^{(1)}\right)$ and Jovin[161] $\left(W_{00}^{(2)}\right)$ functions versus the channel location $x$ at $t = 0$ and $t = 1.1$ s, respectively. Simulations were performed using the Simul software,[144] with parameters: capillary length of 3 mm, driving voltage of 50 V, $x$-coordinate mesh consisting of 1000 nodes. Figure is reproduced with permission from Ref. [132]. Copyright 2007 John Wiley and Sons.

Though the theory presented in **Sections 4.1-4.3** is applicable only to univalent weak (and strong) electrolytes within the safe and moderate pH range, we note this covers most practical microfluidic ITP applications. We refer to Refs.[87,132,140,169] for modified conservation principles governing ITP systems which are acidic or alkaline (outside the safe pH range), including systems containing ampholytes and significant ionic strength effects. Note that systems involving multivalent weak electrolytes often require additional conservation laws which can in principle be formulated as solutions to an eigenvalue problem derived from the species conservation equations.[132] However, there are currently no generally applicable analytical solutions for such systems, and they often do not admit any Kohlrausch-type, or Jovin and Alberty-type conservation principles. ITP systems which do not admit such conservation are also referred to as *non-conservative*.[132] An example of such system is a buffer consisting of sodium (titrant) and oxalate (buffering ion) within certain ranges of concentration.[170]

We also note that the physicochemistry of ITP summarized here is strictly only applicable for ionic strengths lower than about 20 mM. Higher ionic strengths can require additional physical considerations due to the effects of ionic strength on ion mobility.[169] Such effects primarily cause a reduction in the mobility of ions. Ionic strength also (and independently) affects p$K_a$ values of weak electrolytes. There exist several models of varying degree of complexity to capture such non-ideal effects on ITP.[169,171–174] In addition, effects of Joule heating/temperature differences between zones can also affect the physicochemical properties of ions and in turn affect its ionization state and effective mobility of the ions.[162,163,175,176] Prediction of ITP dynamics in such systems often requires a computational approach (see **Section 9**). Nevertheless, the relations presented here are still very useful in design of ITP systems including back-of-the-envelope estimates and buffering requirements with reasonably good approximation for ionic strengths less than or about 100 mM.

## 5. Peak mode ITP dynamics

In this section, we review simple theory and models useful in prediction of analyte accumulation rate and the focused species shape in peak mode ITP. Peak mode is a useful mode of ITP wherein one or more trace species are accumulated at an interface between a pair of (or multiple pairs of) plateau regions (see **Fig. 1**).[97] Peak mode ITP is best suited for trace (i.e. very low initial concentration) analytes and is an excellent focusing (or pre-concentration) technique. Furthermore, peak mode may be necessary if working with a sample which has limited solubility. In its simplest form, peak mode focuses a single targeted analyte between the LE and ATE zones. If the analyte effective mobility is sufficiently larger (smaller) than the trailing (leading) plateau, then the analyte ion tends to be focused within a very narrow peak approximately Gaussian in shape. In peak mode, the analyte contributes insignificantly to ionic current and so local-ionic current and electric field distribution is dominated by the effects of the two neighboring plateau zones. Peak mode is generally associated assays wherein the analyte is directly detectable. Example applications of this include selective focusing and direct detection of fluorescent species,[27,30,110] extraction and purification of a target species (most notably nucleic acids but also proteins from blood, tissues or cell lysate),[31,32,77,109,177] and preconcentration and co-focusing of reactants in order to accelerate chemical reactions (see **Section 7** and **11**).[39,111] Of course, the very early stages of the focusing of any ion at virtually any initial concentration (e.g. on the order of or less than the concentration of the ATE) is a peak-mode focusing which, if allowed to proceed sufficiently long, the focused analyte will eventually transition into a plateau and displace and separate the adjoining plateaus.

### 5.1 Analyte accumulation rate

We here present a control volume (CV) analysis useful in deriving relations for the accumulation rate of trace analytes in peak mode ITP. For this, we choose a CV that moves with the ITP zone at $V_{ITP}$ and extends from the adjusted TE region to the LE region across the interface, as depicted by the dotted lines in **Fig. 1B**. We denote the sample (a.k.a. analyte) ion of interest by $s$ and, for simplicity, consider constant current ITP operation in a channel with uniform cross-sectional area $A$ (see Bahga et al.[121] for analytical theory applicable to ITP in channels of varying cross section). For simplicity, we further assume negligible pressure driven flow and electroosmotic flow (see for example Garcia-Schwartz et al.[61] for the effects of internal pressure gradients due to electroosmotic flow on peak-mode ITP). For this simple peak-mode case, the species transport equation for the focusing sample in the moving frame (based on Eq. (3)) as[39]

$$\frac{\partial c_S}{\partial t} = \frac{\partial}{\partial x}\left(-\mu_S^{T'} E^{T'} c_S^{T'} + V_{ITP} c_S^{T'} + D \frac{\partial c_S}{\partial x}\right). \tag{82}$$

Here, the Galilean reference frame is moving at $V_{ITP}$ and $x$ is measured along the direction of travel of the anionic ITP zone. The subscript $S$ refers to some ionic sample species that focuses between the TE and LE zone. Here, we have introduced the primed quantities to indicate the local evaluation in our control volume and within the adjusted TE zone. We now consider the case of a semi-infinite injection of the sample where it is homogenously mixed with the TE buffer (including in an upstream reservoir). We then perform an integration of this equation over a CV including a large portion of the adjusted TE and bound by the ITP interface. We apply the divergence theorem to the right hand side and derive

$$\int \frac{\partial c_S}{\partial t} dV = \int \left(-\mu_S^{T'} E^{T'} c_S^{T'} + V_{ITP} c_S^{T'} + D \frac{\partial c_S}{\partial x}\right) dA. \tag{83}$$

The left-hand side of Eq. (83) is simply the rate of accumulation of total moles of species $N_S$ within the ITP peak (e.g., in units of mol/s). The right-hand side of Eq. (83) involves the summation of fluxes over the control surfaces which include the top and bottom channel walls (where sample flux is zero) and the left (adjusted TE) and right (within the LE, just to the right of the TE-to-LE interface) faces of the CV. On the right face of the CV, sample concentration is locally zero for our case, so the corresponding integrand is zero. On the left face of the CV (well within the adjusted TE region as per our CV definition), the sample concentration is locally uniform (i.e., no diffusive flux). Therefore, Eq. (83) simplifies to

$$\frac{dN_S}{dt} = \left(\mu_S^{T'} E^{T'} - V_{ITP}\right) c_S^{T'} A = \left(\mu_S^{T'} - \mu_T^{T'}\right) E^{T'} c_S^{T'} A = p_{S,TE} V_{ITP} c_S^{T'} A. \tag{84}$$

where we have used $V_{ITP} = \mu_T^{T'} E^{T'}$ and introduced the so-called separability[178,179] factor $p_{S,TE} = (\mu_S^{T'}/\mu_T^{T'}) - 1$. Note that $c_S^{T'}$ is the adjusted concentration of the sample in the adjusted TE zone which is equal to the product of the initial sample concentration in the reservoir $c_S^W$ and the ratio of electric fields in the TE and adjusted TE zones.[64] Eq. (84) shows that the sample accumulation rate in ITP is proportional to the initial sample concentration in the reservoir and the mobility difference (equivalently, the separability) of the sample and TE co-ions in the adjusted TE zone. The terms on the right-hand side of this equation are (for constant current) constant (in time), so the total accumulated sample amount as a function of time is simply given by

$$N_S(t) = \left(\mu_S^{T'} - \mu_T^{T'}\right) E^{T'} c_S^{T'} A\, t = p_{S,TE} V_{ITP} c_S^{T'} A\, t. \tag{85}$$

Further, for the present case where the effects of bulk flow are negligible, we have $l = V_{ITP} t$. Here $l$ is the distance swept by the ITP peak along the channel in time $t$. Thus, the amount of accumulated sample can be expressed in terms of $l$ as

$$N_S(x) = p_{S,TE} c_S^{T'} A l = p_{S,TE} c_S^{T'} \forall, \tag{86}$$

where $\forall$ is the volume swept by the ITP zone as it migrates through the channel (a concept very useful in the more complex case of channels of varying cross-sections[121]). Eqs. (85) and (86) show the total moles of sample accumulated in the ITP peak increases proportionally with time and with the distance swept by the ITP peak along the channel. We note that our formulation above is valid in the limit of negligible depletion of the trace analyte from the TE reservoir, e.g., when the reservoir to channel volume ratio is high, so the concentration of the sample in the TE reservoir is approximately constant. Refer to Eid and Santiago[64] for derivation of accumulation rates in ITP in the regime of significant depletion of sample from the reservoir. Note further that the expressions for peak mode accumulation rate (Eqs. (84)-(86)) were obtained for the case when the sample was initially dissolved in the TE. Similar expressions can be obtained for the case when the sample is dissolved in LE by simply replacing $p_{S,TE}$ and $c_S^{T'}$ in these equations with $p_{S,LE} = 1 - (\mu_S^L/\mu_L^L)$ and $c_S^L$, respectively.[64]

We next present below simple heuristics to approximate the effects of bulk liquid flow velocity $u_b$ (area-averaged velocity of bulk solvent) in the above control volume analysis. Bulk flow can be caused by pressure-driven flow, electroosmotic flow, or both.[61] For this discussion, we assume the diffusion time for the species along the channel transverse direction (e.g., radius in a cylindrical channel) is significantly smaller than the transport time of interest. Therefore, the local effect of bulk flow can be described by a simple frame of reference change given by the area-averaged bulk flow velocity. Non-uniform velocities across the cross-sectional area of the channel (e.g., due to the no-slip condition) would also tend to disperse the sample, but we assume the control volume axial dimension is longer than the dispersed length of the ITP zone. In this regime, the distance swept by the ITP peak is $l = (V_{ITP} + u_b) t$. Importantly, since bulk motion does not influence the rate of focusing (only drift velocities can focus species, and not the divergence-free bulk flow), the rate of focusing $dN_S/dt$ is the same as in the case with no bulk flow (Eq. (84)). However, bulk flow affects the relation between spatial location $l$ and amount of sample accumulated as follows:

$$\frac{dN_S}{dx} = \frac{dN_S}{dt}\frac{dt}{dx} = \frac{dN_S}{dt}\frac{1}{V_{ITP} + u_b} = \frac{\left(\mu_S^{T'} - \mu_T^{T'}\right) c_S^{T'} E^{T'} A}{\left(\mu_T^{T'} E^{T'} + u_b\right)}. \tag{87}$$

where we have used $V_{ITP} = \mu_T^{T'} E^{T'}$. Integrating Eq. (87), we have

$$N_S(x) = \frac{\left(\mu_S^{T'} - \mu_T^{T'}\right) c_S^{T'} E^{T'}}{\left(\mu_T^{T'} E^{T'} + u_b\right)} Al = \frac{p_{S,TE} V_{ITP} c_S^{T'}}{(V_{ITP} + u_b)} Al. \tag{88}$$

Note that the denominator $V_{ITP} + u_b$ in Eqs. (87) and (88) can tend to very small values implying a large amount of analyte for very long times. One example of this is the case of counter-flow ITP where pressure-driven flow is used to counteract the electromigration of the ITP zone[60,62,180] (see **Fig. 9**).

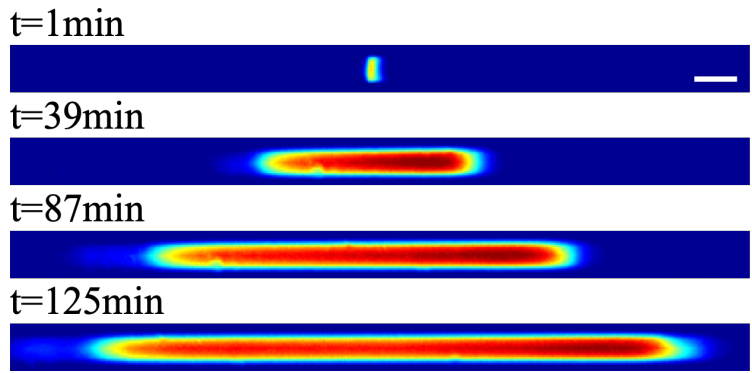

t=1min

t=39min

t=87min

t=125min

**Figure 9**. Experimental images showing accumulation of 10 nM AF488 dye versus time in counterflow ITP. A 1kV voltage is applied in a 5 cm capillary. The leading and trailing electrolyte are respectively 750 mM Tris-HCl and 10 mM Tris-HEPES. Pressure-driven flow is used to counteract ITP electromigration. Scale bar: 50 μm. Results shown here are from an unpublished study performed by J. Sellier, F. Baldessari, A. Persat, and J.G. Santiago in the Santiago lab at Stanford University.

For the special case when bulk flow is caused due to electroosmotic flow alone, $u_b = \langle \mu_{eof} E \rangle$, where $\mu_{eof}$ and $E$ are the local electroosmotic mobility and electric field, respectively, and $\langle \; \rangle$ is an axial average operator. This type of axial average is discussed by, for example, Bharadwaj and Santiago[35] and Garcia-Schwarz et al.[61]. In the case of EOF alone, $\langle \mu_{eof} E \rangle$ is a function of the location of the ITP zone and it tends to increase as more of the channel is occupied by TE since the TE typically has lower ionic strength, slightly higher pH (for anionic ITP), and higher electric field (see Garcia-Schwarz et al.[61]).

We note that the heuristic solution presented above should be derivable from a Taylor-Aris type analysis of the transport. Such analysis would assume that the diffusion time across the cross-section of the channel is significantly smaller than the advection times of interest. The analysis would then proceed with an area-averaging of the transport equations and a Galilean transfer to a reference frame moving at velocity $(V_{ITP} + u_b)$.

We note also that the analyses presented in this section assumed that the sample concentration flowing into the control volume was constant. This assumption will not hold indefinitely, for example, for channels of significant volume relative to the channel volume. Eventually, the sample in a TE reservoir is depleted and this depletion can result in a decaying rate of accumulation. A model for these dynamics was proposed by Rosenfeld et al.[98].

Lastly, the analyses in this section were limited to a semi-infinite injection configuration. Consistent with this, we derived relations for the total amount (in moles) of accumulated species which increased in proportion to time. However, the analysis presented here could be extended to the case of finite injection. In the case of finite injection, the accumulated sample would initially increase in proportion to time and then achieve a steady-state value as the sample is depleted from the (finite) injection zone.

### 5.2 Focused analyte shape in peak mode

The shape of the focused analyte in peak mode ITP is important in determining the relations among maximum peak concentration, accumulated amount, and the area under the concentration versus location curve. Peak shape can also strongly influence the sensitivity and resolution of detection.[97,181] In the limit of low electric fields, negligible electroosmotic flow, and no applied pressure-driven flow, ITP interfaces (including zone-to-zone interfaces $\delta$ and the width of the focused analyte peak $\delta_a$ in peak-mode ITP) are predicted well by a one-dimensional balance between non-uniform electromigration and axial molecular diffusion.[16,61,97] In peak-mode ITP, this regime yields an ITP focused zone which is small ($\delta \approx \delta_a$ and is on the order of the smallest channel cross section dimension), approximately Gaussian (provided the mobility of the analyte is significantly different from the neighboring plateau co-ions[61]), and inversely proportional to the current density (c.f. **Section 3.2**). In practice, this ideal one-dimensional picture of ITP is most often disrupted by one or more of the following four effects: Tailing of the ITP peak due to a focused species mobility approaching that of a neighboring plateau (an effect different than Taylor-

Aris type dispersion[59,61]), dispersion of the ITP peak due to external pressure gradients,[62] dispersion due to residual electroosmotic flow,[61,139] and electrokinetic instabilities[182].

We first discuss the first of the four aforementioned effects. For this case, the LE-to-TE interface width $\delta$ is typically smaller than the width $\delta_a$ of the analyte distribution. For simplicity, we consider tailing behavior for a focused sample peak under ideal, dispersion-free conditions. This situation represents the common case of peak mode ITP experiments performed at moderate electric fields, and for which EOF is significantly suppressed and there is no external pressure driven flow. EOF suppression can be achieved by both operating at sufficiently high buffer concentrations (e.g. at 50 mM and above) and by introducing additives which suppress EOF such as PEO or PEG (c.f. **Section 8** for further discussion).[183–189]

Garcia-Schwarz et al.[61] was the first study to take into account an analyte's physical properties to predict non-ideal sample distributions (e.g., tailing, dispersion) in peak mode ITP. They presented a one-dimensional, area-averaged, three-zone model to capture the shape of sample zones of fully ionized species focused in peak mode ITP (including the effects of Taylor-Aris-type dispersion due to induced secondary flows). Here we present a simplified version of the analysis of Garcia-Schwarz et al.[61] (which included Taylor-Aris analyses that describe how dispersion couples with tailing). The steady state behavior of analyte concentration can be obtained from Eq. (82) (which is applied in the frame of reference of the moving ITP interface) and is described by the following equation:

$$-\mu_S E c_S + D \frac{dc_S}{dx} + V_{ITP} c_S = 0. \tag{89}$$

Eq. (89) is obtained by integrating the right-hand side of Eq. (82) and noting that the constant of integration is zero (obtained by evaluating it far away from the ITP interface). Since the analyte concentration distribution spans the TE and LE zones, we do not use superscripts in Eq. (89). We will instead assume the analyte has approximately the same mobility in the LE and ATE. Note in Eq. (89), we assume that $D$ is simply the diffusion coefficient of the focused species while Garcia-Schwarz et al.[61] assume an effective Taylor dispersion diffusivity to model additional effects due to EOF-driven dispersion (analytical result shown as black dashed lines in **Fig. 10**). In the dispersive case, dispersion effectively acts to smear and widen the already tailing dispersion-free distribution (the blue dotted curves in **Fig. 10** are the distributions of the diffusion-only case). Assuming fully ionized mobilities, Eq. (89) can be integrated directly to yield

$$c_S = \kappa \exp\left(\int \frac{\mu_S E - V_{ITP}}{D} dx\right), \tag{90}$$

where $\kappa$ is a constant. This compact and intuitive result is implicit in terms of the spatial integral involving the unknown electric field distribution but describes the shape of the sample peak.

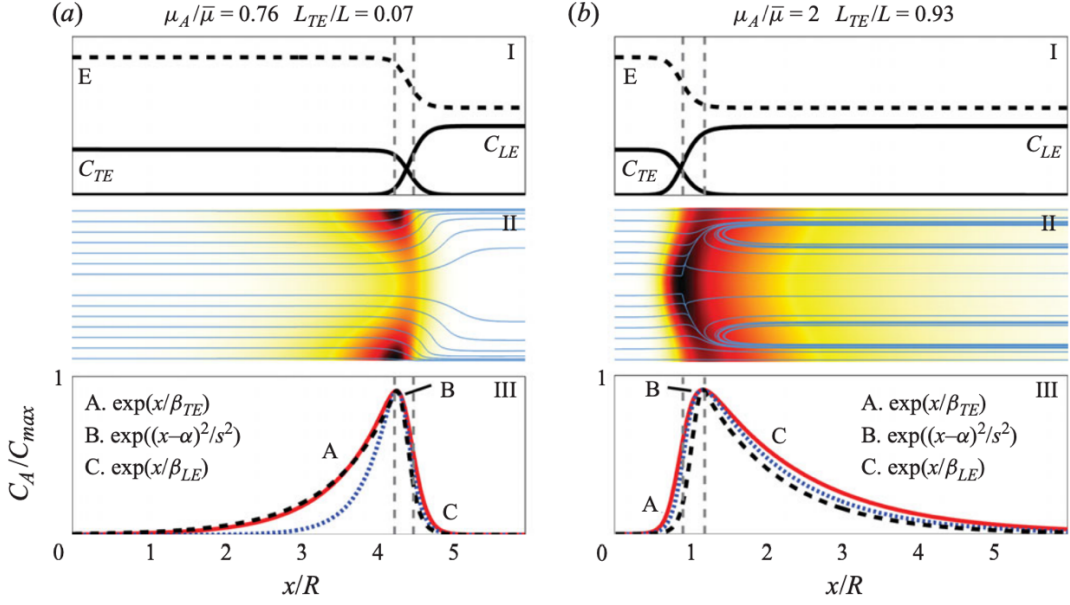

**Figure 10**. Simulation results for sample concentration fields in peak mode ITP under conditions where the sample tails into the (A) adjusted TE region, and (B) LE region. (Top row) Axial concentration profiles of the leading, trailing co-ions and the common counterion. (Middle row) 2D concentration profiles of the sample obtained from simulation. Superimposed lines are tangents to the combined electromigration and convective mass flux components. (Bottom row) Area-averaged distributions corresponding to the dispersed and non-dispersed simulations are shown as red (solid) and blue (dot) curves, respectively. Analytical model predictions are shown by the dashed curve. A uniform EOF mobility is assumed in this figure. We refer to Garcia-Schwarz et al.[61] for details about the simulation and model parameters. Figure is reproduced with permission from Ref.[61]. Copyright 2011 Cambridge University Press.

Consider the interface region in which the analyte is focused as shown in **Figures 10A** and **10B**. The magnitude of the sample concentration in **Fig. 10** is typically everywhere much lower than that of the adjoining trailing and leading buffering ions (**Fig. 10**). The tail of the sample distribution into the adjusted TE (LE) region to the left (right) is caused by a sample mobility near to that of the trailing (leading) ion. As discussed in Garcia-Schwarz et al.[61], it is useful to conceptually divide the interface region into three zones: $A$, $B$, and $C$. As a heuristic approximation, regions $A$ and $C$ can be treated as regions of uniform electric field, and region $B$ as one with a constant (and uniform) electric field gradient. Under this heuristic, the integral equation in Eq. (90) can be explicitly solved for $c_s$ as

$$c_S = \begin{cases} a \exp\left(\dfrac{x}{\beta_{TE}}\right); & \text{zone A: sample tailing into the TE} \\ b \exp\left(-\left(\dfrac{x+\alpha}{\sigma}\right)^2\right); & \text{zone B: Gaussian for linear E field} \\ c \exp\left(-\dfrac{x}{\beta_{LE}}\right); & \text{zone C: sample tailing into the LE} \end{cases} \quad (91)$$

The constants $a$, $b$, $c$, $\alpha$ and $\sigma$ in Eq. (91) can be found in Garcia-Schwarz et al.[61] and these are obtained by ensuring that $c_s$ values and its gradients match at the $A$-$B$ and $B$-$C$ interfaces, and for a specified integrated amount of focused sample. The three sections of the $c_s$ curve are well approximated by decaying, first-order exponential tails in regions $A$ and $C$, and a local Gaussian "peak" region in region $B$. The constant $\sigma$ scales the Gaussian width and is well approximated by 0.9 $\delta$, where $\delta$ is the LE-to-TE interface width. The length scales $\beta_i$ describe the length of the sample peak "tail" into $i$ = TE ($T$) or LE ($L$) zones, and this can be expressed as

$$\frac{\beta_{TE}}{\delta} \sim \frac{\mu_i}{\mu_L} \frac{\mu_L - \mu_T}{\mu_S - \mu_T} \quad \text{and} \quad \frac{\beta_{LE}}{\delta} \sim \frac{\mu_i}{\mu_T} \frac{\mu_L - \mu_T}{\mu_L - \mu_S}. \tag{92}$$

Thus, the effective width of the focused sample $\delta_a$ scales as $\delta + \beta_{TE} + \beta_{LE}$. For the special case of analyte mobility $\bar{\mu} = 2\mu_L\mu_T/(\mu_L + \mu_T)$, the concentration distribution of the focused sample is symmetric. Further, for $\mu_S > \bar{\mu}$ and $\mu_S < \bar{\mu}$ the sample will have a longer penetration length in the LE and TE, respectively. In other words, analyte mobilities $\mu_S$ near those of the TE (LE) can cause significant tailing of analyte into the TE (LE). These trends predicted by the model were validated with numerical simulations and experiments by Garcia-Schwarz et al.[61].

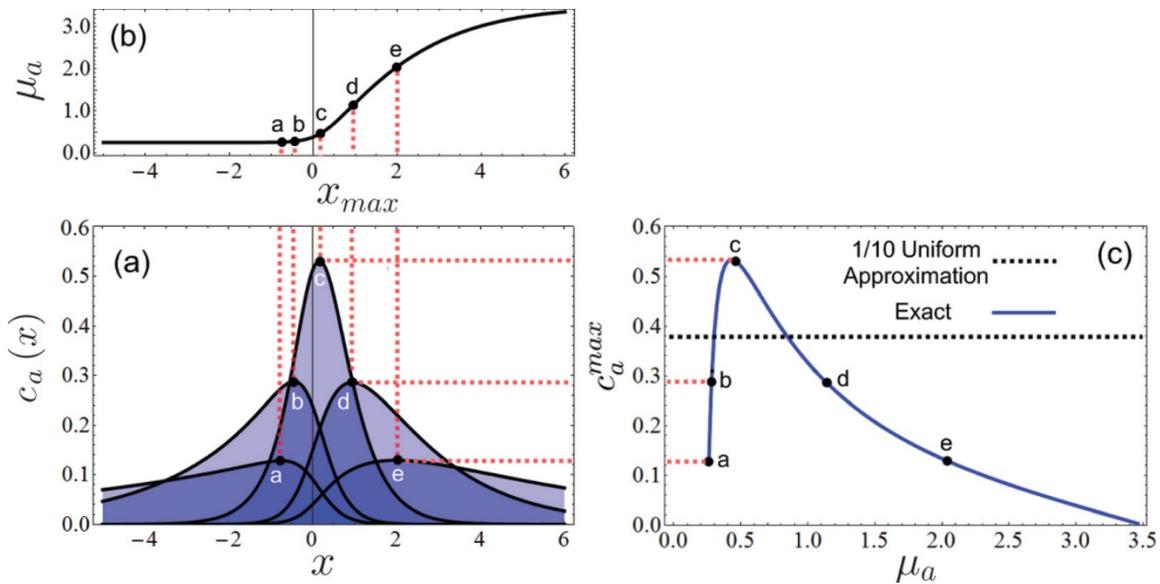

**Figure 11.** Analytical model predictions for sample distribution in peak mode ITP. Effect of sample ion mobility on (A) sample concentration profiles $c_a$ ($c_s$ in our notation) versus $x$, and (B) the curve relating the maximum sample concentration (and its axial position). (C) Peak concentration as a function of the sample mobility. Figure is reproduced with permission from Ref. [59]. Copyright 2014 AIP Publishing.

Rubin et al.[59] obtained exact analytical solutions and closed form engineering approximations for the cross section averaged, one-dimensional analyte distribution in peak mode ITP (diffusion-dependent, and no EOF or pressure driven flow) involving systems of weak electrolytes. Specifically, they obtained exact solutions for leading-, trailing-, and counter- ion concentration profiles and the electric field distribution, which were used to derive the focused sample concentration profile. The exact solution involves the hypergeometric function and is more complex to implement in practice. However, the useful engineering approximation for the sample concentration, as derived in Rubin et al.[59], is given by

$$c_S(x) \approx \frac{N}{A\delta} \frac{\sin(\pi b)}{\pi} \frac{\exp\left(\frac{x}{\delta_a}\right)}{1 + \exp\left(\frac{x}{\delta}\right)}, \tag{93}$$

where $b = \delta/\delta_a$, $\delta^{-1} = \mu_T^{-1} - \mu_L^{-1}$, and $\delta_a^{-1} = \mu_T^{-1} - \mu_A^{-1}$ for a monovalent analyte. In this formulation, the length scales $\delta$ and $\delta_a$ are normalized by a characteristic ITP length scale $RT\mu_c/FV_{ITP}$, and the mobilities are normalized by the counterion mobility. Importantly, Eq. (93) provides the dependence of the concentration profile on the mobility of the analyte $\mu_A$ (normalized by the counterion mobility). Eq. (93) is plotted in **Fig. 11A** for various values of the analyte mobility. Note that the maximum value of the focused sample concentration and its corresponding spatial location are strong functions of the analyte mobility (relative to that of the adjoining plateaus); see **Figs. 11B** and **11C**.

Note that the models reviewed above[59,61] are experimentally validated, one dimensional, area-averaged sample concentration distributions under ideal peak mode ITP conditions (i.e., with no dispersion).

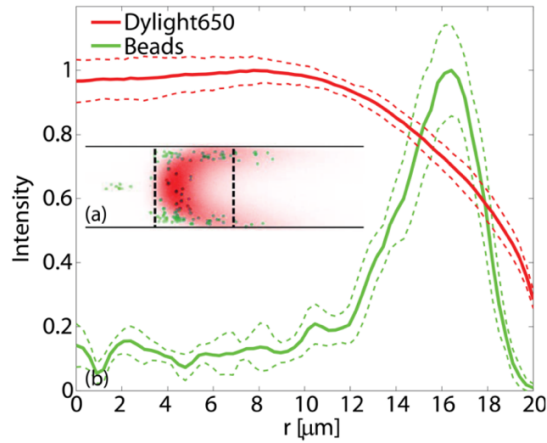

**Figure 12**. Experimental data showing effects of diffusion-dependent, radial separation of two co-focused species (DyLight650 dye and 2.8 μm magnetic beads) in counterflow ITP. The two species (green vs. red in the figure) have similar mobilities but significantly different diffusivities. Figure is reproduced with permission from GanOr et al.[60] Copyright 2015 AIP Publishing.

We next mention some work on analyses of dispersed ITP peaks. See Refs.[60–62,143] for detailed discussions around models which include dispersion effects on sample zone dynamics due to EOF and externally applied pressure gradients (e.g., in counterflow ITP). For readers interested in the effects of dispersion due to EOF or pressure driven flow on the LE-to-TE interface itself (independent of the focused sample dynamics), we refer to Refs.[8,139,190–192]. Quantitative studies of the axial distribution of ITP peak mode under dispersive conditions typically employ Taylor-Aris-type dispersion models and hence describe area-averaged, one-dimensional concentration fields.[61,192]

Lastly, we briefly mention some studies which attempted to consider two- or three-dimensional ITP shapes under dispersion. The source of the dispersion can be an externally applied pressure driven flow (e.g., a pressure difference between the LE and TE reservoirs) or the dispersive velocities associated with axially non-uniform EOF (which causes internally generated and non-uniform pressure-driven flows). For example, Schönfeld et al.[139] and Garcia-Schwarz et al.[61] (see **Fig. 10**) each observed that time varying, internal pressure gradients and secondary flow generated due to non-uniform electroosmotic flow in microchannels can lead to dispersed and axially distorted LE-to-TE interfaces and focused sample concentration profiles (see **Fig. 10**). Schönfeld et al.[139] reported in qualitative observations that the LE-to-TE boundary in a microchannel ITP experiment in the presence of EOF transitioned from a ⊂-shape (concave right) to ⊃-shape (concave left) when a larger portion of the channel was occupied by the TE. They also correctly identified that the ratio of EOF velocities in the LE and TE determined the degree of dispersion. Later, Garcia-Schwarz et al.[61] significantly expanded on the analysis of Schönfeld et al.[139] and analyzed the dynamics of a focused sample in peak mode ITP in the presence of EOF. The latter was the first quantitative study of radially non-uniform distribution of focused sample in ITP. They observed focusing regimes where the sample concentrated either near the channel walls or near the channel center (sometimes achieving concentration values higher than the non-dispersed case!) depending on the mobility of the analyte and position of the ITP peak within the channel.

Three-dimensional sample concentration profiles have also been reported for peak mode ITP in the presence of externally applied pressure gradients. For example, GanOr et al.[60] observed diffusion-dependent sample focusing regimes for counterflow ITP. They reported an approximate, heuristic model for the shape of the ITP interface. Further, they identified two parameters, namely, the ratio of the analyte to TE ion mobility and the ratio of ITP velocity to analyte diffusivity, which determined different radial sample focusing regimes. However, note this simple model assumes only axial velocities for electromigration and hence does not take into effect the full coupling between electric field and non-uniform (along cross-section) conductivity field. GanOr et al.[60] hypothesized that a sufficient condition for radial focusing of the sample is the existence of a non-uniform axial velocity field. They also observed that focused species with nearly the same mobility but significantly different diffusivities (e.g., small molecules versus micron-size beads) can exhibit very different radially dispersed and focused concentration profiles (see **Fig. 12**), an effect they attributed to the difference in balance between axial electromigration and axial diffusion among the focused species. The effects of pressure driven flow on the LE-to-TE interface and sample zone dynamics in peak mode ITP was also studied in using numerical simulations by Bhattacharyya et al.[62] and Gopmandal and Bhattacharyya[143]. To date, we know of no fully three-dimensional models of peak mode ITP concentration fields taking into account full coupling of electric body forces and bulk flow velocities.

## 6. Plateau mode ITP theory for separations and sample identification

Plateau mode ITP is well suited for applications around separation and identification of chemical species. Co-ionic species are most typically segregated into adjoining zones between the LE and ATE zones. The order of these zones is related to the effective mobilities of the analytes (as influenced by local pH). Plateau mode ITP is described schematically for the cases of semi-infinite and finite ITP injections in **Figures 1C** and **4C**, respectively (c.f. **Sections 1** and **2**). ITP focusing of low concentration species requires that sufficiently large volume of sample be processed to ensure accumulation of each species into a plateau. In semi-infinite injection, plateau zones form but only the first zone (adjoining the LE) will be purified as species are constantly entering the various train of plateaus. For example, the plateau of a species adjoining the ATE zone will contain the maximum concentration of that species and also small concentrations of all other sample species which must cross that plateau as they accumulate in their respective plateau.

In finite injection, we typically eventually reach a steady state where each plateau is a locally purified. For simplicity, we will here concentrate on this finite injection case.

Plateaus resulting from finite injection are regions of locally uniform concentrations which (together with counter ion) control local conductivity. Recall from our discussion in **Sections 3** and **4** that plateau zone sample concentrations are governed by the adjusted/regulated values obtained from the Kohlrausch regulating function for strong electrolytes (c.f. **Section 3**), and the Jovin and Alberty relations for weak electrolytes (c.f. **Section 4**). In this way, the LE zone helps determine the concentration of each plateau zone. For well-buffered ITP processes where each plateau pH is near that of the LE, the LE zone also largely determines the effective mobility of each species and therefore the order of the zones. Note that, to achieve plateau mode, an ion must be soluble at its respective plateau-value concentration. As an example, consider two sample species $S_1$ and $S_2$ focusing into a sequence of plateaus, as shown in **Figure 4C** (c.f. **Section 2**). The plateau ions typically have concentrations that are on the same order of magnitude and very close to (but slightly lower than) the leading ion concentration. Note this is in contrast with peak mode where the focused sample ion concentration is most often orders of magnitude lower than the leading ion concentration. The conductivity (and, therefore the electric field) in each plateau is controlled by the local ion density of the sample (and associated counter ion concentration). Essentially, each sample ion "displaces" others to form its own plateau, and these sample ions segregate into plateau zones behind the leading zone in the decreasing order of the absolute value of effective mobilities. The latter sentence, of course, assumes our system does not contain so-called "shared zones" where a single plateau is made up of two co-ionic species with approximately the same mobility.[193]

The regime of plateau mode ITP can therefore be summarized as follows: (i) High initial concentrations of species and sufficiently long focusing times, (ii) separation of species (e.g. in preparation for either direct or indirect detection), and (iii) leveraging of the displacement physics of ITP (i.e., where sample ions displace other ions and form locally purified zones where sample ions control current). The displacement physics of ITP can also be used to effect indirect detection methods wherein exogenous fluorescent species are introduced in order to detect and quantify species that are not directly detectable. The latter includes fluorescent markers, non-focusing tracers, and spacers (c.f. **Section 10**).

In this section, we extend the theory presented previously for plateau mode ITP in **Sections 3** and **4** and provide expressions for the analyte accumulation rate and the length of plateau zones. Much of the theory presented in this section has been covered extensively and in great detail in several references.[1,2,73,194,195] We here only highlight key features of the plateau mode ITP theory and discuss parameters useful in optimizing separation processes.

### 6.1 Accumulation rate and length of plateau zones

We here analyze the case when the sample ions are mixed with and dissolved in the TE prior to segregation in plateau mode ITP. For simplicity, we will here assume constant mobilities for the sample and TE ions (equal to fully dissociated values, i.e., strong electrolyte assumption). The formulation of the ITP accumulation rate for the case of sample accumulating into a plateau directly from the trailing zone follows the same approach as in the case of peak mode ITP (c.f. **Section 5**). The sample accumulation rate in plateau mode ITP is

$$\frac{dN_S}{dt} = \left(\mu_S E^{T'} - V_{ITP}\right) c_S^{T'} A = (\mu_S - \mu_T) E^{T'} c_S^{T'} A \,. \tag{94}$$

Note that the expression for the accumulation rate in Eq. (94) is mathematically the same as peak mode ITP (Eq. (84)). The important difference of the current case is that initial sample concentrations $c_S^T$ are typically much higher than peak mode sample concentrations and are locally uniform within the plateau. We next present expressions for estimating the length of the plateau zone for a sample $S$ initially loaded in TE in plateau mode ITP.

From the accumulation rate into a plateau from the trailing zone given by Eq. (94), we can define a length for the plateau $\Delta x_S$ ignoring diffusive effects and assuming univalent ions and pure zones as

$$\Delta x_S = \frac{N_S(l)}{A c_S^S} = \frac{N(l)}{A c_L^L}\left(\frac{\mu_L}{\mu_S}\right)\left(\frac{\mu_C - \mu_S}{\mu_C - \mu_L}\right). \tag{95}$$

where $N(l)$ is the accumulated amount of sample in moles as a function of the distance $l$ along the channel swept by the ITP zone. This expression works for both finite and semi-infinite type injection. In the former case, $\Delta x_S$ at first grows linearly in time and then saturates while, in the latter case, $\Delta x_S$ grows linearly for as long as you run the ITP process (assuming no significant depletion of sample in the reservoir). Note we have used KRF (Eq. (59)) to calculate the sample concentration in terms of the LE co-ion concentration. At first glance, Eq. (95) seems to have an inverse relation to LE concentration, but as we show below, $N_S$ is also proportional LE concentration, so $c_L^L$ will cancel out and we shall see that $\Delta x_S$ is relatively insensitive to $c_L^L$.

Similar to peak mode ITP (Eq. (86)), the accumulated sample can be cast in terms of the distance $l$ traveled by the LE zone's trailing edge along the channel as

$$N_S(l) = p_{S,TE} c_S^{T'} A l = \left(\frac{\mu_S}{\mu_T} - 1\right) c_S^{T'} A l. \tag{96}$$

Relating the sample accumulation in the ITP zone to the entrance of channel (e.g., from the reservoir well, $W$) via species conservation, we have $c_S^{T'} E^{T'} A = c_S^W E^W A$. We have assumed that the well liquid presents itself slightly into the channel (so the areas in the latter equation cancel). Thus, the accumulated sample concentration can be rewritten as

$$c_S^{T'} = c_S^W \frac{E^W}{E^{T'}} = c_S^W \frac{\sigma_{T'}}{\sigma_W} = c_S^W \frac{c_T^{T'}}{c_T^W} = \left(\frac{\mu_T}{\mu_L}\right)\left(\frac{\mu_C - \mu_L}{\mu_C - \mu_T}\right) c_S^W \frac{c_L^L}{c_T^W}. \tag{97}$$

Here we have used current continuity and KRF. Substituting Eqs. (96) and (97) into Eq. (95), we have

$$\Delta x_S = \underbrace{\left[\left(\frac{\mu_S}{\mu_T} - 1\right)\left(\frac{\mu_T}{\mu_S}\right)\left(\frac{\mu_C - \mu_S}{\mu_C - \mu_T}\right)\right]}_{f(\mu_S)} \frac{c_S^W}{c_T^W} l \equiv f(\mu_S) \frac{c_S^W}{c_T^W} l, \tag{98}$$

where $f$ is the function of mobilities of the sample, TE co-ion, LE co-ion, and the counter ion ($f$ is the quantity in square brackets in Eq. (98)), and here $l$ is the channel length between the reservoir and detection region. From Eq. (98), we see that a longer distance $l$ and lower TE concentration (in well) each increase zone length $\Delta x_S$. Further, the zone length $\Delta x_S$ is directly proportional to the initial (reservoir) concentration of the sample. Also importantly $\Delta x_S$ is independent of concentration of LE.

Note we have assumed in the derivation above that the focusing species are strong electrolyte species. Similar analysis (and scaling) can be extended to weak electrolyte systems. In this context, we refer to the work of Gebauer and Boček[196] who introduced the concept of zone existence diagrams (ZED). ZED plots the zone pH versus the effective mobility using the fully ionized mobilities and dissociation constants of the species as parameters. ZEDs are useful in designing buffers and understanding the migration and zone orders for ITP separations involving weak acids and bases. This concept has been used to develop strategies to choose electrolyte systems for ITP.[197,198]

## 7. Reactions controlled and accelerated using microfluidic ITP

We here review simple models for microfluidic ITP systems where ITP is used to initiate, control, and accelerate biochemical reactions involving one or more reactants which focus(es) in ITP.[111] We discuss both homogeneous and heterogeneous reaction systems. Example applications include reactions involving one or more of nucleic acids, proteins, and cells (c.f. **Section 11.4**). Most studies reviewed here use peak mode ITP for reaction acceleration. We refer to Eid and Santiago[38] and Khnouf and Han[66] for recent review articles on this topic.

### 7.1 Theory and models for homogenous reactions

We briefly review homogeneous reactions in ITP. We define "homogenous" reactions in the context of ITP as those wherein all reacting species are solutes in solution, and at least one of the reacting species focuses and preconcentrates in ITP. Consider the following standard, well-mixed, second-order bimolecular chemical reaction between species $A$ and $B$ resulting in product $AB$ and which has forward and reverse reaction rate constants $k_{on}$ and $k_{off}$, respectively:

$$A + B \underset{k_{off}}{\overset{k_{on}}{\rightleftharpoons}} AB, \qquad (99)$$

For simplicity, consider the limit where one of the species, say $c_A$, is present initially in significant abundance relative to the other, $c_B$, and where the equilibrium constant is sufficiently low ($K_{eq} \ll c_A$). The reaction in Eq. (99) is then well described by pseudo-first-order kinetics. In this limit, the fraction $f_{\text{std}}$ of the limiting reactant $B$ which has reacted to form $AB$ versus time $t$ is given by

$$f_{\text{std}} = \frac{c_{AB}}{B_0} = 1 - \exp(-k_{on} A_0 t), \qquad (100)$$

where, $c_{AB}$ is the concentration of the product, $B_0$ and $A_0$ are the initial concentrations of $c_B$ and $c_A$ respectively. The relation shows a pseudo-first-order reaction time scale of the form $\tau_{\text{standard}} = \ln(2)/(k_{on} A_0)$. We shall see that the primary effect of ITP is to decrease reaction time by increasing abundance species concentration and to continuously increase available concentration of the low abundance species (hence increasing production rate).

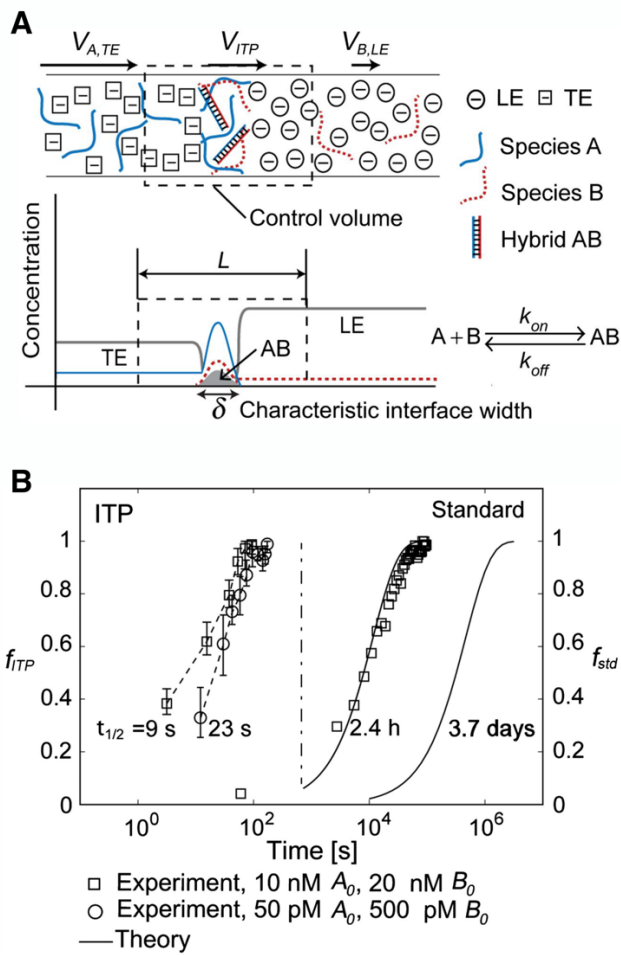

**Figure 13**. (A) Schematic of a model homogeneous reaction system in ITP consisting of two co-focused, chemically reacting species in peak mode ITP. (B) Experimental demonstration of up to 14,000-fold reaction acceleration of DNA-DNA hybridization reaction in ITP. The latter reaction is an example of a second-order homogeneous reaction. Shown are the fraction of hybridized reactants versus time for standard (well-mixed) conditions and ITP. Symbols represent experimental data, and solid lines are model predictions. Figure is reproduced with permission from Ref.[39]. Copyright 2012 National Academy of Sciences.

For simplicity, consider an ITP system where both reactants *A* and *B* focus and preconcentrate in ITP. Bercovici et al.[39] presented a system of first-order ordinary differential equations to describe such homogeneous reactions controlled and accelerated by ITP. They considered an ITP system where species *A* and *B* are initially placed in the TE and LE using semi-infinite injection, respectively (see **Figure 13A**). The reactants focus in peak-mode ITP, and are assumed to be co-located within one-dimensional Gaussian concentration profiles $c_i$. A volume-averaged species concentration $\bar{c}_i$ within the ITP peak was obtained by averaging the concentration profiles over the ITP peak axial width $\delta$. Bercovici et al.[39] derived a set of volume-averaged, electromigration-reaction conservation equations for the reactants and products in ITP as

$$\frac{d\bar{c}_A}{dt} = \frac{Q_A}{S\delta} - \frac{1}{\delta}\frac{d\delta}{dt}\bar{c}_A - \frac{3}{\sqrt{\pi}}k_{on}\bar{c}_A\bar{c}_B + k_{off}\bar{c}_{AB} \tag{101}$$

$$\frac{d\bar{c}_B}{dt} = \frac{Q_B}{S\delta} - \frac{1}{\delta}\frac{d\delta}{dt}\bar{c}_B - \frac{3}{\sqrt{\pi}}k_{on}\bar{c}_A\bar{c}_B + k_{off}\bar{c}_{AB}$$

$$\frac{d\bar{c}_{AB}}{dt} = -\frac{1}{\delta}\frac{d\delta}{dt}\bar{c}_{AB} + \frac{3}{\sqrt{\pi}}k_{on}\bar{c}_A\bar{c}_B - k_{off}\bar{c}_{AB}$$

where $Q_A$ and $Q_B$ are the influx rates of reactants $A$ and $B$ into the ITP peak region, which are given by[39,64]

$$Q_A = (U_A^{TE} - V_{ITP})Sc_A^{T'} = \left(\frac{\mu_A}{\mu_T} - 1\right)V_{ITP}S\beta A_0, \text{ and}$$
$$Q_B = (V_{ITP} - U_B^{LE})Sc_B^{L} = \left(1 - \frac{\mu_B}{\mu_L}\right)V_{ITP}SB_0. \tag{102}$$

$\beta$ is the ratio of the TE ion concentration in the adjusted TE and TE zones (the latter is assumed to be the TE reservoir), and $\mu_i$ is the mobility of species $i$, and $S$ is the cross-section of the channel. To simplify analysis, Bercovici et al.[39] further assumed constant ITP width $\delta$ and obtained from Eqs. (101) and (102) the following conservation equations for species $A$ and $B$ in ITP:

$$\bar{c}_A + \bar{c}_{AB} = \frac{Q_A}{S\delta}t, \text{ and } \bar{c}_B + \bar{c}_{AB} = \frac{Q_B}{S\delta}t. \tag{103}$$

In contrast, the standard (non-ITP), well-mixed reaction obeys

$$\bar{c}_A + \bar{c}_{AB} = A_0, \text{ and } \bar{c}_B + \bar{c}_{AB} = B_0. \tag{104}$$

Eqs. (103) and (104) highlight how, in ITP, there is no longer a steady state solution and instead species continuously focus and produce a product whose growth is enhanced by the rate of accumulation and focusing.

Eqs. (101)-(103) can be simplified for the case when one of the species (say, $A$) is present in excess concentration within the ITP interface and for sufficiently low equilibrium constant. This yields a pseudo-first-order approximation for the concentration of product in ITP as[39]

$$\bar{c}_{AB} = \frac{b}{2a}\left[t - \frac{1}{2}\sqrt{\frac{\pi}{a}}e^{-at^2}\text{erfi}(\sqrt{a}t)\right] \approx \underbrace{Q_B t}_{B_{0,ITP}}\left(1 - e^{-\frac{3}{2\sqrt{\pi}}k_{on}\frac{Q_A}{S\delta}t^2}\right), \tag{105}$$

where

$$a = \frac{3}{\sqrt{\pi}}\frac{k_{on}}{2}\frac{Q_A}{S\delta}, \text{ and } b = \frac{3}{\sqrt{\pi}}k_{on}\frac{Q_A}{S\delta}\frac{Q_B}{S\delta}. \tag{106}$$

Thus, the fraction $f_{ITP}$ of reactant $B$ (low abundance) within the ITP interface which has reacted to form $AB$ is

$$f_{ITP} = \frac{\bar{c}_{AB}}{B_{0,ITP}} \approx 1 - \exp\left(-\frac{3}{2\sqrt{\pi}}k_{on}\frac{Q_A}{S\delta}t^2\right). \tag{107}$$

Based on Eqs. (105) and (107), Bercovici et al.[39] derived a pseudo-first-order time scale associated with ITP-mediated reactions as

$$\tau_{ITP} \approx \sqrt{\frac{\ln(2)S\delta}{k_{on}Q_A}}. \tag{108}$$

Eqs. (105)-(108) depict two key features of ITP-accelerated reactions. First, the pre-factor $B_{0,ITP} = Q_B t$ in Eq. (105) creates what is called a quasi-steady condition for $t \gg \tau_{ITP}$ The system has no true steady state but, in this regime, it tends toward a balance where the production rate is precisely limited by the accumulation rate of the low-abundance species. This results in a linear "quasi steady" increase in the concentration of the product. Second, from Eqs. (*105*) and (*107*), reactions in ITP have a $t^2$ in the exponent (compared to $t$ for standard case). This $t^2$ dependence is due to the simultaneous effects of the (standard) pseudo-first-order reaction and the ITP-aided, linear increase of the abundant species concentration (here, $A$) with time. The ratio of time scales associated with reaction completion for the standard and ITP reaction conditions is then[39]

$$\frac{\tau_{\text{standard}}}{\tau_{\text{ITP}}} \approx \sqrt{\frac{\ln(2)\, Q_A}{k_{on} A_0^2 S \delta}}. \tag{109}$$

From Eq.(109) we see that ITP significantly decreases the time for reaction to occur, and especially in the challenging regime of low $k_{on}$ and $A_0$. Bercovici et al.[39] reported $\tau_{\text{std}}/\tau_{\text{ITP}}$ of ~10,000 for $A_0$ of 100 pM and $k_{on}$ of $10^3$ M$^{-1}$s$^{-1}$, using typical ITP experimental conditions. They validated the ITP reaction kinetics model with experiments using DNA hybridization reactions involving ssDNA oligos and molecular beacons (MB) and demonstrated ~14,000-fold acceleration of reactions using ITP; c.f. **Figure 13B**. Specifically, for 50 pM MB and 500 pM target ssDNA, the time scale to complete 50% of the reaction in the standard case was ~3.7 days, while, for ITP, the same time scale was ~23 s for the same initial concentration. This corresponds to a ~14,000-fold faster reaction time scale in ITP compared to the standard case.

Subsequently, Rubin et al.[59] significantly expanded on the model of Bercovici et al.[39] to include effects of asymmetric, non-overlapping, non-Gaussian concentration profiles of reactants in ITP. Such situations can arise when the focused analytes have mobility values near those of the TE or LE ions which leads to penetration of the analyte into the TE or LE, respectively, and results in asymmetric peak-mode concentration profiles.[61] Rubin et al.[59] derived closed-form, analytical expressions for such dispersed concentration profiles and studied the effects of sample dispersion on reaction kinetics in ITP. They identified a novel form factor parameter $k_{form}$ which accounts for the degree of spatial overlap between reacting species in ITP, and used it to obtain an effective on-rate of the reaction $k_{on}^{(eff)}$ as

$$k_{on}^{(eff)} = k_{on}\, k_{form}, \tag{110}$$

where,

$$k_{form} = \frac{1}{\pi S \delta} \frac{1 - (y_A + y_B)}{\cot(\pi y_A) + \cot(\pi y_B)}. \tag{111}$$

$y_A$ and $y_B$ are functions of the mobilities of the reactant species and LE and TE ions, and are given by

$$y_A = \frac{\mu_T^{-1} - \mu_A^{-1}}{\mu_T^{-1} - \mu_L^{-1}}, \text{ and } y_B = \frac{\mu_T^{-1} - \mu_B^{-1}}{\mu_T^{-1} - \mu_L^{-1}}. \tag{112}$$

An important consequence derived from the work of Rubin et al.[59] is that for non-symmetric concentration profiles of the reactants, maximum reaction rate is achieved for imperfect overlap where the concentration maxima are not aligned (**Fig. 14**).

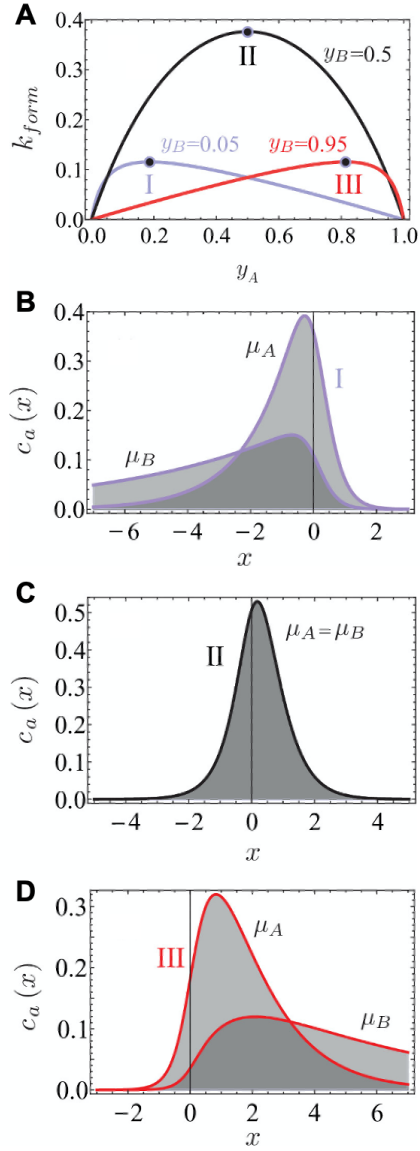

**Figure 14**. Analytical model predictions for the effect of reactants mobility on the rates of reaction accelerated using ITP. (A) Variation of the form factor $k_{form}$ as a function of the mobility parameters of reactants $y_A$ and $y_B$. (B)-(D) Concentration profile of reactants versus $x$ corresponding to parameters of the points I, II, and III in (A). Figure is adapted with permission from Ref.[59]. Copyright 2014 AIP Publishing.

In addition, Rubin et al.[59] also identified the time scale associated with ITP-aided reactions when the equilibrium constant is not negligible. In this case, the time to complete half of the reaction $\tilde{\tau}_{\text{ITP}}$ scales as[59]

$$\tilde{\tau}_{\text{ITP}} \sim \frac{k_{off}}{(Q_A + Q_B) k_{on}^{(eff)}} \approx \frac{k_{off}}{k_{on}^{(eff)} Q_A} \text{ (if } Q_A \gg Q_B) . \tag{113}$$

Note this scaling is different from $\tau_{ITP}$ for the case when $k_{off} \approx 0$ (c.f. Eq. (108)). In the latter case, $\tau_{ITP} \sim 1/\sqrt{k_{on}Q_A}$. The effect of $k_{off}$ can be significant, especially at low analyte concentrations. For example, for standard DNA hybridization reactions with $k_{on} \sim 10^4$ M$^{-1}$s$^{-1}$ and analyte concentrations of order 1 fM, Rubin et al.[59] found that reaction timescales for typical ITP conditions can be ~10-fold lower when $k_{off}$ takes even a small but non-negligible value of $10^{-4}$ s$^{-1}$ compared to the case when $k_{off} = 0$.

Eid and Santiago[64] expanded aspects of the model of Rubin et al.[59] and explored the design of ITP systems which accelerate chemical reactions. They studied the effects of initial sample placement (e.g., in LE vs TE) on the influx rates of the reacting species and the associated product formation rates. Importantly, they identified that the amount (in moles) of product formed in ITP $N_{AB}$ for timescales larger than the ITP reaction time scale $\tilde{\tau}_{ITP}$ is limited, in general, by the influx rates of the reacting species. Specifically,

$$\bar{N}_{AB}^{\infty} = \frac{N_{AB}(t \gg \tilde{\tau}_{ITP})}{Q_B^j t} = \frac{Q_A^j}{Q_A^j + Q_B^j} \to 1 \text{ (if } Q_A^j \gg Q_B^j \text{)}. \tag{114}$$

$Q_A^j$ and $Q_B^j$ represent the influx rates of $A$ and $B$ into the ITP interface when initially loaded in the zone $j$ (LE or TE). See Eid and Santiago[64] for a detailed discussion on parameters that govern influx rates in ITP. In the limit where species $A$ is in excess compared to $B$, $\bar{N}_{AB}^{\infty}$ approaches unity. In other words, the quasi-steady state (i.e., for sufficient reaction time) product formation rate in ITP is generally limited by the influx rate of the less abundant (limiting) reactant into the ITP interface.[64] Recently, Qi et al.[199] further expanded on the work of Rubin et al.[59] and Eid and Santiago[64] and studied the combined effects of species abundance and reaction off-rates on product formation dynamics in ITP. Specifically, they identified two key non-dimensional parameters $\epsilon$ and $\nu$, which governed reaction dynamics in ITP and studied regimes where the reaction was reactant or/and kinetics limited. Physically, $\epsilon$ represents the ratio of the reverse and forward reaction rates in ITP, and $\nu$ represents the relative abundance (accumulation rate) of the reacting species within the ITP peak. Importantly, Qi et al.[199] observed that the amount of product formed in ITP under quasi-steady state solely depends on $\nu$, while the time to reach quasi-steady state is proportional to $\epsilon$.

We have so far discussed homogeneous ITP reactions where both reactants focus in ITP. Eid et al.[200] studied the case of homogeneous reactions in ITP where only one of the reacting species focuses and preconcentrates in ITP. A key non-dimensional parameter that resulted from their analysis was a Damköhler-type number $\lambda$ which represented a tradeoff between reducing assay time and increasing product formation. $\lambda$ is given by[200]

$$\lambda = \frac{\left(\frac{L_0}{V_{ITP}}\right)}{\left(\frac{1}{k_{on}A_0}\right)} \sim \frac{\tau_{adv}}{\tau_{rxn}}, \tag{115}$$

where $L_0$ is the length of the channel. $\lambda$ can be interpreted as a ratio of the advection time scale $\tau_{adv}$ resulting from the electromigration of the ITP peak (which contains one of the reacting species) and the reaction time scale $\tau_{rxn}$ (which is governed by the abundant species, $A$). At low $\lambda$ (i.e., when $\tau_{adv} \ll \tau_{rxn}$), the ITP peak migrates fast (i.e., low assay times), and the advective time scale is much smaller than the time scale for reaction completion, leading to minimal product formation in ITP. In contrast, for high values of $\lambda$ (i.e., when $\tau_{adv} \gg \tau_{rxn}$), the reaction proceeds to near-completion over a time scale much smaller than ITP electromigration across the channel, leading to minimal effects of ITP on reaction kinetics.[200] Thus, desirable operating conditions should balance reaction completion and assay time such that advective and reaction time scales are comparable (i.e., $\tau_{adv} \approx \tau_{rxn}$). We here also note the work of Zeidman Kalman et al.[201] who developed a detailed experimentally validated model for association-dissociation dynamics when only one of the reactants focuses in ITP and when the reverse reaction rate constant $k_{off}$ is non-negligible. The unique dynamics associated with such reactions (for varying $k_{off}$) can be used to discriminate among specific versus non-specific reaction products (e.g., for nucleic acid hybridization).

## 7.2 Theory and models for heterogenous reactions

We here discuss heterogeneous chemical reactions in ITP which we define as those which involve at least one reactant immobilized to a surface or matrix within a microchannel, while the other (solute) reactants electromigrate and focus in ITP. To a simple approximation, the reaction kinetics in such heterogeneous ITP reaction systems is governed by the interplay among transverse diffusion (diffusion to and from the surface), ITP peak advection, kinetics time scale, and depletion effects. Consider that in such systems, the immobilized reactant concentration is prescribed as a number density on the surface $b_m$ with dimensions of moles per unit surface area, while the reactant which focuses in ITP has a concentration in dimensions of moles per volume $c_0$. The ratio of $b_m$ and $c_0$ naturally provides a length scale which can be compared to the length scale normal to the immobilized reactant surface. We here suggest that a useful quantity is then the length scale of the dissolved-species reactant cloud which must be "depleted" to completely saturate the surface. For a dissolved species in global abundance (relative to surface reactant), this length scale is less than the channel dimension normal to the reaction surface, the channel depth $h$. For a low-abundance dissolved species, the maximum length scale for diffusion is the channel height, $h$.

Karsenty et al.[202] presented a model and scaling analysis for ITP systems involving heterogeneous reactions. They studied the reaction between target molecules focused in ITP and probe molecules immobilized on the surface of a microchannel (e.g., see **Fig. 15**). They considered the limit where the diffusion time $\tau_D (= h^2/D$, where $D$ is the diffusivity) across the channel depth was much smaller than the residence time of the ITP peak over the reaction surface (with axial scale $\delta$) $\tau_{adv} (= \delta/V_{ITP})$, i.e., $\tau_D \ll \tau_{adv}$. Further, they argued that the concentration of target molecules $c_a$ in ITP remained nearly constant which electromigrating over the length $l$ of the surface probes (i.e., $\Delta c_a/c_a \sim k_{on}b_m l/hV_{ITP} \ll 1$), and the system was kinetics-limited (not diffusion-limited).[202–204] These scaling arguments allowed them to model the concentration profile of ITP focused targets as a top-hat function with a characteristic target concentration value (determined by the ITP preconcentration factor) and a characteristic width $\delta$, traveling at $V_{ITP}$.[202–204] Paratore et al.[204] later termed this simple operation method wherein the ITP focused target molecules as they electromigrated over surface probes as the "pass-over" (PO) mode. Karsenty et al.[202] described the reaction kinetics for PO mode in terms of the following differential equation for the concentration of surface probes bound by the target $b(t)$:

$$\frac{db(t)}{dt} + \frac{b(t)}{\tau_R} = \frac{b_m}{\tau_{on}}, \tag{116}$$

where, $\tau_{on}^{-1} = \alpha c_0 k_{on}$ ($\alpha$ is the ITP preconcentration factor and $c_0$ is the reservoir target concentration) and $b_m$ is the total number of surface probes (free and bound). The reaction time scale $\tau_R$ in Eq. (116) is given by

$$\tau_R = \frac{1}{k_{on}(\alpha c_0 + K_d)}, \quad (117)$$

where, $K_d = k_{off}/k_{on}$. Note that $\alpha = 1$ corresponds to the standard flow conditions without ITP. For the PO mode, the fraction of the bound probes ($b_{PO}/b_m$) versus ITP residence time and kinetic rates is obtained from Eq. (116) as[202–204]

$$\frac{b_{PO}}{b_m} = \frac{\alpha c_0 k_{on}}{k_{off} + \alpha c_0 k_{on}} \left[1 - \exp(-(k_{off} + \alpha c_0 k_{on})\tau_{PO})\right], \quad (118)$$

where $\tau_{PO}$ is the duration that the ITP peak passes over the surface probes. Defining $\tau_{tot}$ as the total assay time (includes focusing, preconcentration, and reaction), Karsenty et al.[202] obtained an enrichment factor from ITP compared to standard flow conditions as

$$R = \alpha \frac{k_{off} + c_0 k_{on}}{k_{off} + \alpha c_0 k_{on}} \frac{1 - \exp(-(k_{off} + \alpha c_0 k_{on})\tau_{PO})}{1 - \exp(-(k_{off} + c_0 k_{on})\tau_{tot})}. \quad (119)$$

An upper bound for ITP enrichment $R$ is obtained at low concentrations and for $\tau_{PO}, \tau_{tot} \ll \tau_{off}$, and the enrichment factor is equal to $\alpha \tau_{PO}/\tau_{tot}$. In the limit of high concentrations compared to the equilibrium constant, $R$ approaches unity and ITP is not beneficial. A major limitation of the PO mode is that only a small fraction of target molecules (for example, Karsenty et al.[202] report ~1%) reacts with surface probes due to the short incubation time of the ITP peak over the probes relative to the advection of the ITP peak.

Subsequently, Paratore et al.[204] presented novel operation methods and associated models to improve overall reaction efficiency of ITP heterogeneous reactions (c.f. **Fig. 15**). They introduced two new operating methods for ITP reactions which they termed "stop and diffuse" (SD) and "counterflow" (CF).

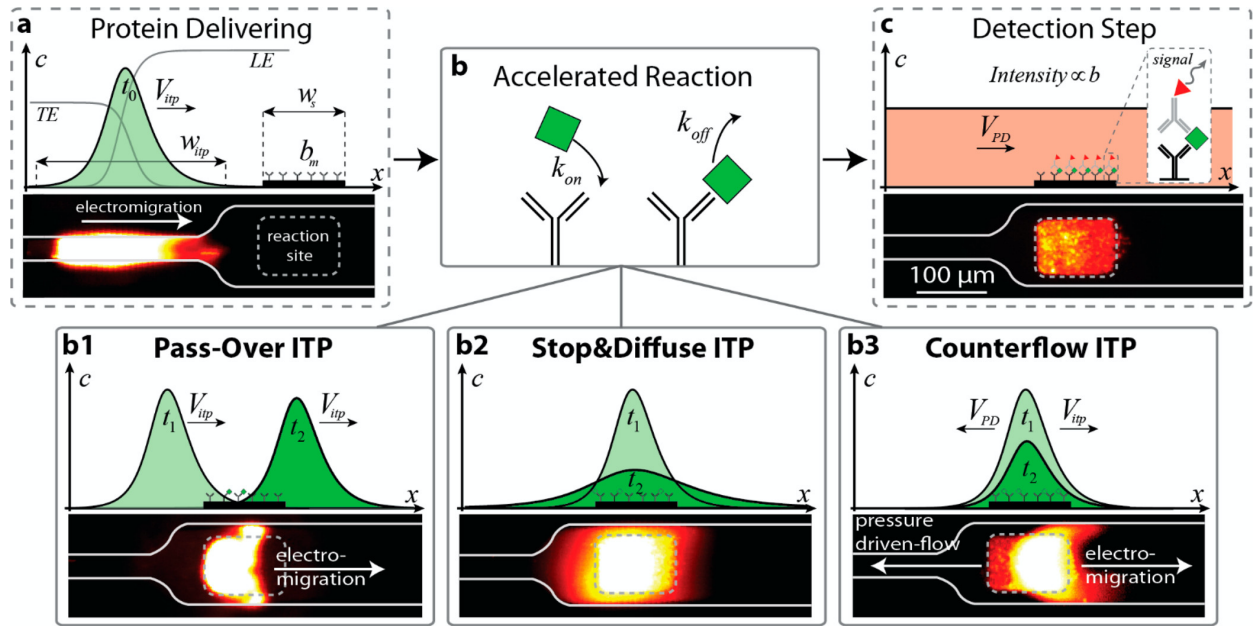

**Figure 15**. Schematic representation and raw experimental images of the ITP-based surface immunoassay of Paratore et al.[204]. (A) ITP focuses target proteins and delivers them to the site where surface reaction occurs. (B) Acceleration of reaction between surface-immobilized capture antibodies and the proteins focused (in solution) in ITP, achieved using three modes: In Pass-Over ITP, the ITP

peak electromigrates continuously past the surface reaction site. In the Stop and Diffuse mode, the electric field is temporarily turned off when the ITP peak is over the reaction site and the peak is allowed to diffuse away from the reaction site with time. In Counterflow ITP, pressure-driven counterflow is used to hold the ITP peak stationary over the reaction site. (C) Detection is achieved by measuring the labeled antibodies bound to the reacted targets. Figure is reproduced with permission from Ref.[204]. Copyright 2017 American Chemical Society.

In the SD method, the applied voltage is turned off when the ITP peak is over the surface probes and the focused target is allowed to diffuse (and broaden) while reacting with the probes for some long duration. Paratore et al.[204] studied the limit where the diffusion timescale across the depth is much smaller than the depletion time due to the surface reaction, an ITP peak width which is smaller than the reaction site, and a kinetic regime where the surface probes are far from saturation and where the on-rate dominates the reaction kinetics. For this regime, Paratore et al.[204] found an analytical expression for the fraction of surface probes bound to the target in SD mode ($b_{SD}/b_m$) as

$$\frac{b_{SD}}{b_m} = k_{on}\tau_{dep}\left[\chi - \left(32\frac{t}{\tau_{Dx}} + 1\right)^{-\frac{1}{2}} \exp\left(-\frac{t}{\tau_{dep}}\right)\right]\alpha c_0 \qquad (120)$$

where

$$\chi = \left(1 - \text{erf}\left(\sqrt{\frac{1}{32}\frac{\tau_{Dx}}{\tau_{dep}}}\right)\right)\sqrt{\frac{\pi}{32}\frac{\tau_{Dx}}{\tau_{dep}}}\exp\left(\frac{1}{32}\frac{\tau_{Dx}}{\tau_{dep}}\right). \qquad (121)$$

In Eqs. (120) and (121), $\tau_{dep} = h/b_m k_{on}$ and $\tau_{Dx} = \delta^2/D$, where $h$ is the depth of the channel and $\delta$ is the ITP peak width. For long times $t \gg \tau_{dep}, \tau_{Dx}$, the bound fraction of the surface probes approaches a value of $\chi\alpha c_0$. The SD mode is especially beneficial for the binding of large molecules with low diffusivity such as proteins or antibodies, since low diffusivity species retain the higher bulk concentration (due to ITP preconcentration and low diffusivity) for a longer time.[204]

To further improve reaction efficiency, Paratore et al.[204] considered counterflow (CF) ITP wherein the electromigration velocity of the ITP peak containing the targets is balanced by pressure driven flow in the opposite direction so that the ITP peak remains stationary above the reaction site. They modeled reaction kinetics for a dispersion-free CF mode and derived an expression for the fraction of bound surface probes ($b_{CF}/b_m$) for time scales greater than the depletion time scale as

$$\frac{b_{CF}}{b_m} = \left(\sqrt{\frac{\pi}{8}}\alpha\frac{\tau_{dep}}{\tau_{acc}}\right)k_{on}c_0 t \qquad (122)$$

where $\tau_{acc} = x_s/V_{ITP}$ is the electromigration time of the target from the reservoir to the ITP peak. Note for the standard flow-through reaction, the bound fraction scales as $\sim k_{on}c_0 t$. Thus, the gain factor for CF mode is proportional to $\sim\sqrt{\pi/8}\,\alpha\,(\tau_{dep}/\tau_{acc})$. This gain factor can be tuned by adjusting ITP focusing conditions such as electric field strength, buffer concentrations and pH, and channel geometry.

Lastly, Shkolnikov and Santiago[205] developed a numerical and analytical model for volumetric homogeneous reactions involving the coupling of ITP and affinity chromatography (using a custom, functionalized porous column). They used the model to predict spatiotemporal dynamics association with the coupled electromigration, diffusion, and affinity reaction problem, and to evaluate key figures of merit including capture efficiency, assay time, and required length of capture region. Importantly, they identified key limiting regimes of operation and obtained three nondimensionalized parameters which collapsed the dynamics. The first parameter was a Damköhler number which described the ratio of electrophoretic time scale and reaction time scale. The second parameter was the equilibrium constant non-dimensionalized by the initial probe concentration. The last parameter represented the peak target concentration in ITP scaled by the initial probe concentration. We note that in the latter case and the aforementioned cases of functionalized gels,[206,207] the timescales of diffusion to and from the immobilized species is negligible. Hence, in such systems, the dynamics tend to be governed by competition between advection time of the ITP zone and the reaction time scale. In these systems, diffusion plays a role only in the balance between non-uniform axial electromigration and axial diffusion inherent in ITP.

## 8. Practical considerations and limitations for microfluidic ITP

In this section, we review several considerations, useful practices, and limitations when designing microfluidic ITP systems. We discuss topics related to interface dispersion and electrokinetic instability, Joule heating, pH buffering, separation capacity, tradeoff between sample volume and sensitivity, operation method, and channel materials.

### 8.1 Interface dispersion and electrokinetic instabilities

As discussed in **Section 5**, an ideal, one-dimensional ITP interface can become distorted (or disrupted) due to dispersion associated with external pressure gradients, unsuppressed residual electroosmotic flow (EOF), and electrokinetic instabilities (typically, at high electric fields). These effects can strongly influence resolution and sensitivity of ITP systems. We here briefly review these effects and approaches that can be used to mitigate such non-ideal behaviors.

Residual EOF in the channel often causes non-uniform EOF between the migrating LE and TE zones, which leads to the generation of internal pressure gradients and interface distortion. In systems where the EOF is significant, adding a high molecular weight neutral polymer such as PVP, HPMC, or PEO in the LE and/or TE buffers suppresses EOF via dynamic coating of channel walls and results in a stable ITP interface.[183–189] Among the polymers, PVP at 0.5 to 1% concentration is the most commonly used approach for glass and plastic-based microfluidic channels. In case addition of PVP results in undesirable sieving effects, silanol coating such as Sigmacote in addition to a surfactant such as Triton-X 100 is an effective alternative solution to suppress EOF, especially for borosilicate glass channels.[34] See Dolník[189] for a comprehensive review on various dynamic and permanent wall coating approaches used in microfluidics to suppress EOF including for channels fabricated using PMMA, polycarbonate, PDMS, and glass materials.

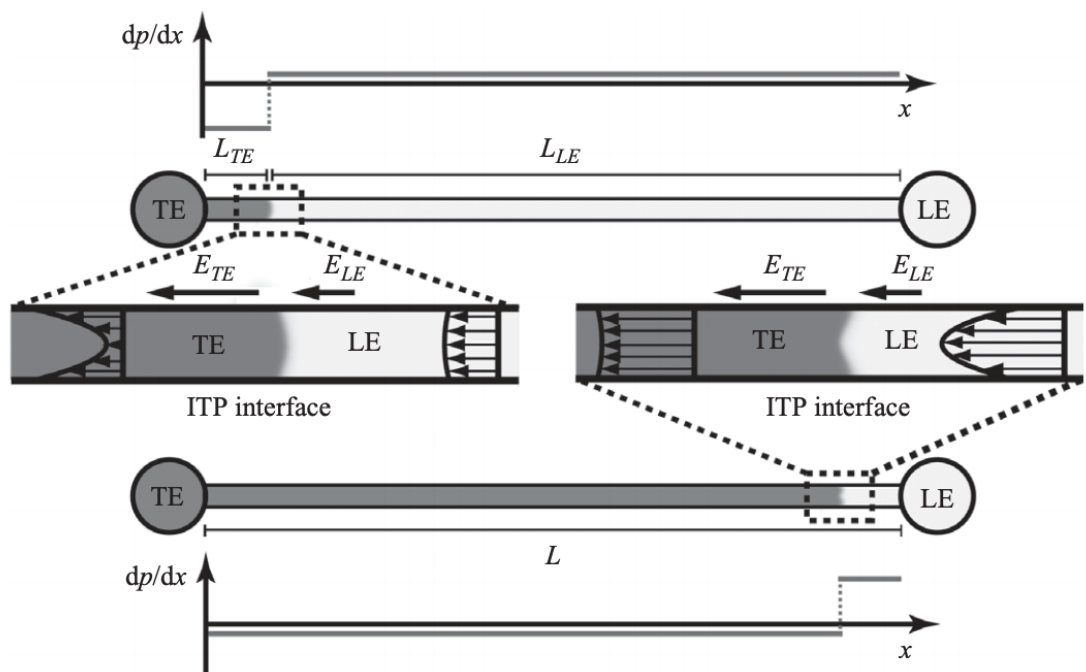

**Figure 16**. Schematic depicting the curvature of the interface between the LE and adjusted TE zones at two different locations in the microchannel. An axially non-uniform electric field results in non-uniform EO slip velocities in the two zones. The latter leads to the generation of internal pressure gradients which can lead to a distorted ITP interface. Figure reproduced with permission from Ref.[61]. Copyright 2011 Cambridge University Press.

In addition to EOF, pressure driven flow can also disperse the ITP interface. This can arise, for example, from residual pressure driven flow due to imbalances in the volume of liquid in reservoirs residual pressure driven flow, or due to processes which are designed to include counter pressure-driven flow in ITP.[60–62] Another source of pressure driven flow can be due to random bubble formation processes at the electrode tips in the reservoirs due to electrolysis and the resulting differential gas generation at the two electrodes.[208] Consider that 2 moles of hydrogen gas are produced at the cathode for 1 mole of oxygen produced at the anode. Effects due to such pressure differences can be mitigated by the use of gel-based plugs in the reservoirs composed of buffer in agarose[162] or Pluronic F-127[203,209]. Agarose melts at high temperatures and Pluronic F-127 is a liquid below 4°C [210], and both solidify and form gels when brought to room temperature. Next, note that instabilities in ITP can also occur due to differential Joule heating between LE and TE and internally generated temperature gradients – factors which result in three dimensional temperature field[162] and secondary flow fields across the interface. Joule heating effects are discussed in **Section 8.2**.

ITP as an electrophoretic process is characterized by an ion-concentration shock wave with self-sharpening gradients in ionic conductivity and ion concentrations which are primarily collinear with an applied electric field.[142,211,212] The sharp gradient in conductivity results in the accumulation of free charge at the moving ITP interface, and this in turn couples with the local electric field and generates destabilizing electric body forces. This destabilizing force can disrupt the stability of the ITP interface and result in dispersion of focused samples which can lower the sensitivity of detection. Electric fields in microfluidic ITP are typically on the order of ~1 to 10 kV/m within channels of dimensions on the order of 10 to 100 μm. Typically, high values of electric fields are desired for ITP applications in order to improve sensitivity and assay speeds. However, overly high electric fields can result in electrohydrodynamic instabilities,[70,213–215] and such effects can disrupt the sharp ITP interface and adversely affect sensitivity.

Finite (including non-uniform) electroosmotic flow and pressure driven flow (including counterflow) in can cause dispersion of the ITP interface and lead to asymmetric profiles, and existing numerical models and theory can predict some aspects of this behavior.[59–62] At high electric fields, the ITP interface can get significantly distorted and dispersed due to the onset of electrokinetic instabilities (EKI, a regime of electrohydrodynamic (EHD) instabilities; see **Fig. 17**) .[124,216] The latter regime is characterized by unsteady, three-dimensional fluctuations of concentration, conductivity, and electric fields near the ITP interface, which can lead to time-varying body force fields.

In general, EKI occurs when electric body forces dominate viscous forces in the flow. Posner and Santiago[82] summarized various formulations for scaling of electric body forces relevant for EKI in terms of an electric Rayleigh number.[182,217–219] Later, Persat and Santiago[124] presented a scaling for the electric body force $\rho_e E$ at an ITP interface in terms of a modified electric Rayleigh number. Here, $\rho_e$ is the free charge density at the ITP interface and $E$ is the local electric field. Note that for typical EKI studies of symmetric, binary electrolytes, the length scale associated with free charge is determined by convective-diffusive transport (and not on the electric field). In contrast, for ITP, the length scales associated with free charge scales as the width of the ITP interface which varies inversely with electric field. Persat and Santiago[124] thus report the scaling

$$\rho_e E \sim E_{avg}^3 \frac{(\gamma - 1)}{(\gamma + 1)^2}, \tag{123}$$

where, $E_{avg} = (E^L + E^T)/2$ is the average of the electric field in the leading and trailing zones, and $\gamma$ is the ratio of leading and trailing zone conductivities. Importantly, this study strongly suggested an $E^3$-type (stronger) dependence on the electric field for ITP compared to the $E^2$-type scaling of typical other EKI studies.[182,217–219] The study also suggested that the value of $\rho_e E$ based on the above scaling provides empirically an upper threshold value of the electric field which can be used to predict the onset of instabilities.

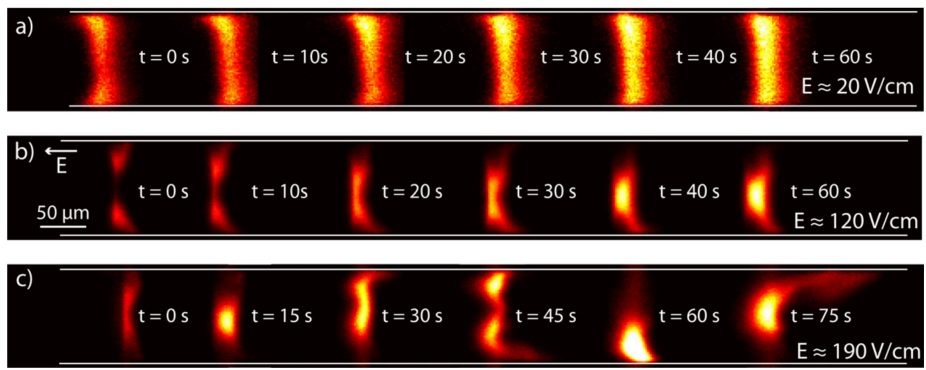

**Figure 17**. Experimental visualization of an ITP peak at different time points in a straight channel, peak mode ITP experiment. Shown are images for three different applied electric fields. Higher electric field strengths can lead to electrokinetic instabilities, which can distort the ITP peak. Figure is reproduced with permission from Ref.[124]. Copyright 2009 IOP Publishing Ltd.

Later, Gaur and Bahga[220] performed a detailed EHD instability analysis of a canonical, single interface ITP in an unbounded domain. They performed a linear stability analysis and showed that electroviscous flow destabilizes ITP while the electromigration of the shock wave has a stabilizing effect. Importantly, to quantify the effect of instability, they obtained a non-dimensional number for ITP in microchannels equal to the ratio of electroviscous flow timescale ($\tau_{ev}$) and restoring electromigration timescale ($\tau_{em}$). The non-dimensional number is expressed as

$$\frac{\tau_{ev}}{\tau_{em}} = \frac{2\eta\mu_L}{\epsilon(\gamma-1)^2 E^L d}, \qquad (124)$$

where $\mu_L$ is the electrophoretic mobility of the leading ion, $E^L$ is the electric field in LE, $d$ is the (shallow) depth of the channel, $\epsilon$ is the dielectric permittivity, and $\eta$ is the viscosity of the fluid. ITP shock wave instability was found to set in for low values of $\tau_{ev}/\tau_{em}$ when electroviscous flow dominates electromigration. Gaur and Bahga[220] hypothesized that the exact value of $\tau_{ev}/\tau_{em}$ below which ITP instability sets in depends on the channel geometry.

From Eq. (124), high values of conductivity ratio, channel depth, electric field promote instability, while high values of fluid viscosity and LE ion mobility have a stabilizing effect on ITP. In addition, the use of an array of posts[221] or porous structures[180] within microchannels can increase hydrodynamic resistance and stabilize the ITP interface. For example, van Kooten et al.[222] optimized and fabricated a microfluidic chip for ITP with an array of cylindrical pillars, where the pillars reportedly served to suppress dispersion of the ITP interface due to potential spatial instabilities. The micropillars reduced the characteristic lateral dimension for viscous effects and conferred overall stability to the moving ITP interface.

Goet et al.[223] also reported instabilities in the context of ITP with emulsions. They studied ITP focusing of oil droplets initially suspended in the LE and noticed that the ITP peak formation exhibited vortical flow structures. They attributed these vortex structures to the body force on the droplets generated by space charge density at the interface. They hypothesized that the electric body force is balanced by the Stokes drag[224] on the droplets within the ITP interface and developed a scaling approximation for the rotational velocity $u_{rot}$ where $u_{rot} \propto E^2$ ($E$ is the applied electric field). The model and scaling of Goet et al.[223] is perhaps applicable to systems involving ITP focusing of micron size particles and beads.[224,225] Interestingly, Shintaku et al.[225] in their bead-based ITP hybridization assay deliberately operated at moderately high electric fields to utilize such secondary flow and vortex structures (see video in the SI of that reference). This improved mixing of reagents and enhanced the hybridization efficiency of the reaction between probe-functionalized beads and nucleic acid targets which co-focused in ITP. The latter study demonstrates that instabilities in ITP at moderate-to-high electric fields can be beneficial, especially for assays that require efficient mixing of focused analytes involved in chemical reactions.

Lastly, note that the distortion of ITP interface due to instability is reversible. Specifically, a stable interface can be quickly reestablished by lowering the electric field.[203,222]

## 8.2 Joule heating and temperature effects

Joule heating is ubiquitous in electrophoresis systems including ITP and it can lead to disruption of separation processes.[107,162,226] Joule heating is also the likely the most significant impediment to scale up microfluidic ITP systems to process large sample volumes.[78,227] Further, the requirement to mitigate temperature rise in the microchannel due to Joule heating in ITP can strongly influence chip design and assay time.[178,222,228–230]

In ITP, the LE and TE zones have high and low conductivity, respectively. The volumetric heating rate (and temperature change) due to Joule heating is inversely proportional to the conductivity for a fixed current density. Further, instantaneous current density is, to very high accuracy, approximately uniform through the channel at any one time. Hence, LE and TE zones necessarily experience different Joule heating rates. The LE region is typically cooler than the TE zone, and there is an associated jump in temperature between the two migrating zones (See **Fig. 18**).[114,162,227,231,232] This feature has been employed for thermometric detection of zone boundaries (c.f. **Section 10.3**). In general, the temperature rise due to Joule heating is function of the applied current, ionic and thermal conductivities of the electrolyte, thermal coupling of the channel to external environment, and the channel geometry. Moreover, the liquid is hotter near the center (interior) of the channel than the surrounding regions near the walls and reservoirs (in absence of external heating), and this results in non-uniform temperature fields within the cross-section (**Fig. 18**). There have been several detailed numerical and theoretical models which predict temperature fields in ITP.[114,231–233] The temperature fields with the LE and TE evolve both spatially and temporally, with high temperature gradients occurring near the ITP interface.[114,162,232] Non-uniform temperature fields within the channel cross-section can also lead to internal, secondary flow (e.g., due to buoyancy and electrothermal effects) which can significantly disperse the ITP interface (**Fig. 18A**).[162]

A passive approach to mitigate Joule heating effects is to use high aspect ratio channels (wide channels with shallow depth) to maximize the contact surface area of the liquid with the top and bottom channel walls and decrease channel center-to-wall distance to enable efficient heat transfer.[233] For channels of aspect ratios of perhaps 3 and higher, the difference in temperature between the center of the channel ($T_{cent}$) and exterior walls ($T_{ext}$) due to Joule heating (in absence of external cooling) under quasi-equilibrium steady state can be approximated as[178,233]

$$\Delta T_i = (T_{cent} - T_{ext})_i = \frac{j^2}{\sigma_i}\left[\frac{h^2}{8k_L} + \frac{hd_{wall}}{2k_{wall}}\right] \propto \frac{j^2}{\sigma_i} \ (= \sigma_i E_i^2), \tag{125}$$

The subscript $i$ refers to the zone (LE and TE), $h$ is the channel depth, $d_{wall}$ is the thickness of the channel walls, $k_L$ and $k_{wall}$ are the thermal conductivities of the liquid and channel walls, respectively. The approximation and scaling in Eq. (125) assume negligible EOF or pressure driven flow, i.e., there is negligible advection of heat in the channel and the temperature field is predominantly governed by a balance between volumetric heating and heat conduction across the channel cross section. In other words, thermal equilibrium within the channel is achieved within the timescale of electromigration.[234] Several studies[114,162,227] have experimentally validated the temperature scaling in Eq. (125) for ITP. For a given microchannel, ITP with higher ionic conductivity solutions (i.e., higher concentrations) and a lower current density results in lower temperature change due to Joule heating. Note that higher current results in faster assay times and is in direct tradeoff with Joule heating effects. Eq. (125) can further be used to obtain an approximate expression for the maximum current $I_{max}$ that can be applied in ITP due to constraints on maximum allowable temperature rise ($\Delta T_{max}$)[178]

$$I_{max} = A_{min}\sqrt{\frac{\Delta T_{max} \ \sigma_{min}}{\left[\frac{h^2}{8k_L} + \frac{hd_{wall}}{2k_{wall}}\right]}}, \tag{126}$$

where "min" subscripts refer to location along the channel where the product $A\sqrt{\sigma}$ is the lowest. Here, $A_{min}$ is cross-section area of the channel and $\sigma$ is the conductivity of the liquid at the "min" location. Eq. (126) assumes negligible changes in material properties due to temperature. The value of $\Delta T_{max}$ depends, of course, on the application. For ITP applications involving proteins (e.g., immunoassays), the maximum operatable temperature is typically less than around 40°C to prevent protein denaturation. For ITP nucleic acids assays (e.g., hybridization reactions or on-chip ITP-mediated isothermal amplification), a higher temperature (e.g., 60°C) might be desirable to provide stringency or to optimize enzyme activity, requiring operation at higher currents. van Kooten et al.[222] and Terzis et al.[162] demonstrated temperatures of ~ 65°C can be achieved in ITP without overly disrupting the ITP interface (merely some additional dispersion). Higher temperatures due to Joule heating can lead to degassing of buffers and significant bubble formation within the channel which can disrupt ITP.

Another consequence of Joule heating is the changes of buffer properties. For example, ion mobilities, ionic conductivity, thermal conductivity (of buffers and chip material), degree of dissociation, pH, and viscosity all change with temperature.[72,163] Of course, higher temperature changes result in larger changes of these thermophysical properties, and consequently affect properties such as ITP electromigration velocity and electrical power required.[162] These changes may sometimes be undesirable in order to preserve design characteristics of the electrolytes chosen for a particular ITP experiment. In addition to good channel geometry design (passive approach), active on-chip cooling mechanisms can be employed (e.g., Peltier cooling or contact with circulating cold water bath) to mitigate Joule heating effects.[235]

Lastly, we note that though Joule heating is usually an impediment in ITP processes, temperature changes and gradients associated with Joule heating in ITP may potentially be useful for certain applications. For example, Bender et al.[236] leveraged the increased temperature (35 to 40°C) in the adjusted TE zone arising from Joule heating to simultaneously perform ITP and isothermal nucleic amplification using a method called recombinase polymerase amplification (RPA); see **Section 11.3** for more details.

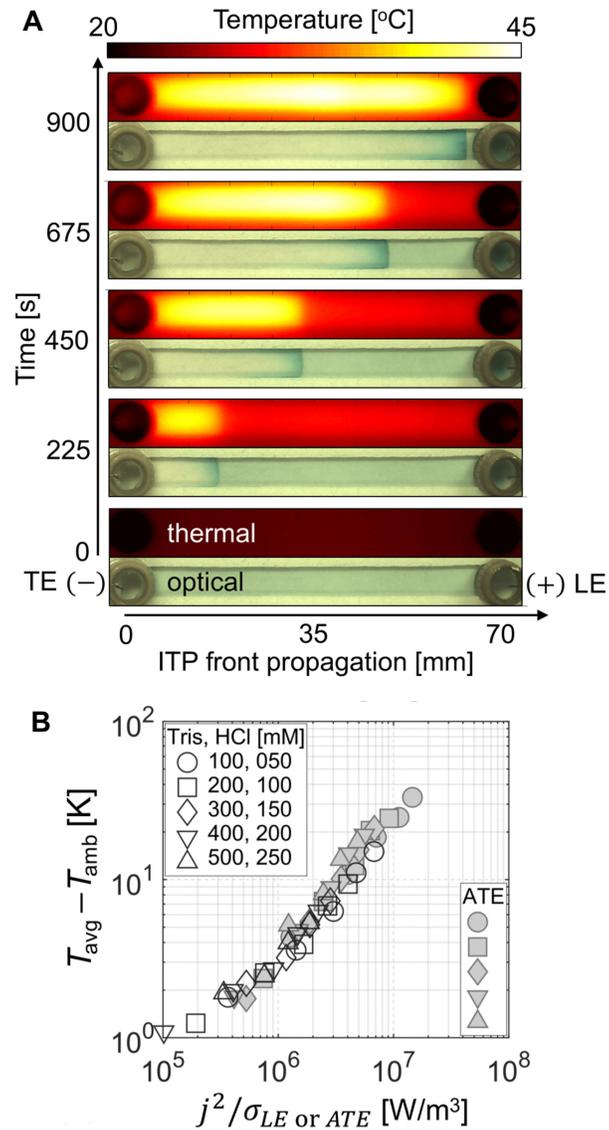

**Figure 18**. (A) Optical and thermal images of the ITP process versus time in a straight microchannel. Optical images show the focusing of a dye in peak mode ITP and helps track the LE-TE interface. Thermal images show spatial temperature fields in the channel. The temperature is significantly higher in the adjusted TE zone compared to the LE. (B) Spatiotemporal averages of quasi-steady state temperature rise from ambient within the LE and adjusted TE zones as a function of the estimated volumetric Joule heating rates (proportional to the square of applied current density). Figure is reproduced with permission from Ref.[162]. Copyright 2020 Elsevier.

### 8.3 Separation parameter, capacity, charge, and distance

Operation conditions and assay design parameters which optimize separations in analytical ITP applications have been well studied in several works.[2,73,168,237–244] An important concept such system design is the so-called separation capacity which plays a key role in regulating the efficiency of ITP process.[238,239] Separation capacity is a measure of the amount of ions (e.g. moles of all ions in a sample volume) which can be processed by a given ITP process. Specifically, separation capacity is a relationship between the amount of electrical charge required to be applied in order to separate and migrate a given amount and composition of the sample mixture. In this section, we will limit our discussion to the analysis of ITP figures of merit and design parameters relevant to purification of sample analytes in peak-mode ITP (**Fig. 19**). We refer to the studies of Refs.[2,73,168,195,238–244] for detailed discussions around optimization and design parameters for ITP processes involving separations in plateau-mode ITP.

The discussion presented here follows the general approach of Boček et al.[239] originally developed for separation of ITP zones (plateau-mode ITP), and which was later expanded by Marshall[178] for peak-mode ITP purification applications and for weakly ionized species. Marshall[178] considered an ITP problem of purification and elution of a dilute analyte ion with a relatively high mobility from a sample containing impurity ions of lower mobility. Marshall[178] considered the system shown in **Fig. 19** where the sample mixture is injected in a finite region between pure LE and TE buffers over a channel length $L_s$ (and cross section area $A_s$) and this section is followed by a separation channel of length $L_{ch}$ (and cross section area $A_{ch}$). Marshall considered a simple ITP process free of applied bulk flows [180,221,243,245–247] Further, for simplicity, Marshall studied a system with five ionic species comprised of a desired trace analyte, the highest mobility trace impurity ion, leading and trailing electrolyte co-ions, and a counter ion common to the LE and TE zones. The analyte and impurities were assumed to be in such low concentrations that they do not significantly alter the zone conductivities or pH. Such systems are common, for example, in ITP applications for purification of nucleic acids or proteins from biological samples which may contain several impurities and inhibitors for downstream analyses.[31] Marshall further proposed a minimum set of design chip aspects for such applications that included air outlet sections to aid finite injection, and extraction and buffering reservoirs. The relations presented below assume that raw sample is diluted in the TE during finite injection. We refer to Marshall[178] for results for the condition wherein sample is diluted in LE.

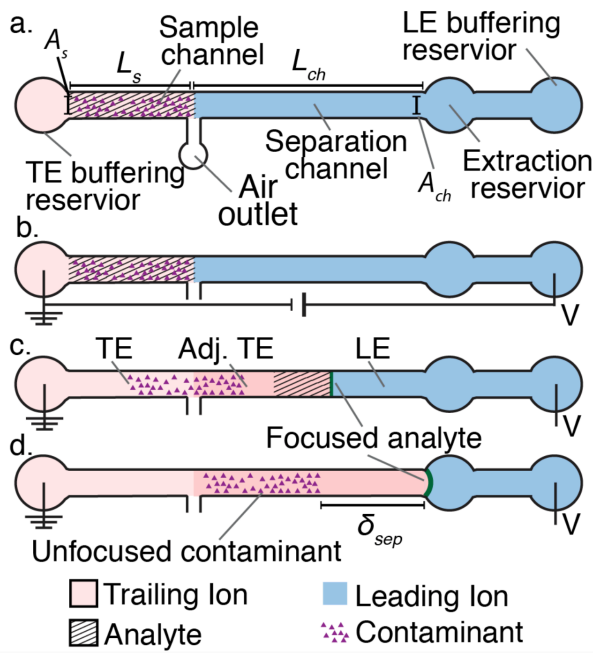

**Figure 19**. Schematic of an example system for ITP purification experiments. (A) Initial sample loading using finite-injection. Sample is diluted in TE prior to loading. (B) Application of electric field in the main channel to initiate ITP. (C) Analytes focus in ITP while contaminants are left behind. (D) The experiment ends when the ITP peak (contained the analytes) reached the extraction reservoir. Purified analytes are eluted from the extraction reservoir. Figure is reproduced with permission from the author of Ref.[178].

A key figure of merit relevant to purification processes is the *recovery efficiency* of the desired analyte from the sample. Recovery efficiency can also be formulated in terms of a *separation capacity* and *separation parameter*.[178,239] Marshall obtained the number of moles of focused analyte $N_A$ per unit applied charge $Q$ as

$$\frac{dN_A}{dQ} = \left[\frac{c_A}{c_T}\frac{\mu_T p_{AT} T_T}{\lambda_T}\right]_{ATE}, \tag{127}$$

where, $c_A$ and $c_T$ are the concentration of the analyte and trailing ion, $p_{AT} = (\mu_A - \mu_T)/\mu_T$ is the *separability* parameter between the analyte with effective mobility $\mu_A$ and the TE co-ion with effective mobility $\mu_T$,[179] and $T_T$ is the transference number of the trailing ion (i.e., the fraction of total current carried by the TE co-ion)[238], and $\lambda_T$ is the molar conductivity of the trailing ion. The molar conductivity $\lambda_i$ and effective mobility $\mu_i$ of species $i$ can be obtained as $\mu_i = \sum_z \mu_{i,z} g_{i,z}$ and $\lambda_i = \sum_z \mu_{i,z} g_{i,z} Fz$, where $\mu_{i,z}$ and $g_{i,z}$ are the absolute mobility and molar fraction of species $i$ in charge state $z$, respectively. Note that the parameters in Eq. (127) are evaluated in the adjusted TE (ATE) zone, which is the newly formed TE zone formerly occupied by the LE. Ions in the ATE zone are subject to the appropriate regulating functions.[87]

The *separation parameter* $Q_s$ is equal to the amount of applied electric charge required to accumulate $N_A$ moles of analyte at the ITP interface.[178,238,239] The total electrical charge $Q_s$ required to extract all analyte ions from the sample can be obtained by setting $N_A$ to be equal to the total analyte content of the raw sample and integrating Eq. (127) to give

$$Q_s = [N_A]_s \left[\frac{c_T}{c_A}\frac{\lambda_T}{\mu_T p_{AT} T_T}\right]_{ATE}, \tag{128}$$

where the subscript $s$ indicates that $N_A$ is evaluated in the initial sample zone. Further, Marshall assumed well-buffered ITP where the pH of all zones is nearly the same, the effective mobility of the ions and the concentration ratio $c_T/c_A$ are nearly the same in both TE and ATE zones.[178,248] Under this assumption, Marshall obtained the separation parameter $Q_s$ in terms of the amount of TE co-ions in the initial sample zone (assuming sample is diluted in TE) $[N_T]_s$ as

$$Q_s = [N_T]_s \left[\frac{\lambda_T}{\mu_T p_{AT} T_T}\right]_{ATE}, \tag{129}$$

Eq. (129) is more useful than Eq. (128) since the number of target analytes is often *a priori* unknown while the TE ion concentration (used to dilute the sample) is a design parameter easily controlled by the user. Further, note from Eq. (129) that the separation parameter $Q_s$ is dependent only on the initial number of TE ions in the sample (or LE ions if sample is diluted in LE[178]) and properties of the ATE (or LE if sample is diluted in LE[178]). Recall that ATE properties are regulated by the LE.

In practice, the charge transferred to the ITP system prior to the end of experiment (i.e., when the ITP interface reaches the elution reservoir) is limited by the time it takes for the ITP interface to traverse the separation channel region. This limits the actual amount of charge that can transferred in an ITP experiment, $Q_L$, which is the *separation capacity*.[178,239] $Q_L$ is best expressed in terms of the LE properties as[178,239]

$$Q_L = \left[\frac{N_L \lambda_L}{\mu_L T_L}\right]_{LE}, \tag{130}$$

where the subscript $L$ refers to the LE co-ion and the variables in the bracket are evaluated in the separation region of the channel which is initially filled with LE. Note from Eqs. (129) and (130), that for a given ITP buffer chemistry, separation capacity and separation parameter are directly proportional to the number of ions in the separation and sample regions of the channel, respectively.[178]

The *recovery efficiency* of ITP purification $\eta$ under finite-injection conditions, defined as the fraction of analyte ions that accumulate in the ITP interface prior to the interface reaching the end of the separation region, is obtained as the ratio of separation capacity and separation parameter as[178]

$$\eta_f = \frac{Q_L}{Q_s} = R_{vol}\frac{[c_L]_{LE}}{[c_T]_s}\frac{\left[\frac{\lambda_L}{\mu_L T_L}\right]_{LE}}{\left[\frac{\lambda_T}{\mu_T p_{AT} T_T}\right]_{ATE}}, \tag{131}$$

where $R_{vol} = \forall_{ch}/\forall_s$ is the volume ratio of the separation and sample channel regions, $[c_L]_{LE}$ is the concentration of the LE co-ion in the separation region, and $[c_T]_s$ is the concertation of the TE co-ion in the sample region. Eid and Santiago[64] later obtained an expression for the recovery efficiency in the case of semi-infinite injection of sample in TE, $\eta_{si}$, as

$$\eta_{si} = 1 - \exp\left(-p_{AT}\beta\frac{\forall_{ITP}}{\forall_{well}}\right), \tag{132}$$

where $\beta$ is the ratio of the TE ion concentration in the ATE and TE zones, $\forall_{ITP}$ is the volume of the channel sept by the ITP peak, and $\forall_{well}$ is the reservoir volume. In the limit when only a small fraction of the sample is processed (e.g., when the reservoir to channel volume ratio is large), then $\eta_{si} \approx p_{AT}\beta\forall_{ITP}/\forall_{well}$.

Next, the *separation time* $t_{sep}$, defined as the duration for the ITP interface to traverse the separation channel, is obtained by dividing the transferred charge by the applied current (in the case of constant current operation). $t_{sep}$ is given by

$$t_{sep} = \frac{Q_L}{I} = \frac{A_{ch}L_{ch}}{I}\left(\frac{\lambda_L c_L}{\mu_L T_L}\right)_{LE}, \tag{133}$$

$t_{sep}$ scales inversely with applied current and is proportional to the LE co-ion concentration and separation channel volume. Recall here the importance of joule heating in limiting applied current (see **Section 8.2**).

Marshall[178] further considered the separation of the focused analyte from trace amounts of background non-focused co-ion impurities in the sample. For simplicity, separation of analyte from the fastest moving background impurity (with effective mobility $\mu_R$, less than $\mu_A$) was considered in the analysis. In that case, the contaminant of interest electromigrates with velocity $U_R$ less than the ITP interface velocity $U_{ITP}$, and is given by

$$U_R = [\mu_R E]_{ATE} = \left[\frac{\mu_R}{\mu_T}\right]_{ATE}U_{ITP}. \tag{134}$$

When the ITP interface reaches the end of the separation channel (e.g., defined by the location of some detector)), the distance of the focused analyte from the contaminant $\delta_{sep}$, defined as the *separation distance*, is obtained as[178]

$$\delta_{sep} = \int_0^{t_{sep}} (U_{ITP} - U_R)\, dt = [p_{RT}]_{ATE} \left( \int_0^{t_{sep}} U_{ITP}\, dt \right) = [p_{RT}]_{ATE}\, L_{ch}\,. \tag{135}$$

Notice when the sample is diluted in TE that $\delta_{sep}$ depends only on the separability of the contaminant and the length of the separation channel. When the sample is diluted in LE, $\delta_{sep}$ depends additionally on the amount of LE ions in the sample and separation channels and the separability of the LE co-ion with respect to the contaminant.[178] Marshall[178] also presented an expression for the *extra charge* $Q_{ex}$ that can be supplied to the ITP system between the time of arrival of the LE-ATE interface and the arrival of the leading contaminant. $Q_{ex}$ is given by

$$Q_{ex} = \left[\frac{\mu_T - \mu_R}{\mu_R}\right]_{ATE} Q_L = [p_{TR}]_{ATE}\, Q_L\,. \tag{136}$$

In other words, $Q_{ex}$ is the maximum additional charge that be applied to the system beyond $Q_L$ to avoid any contaminants in the extraction reservoir. This extra charge may be useful, for example, to help move and guide the purified analyte within the extraction reservoir to enable efficient elution for off-chip analyses.

Another figure of merit for ITP purification is the sample *dilution factor*, $F_{dil}$.[178] $F_{dil}$ is defined as ratio of concentration of the target analyte in the raw sample and the extracted liquid. $F_{dil}$ is obtained in terms of the recovery efficiency $\eta_f$ and the initial sample ($\forall_s$) and extracted ($\forall_{extract}$) liquid volumes as[178]

$$F_{dil} = \frac{\forall_{extract}}{\eta_f \forall_s}\,. \tag{137}$$

In practice, $\forall_{extract}$ depends on the extraction reservoir geometry and design and the liquid handing techniques used in the system.

### 8.4 Well-buffered ITP systems

This section presents some comments around pH buffering of ITP experiments. pH buffering is important to ensure stable and/or functional (e.g., enzymes) biological species. pH buffering is also important for reproducible ITP processes since weak electrolyte effective mobilities and solubilities depend on local pH (and ionic strength and temperature). Of course, species effective mobilities influence conductivity and couple with applied current (or voltage) to determine ITP front velocities. In turn, these quantities determine system separation efficiency and accumulation rates. In this section, we will discuss two main aspects of buffering in ITP. The first is the chemical choice of buffering ion(s) and titrant(s) among the anionic and cationic species of the LE and TE buffers. The second is the requirements (in particular the volume and concentration) of reservoirs required to promote stable and reproducible system performance. Our discussion will assume the reader has a basic working knowledge of pH buffers including buffer capacity and familiarity with common buffering species (e.g., Good's type buffer ions) and common titrants (e.g. chloride and sodium). See Refs.[85,86] for a detailed tutorial on pH buffers including the coupling among ion mobility, ion velocity, and pH. We will also assume typical pH ranges of, say, 4 to 10 and buffers of sufficient ionic strength so that the safe pH and moderate pH assumptions apply.

#### *8.4.1 Choices for buffer species*

Our goal is to ensure pH buffering of the LE and TE zones and, ideally, pH buffering of plateau mode zones. Note we do not need a separate strategy to ensure the buffering of a trace species being focused in peak-mode species as the distribution of pH (and conductivity and electric field) within such peaks is governed by the pH values of the neighboring plateaus. There are several approaches to ensure stable and reproducible pH values in an ITP process. We will here discuss what is, to our knowledge, the most widely applicable and reproducible method: Buffering of all zones using the LE counter ion as the single buffering ion. The counter ion of the original LE buffer is continuously supplied to the system by the LE reservoir. After sufficient time, the counterion from the LE zone displaces whatever counter ion was in the original TE and so, eventually, the LE's counterion is the only counterion in the adjusted TE region. The most straight forward manner to ensure good pH buffering is then to make the LE counterion serve as the system's buffering ion and the LE, adjusted TE, and any plateau-mode co-ionic species act as titrants to this buffering ion. **Figs. 20A** and **20B** demonstrate this arrangement for anionic (top) and cationic (bottom) ITP. Although not shown in the **Fig. 20**, we consider the analyte *X* to be introduced via finite injection. In the top (bottom) schematic, the weak base (weak acid) migrates in the opposite direction of LE anion (cation), eventually replacing the original cation (anion) in the initial, nearby TE zone. Downstream, we observe that the adjusted TE is composed of the co-ionic TE anion (cation) and the weak base (weak acid) which migrates from LE to ATE. To ensure well buffered LE and ATE zones, we can choose an LE counterion with a convenient $pK_a$ as the buffering weak electrolyte. Hence, in creating the LE buffer, we titrate this weak electrolyte using the other LE ion as the titrant. For anionic (cationic) ITP, we titrate the LE weak base (weak acid) using the LE anion (cation) as a titrant. For strong electrolyte univalent ion titrants, strong buffering is achieved when the titrant concentration is roughly half of the concentration of the weak electrolyte. This 1:2 ratio also holds for weak electrolyte titrants with $pK_a$ values of roughly 1.5 pH units greater (lesser) than the $pK_a$ of the buffering weak acid (weak base) (see Persat et al.[85] for design of buffers using weak electrolyte titrants).

As discussed in **Section 4**, the concentration of the ATE co-ion will be on the same order as, and typically just smaller, than the LE co-ion. If we apply the latter section's conservation principles to the counter ion (which we leave as an exercise for the reader), we can show that the counter ion's concentration in the ATE will be roughly the same as its concentration in the LE. Hence, we expect the LE and ATE to each be buffered mixtures with a pH of approximately the same as the $pK_a$ of the weak electrolyte counterion (originating in the LE). Similarly, any plateau zone will be buffered near the $pK_a$ of the counterion if the plateau species is an acid (base) with a $pK_a$ at least about 1.5 pH units below (above) the $pK_a$ of the weak base (weak acid) counterion. In summary, the idea that the TE ion and the ions of any plateau zones in the system act as titrants to the LE buffer's buffering ion. A common example system, applicable to a wide range of biological applications, is an anionic ITP process created using Tris and chloride as the LE buffer and Tris and HEPES as the TE buffer.[31,63,93,162] In creating the LE buffer, the analytical concentration of Tris is set at about twice the value of chloride. Conservation principles (see **Section 2** and **4**) then ensure that Tris enters the ATE at roughly twice the concentration of HEPES. Hence, Tris is the buffering counterion with chloride and HEPES as titrants in the LE and ATE, respectively. The two zones are then well buffered with a pH near the $pK_a$ of Tris. In **Supplementary Information Section S2**, we provide general guidelines and a few recommendations on choosing ions and designing buffers for ITP.

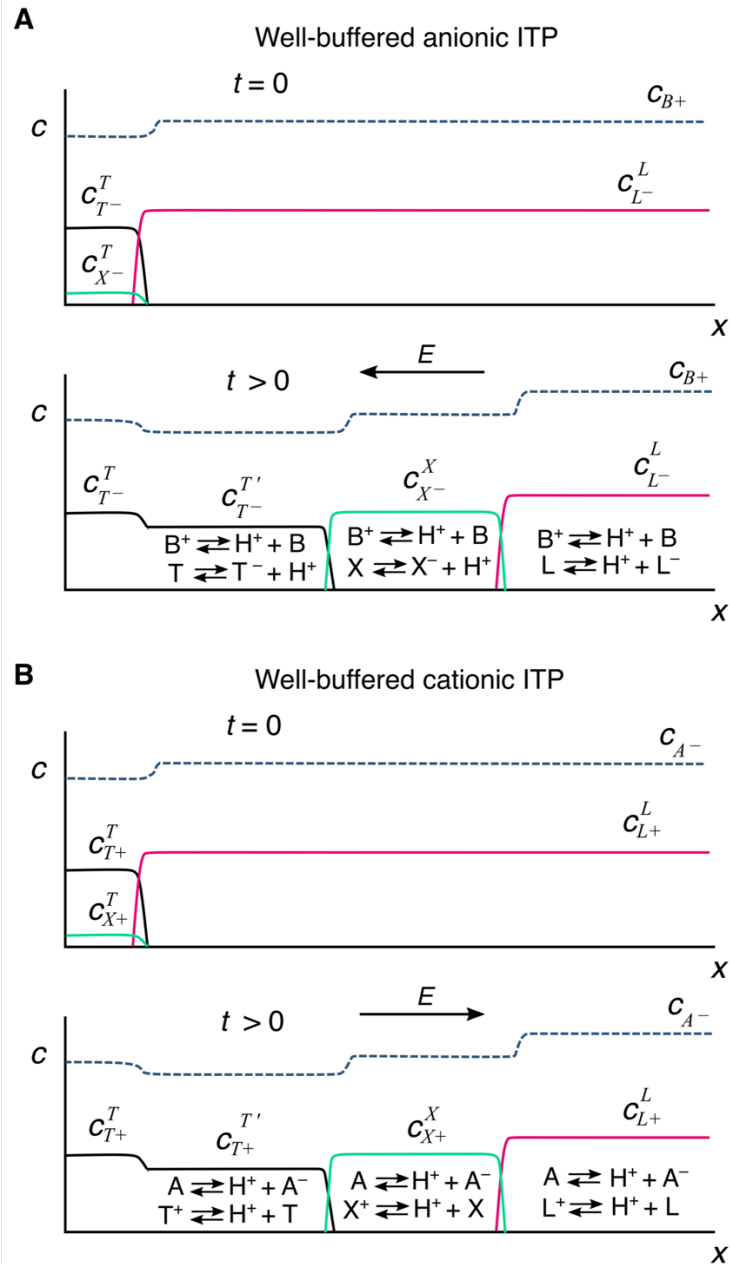

**Figure 20**. Schematic of well-buffered (A) anionic, and (B) cationic plateau mode ITP. Each subfigure depicts the concentration distributions of ions at the initial (top) and steady state separation conditions (bottom). Within each zone, the associated acid/base chemical equilibrium reactions are shown. A common weak base *B* (weak acid *A*) is used to buffer all zones in anionic (cationic) ITP, indicated by dashed lines. Although not shown here, the analyte *X* is considered to be introduced via finite injection.

### *8.4.2 Reservoir buffering*

All ITP is driven by Faraday reactions at electrodes. For microfluidics, these are nearly always electrodes inserted into end-channel reservoirs where electrolysis occurs. These reservoirs are typically open to the atmosphere to accommodate hydrogen (oxygen) gas generation at the cathode (anode). Importantly, the buffering capacity of the reservoir chemistry needs to be sufficiently strong to resist changes in pH associated with electrolysis (i.e., acid and base generation at the anode and cathode, respectively).

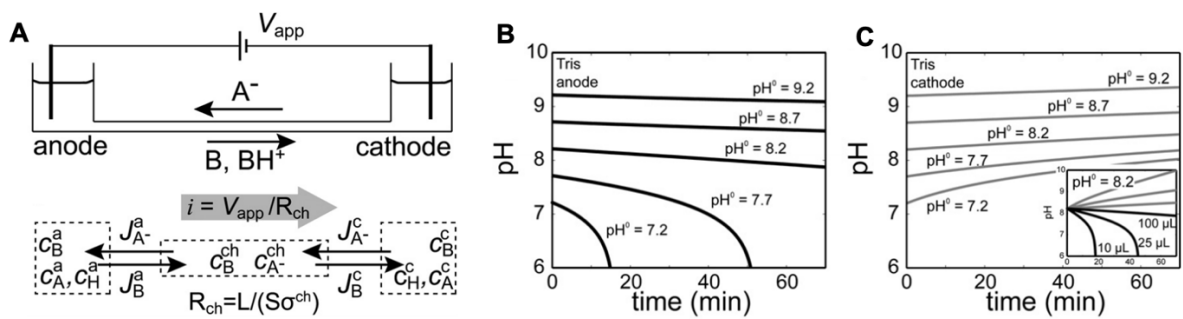

**Figure 21**. (A) Schematic depiction of a generic on-chip electrokinetic system where the electrolyte is a solution of a weak base $B$ titrated with a strong acid $A$. (B) and (C) show predicted variation of the pH at the anodic and cathodic reservoirs versus time, respectively, due to electrode reactions. A Tris-HCl buffer system is considered, and a 100 µL reservoir volume is assumed. Shown are model predictions for various initial titrations of the pH of the buffer within one $pK_a$ unit of the weak base Tris. Figure is reproduced with permission from Ref.[86]. Copyright 2009 The Royal Society of Chemistry.

A properly designed ITP system requires reservoirs with sufficient buffering capacity relative to the amount of charge injected into the system by the electrodes. The relevant proper design of reservoirs is discussed in detail by Persat et al.[86] and are merely summarized here. See the example system of **Fig. 21A**. This figure considers a generic electrokinetic experiment where current is driven by two end-channel electrodes though a microchannel. The species $A$ and $B$ in the diagram refer to generic "acid" and "base" ions of a buffer. As an example, consider $A$ to be chloride ions migrating into the anodic reservoir (which is buffered by Tris). In this anodic reservoir, the pH drops over time as more charge is applied to the system. The drop in pH at the anodic reservoir is rapid if the initial pH is less than the $pK_a$ of the weak base $B$, Tris (c.f. **Fig. 21B**). Likewise, the counterion Tris (weak base $B$ in **Fig. 21A**) migrates to the cathodic reservoir and increases the pH of the cathodic reservoir. The increase in pH at the cathodic buffer (**Fig. 21C**) under the same conditions are, however, more gradual.

These observations lead us to another important consideration for designing buffers in ITP which is related to the buffering capacity (or, buffer index), $\beta$. $\beta$ represents the ability of the buffer to resist changes in pH, and as we discuss below, is proportional to the amount of applied charge.[85,86] $\beta$ can be expressed as the amount of strong base or acid required to induce a small change in pH of a weak electrolyte buffer, i.e., $\beta = -\partial c_A/\partial \text{pH}$ (for the addition of strong acid $A$), or $\beta = \partial c_B/\partial \text{pH}$ (for the addition of strong base B). $\beta$ is a property of the buffer and is a complex function of the ionic strength, $pK_a$, temperature, charge transferred by the electrodes (equivalently, duration of the experiment), and electrode reservoir volume.[85,86] For example, for the common example of a buffer composed of 2:1 molar ratio of weak base and strong acid, $\beta \approx 0.56\, c_B$, where $c_B$ is the concentration of the weak base. For a given buffering capacity $\beta$ and small pH changes (less than one pH unit), the applied charge $Q$ is directly proportional to the change in pH ($\Delta \text{pH}$) and the buffering volume $\forall_{buf}$ as[85,86,178]

$$Q \approx \Delta \text{pH}\, F \beta \forall_{buf}\,. \tag{138}$$

Eq. (138) can be used to estimate the maximum transferable charge $Q_{max}$ for a maximum tolerable pH change in ITP (a design parameter), say $\Delta pH_{max}$, for given a reservoir volume $\forall_{buf}$.[178] Typical recommended values for $\Delta pH_{max}$ are around 0.2 to 0.4. Note further from Eq. (138) that a larger reservoir volume and higher ionic strength allows for more charge transferred to the system while ensuring stable pH (see also inset of **Fig. 21C**).

Furthermore, anodic and cathodic buffers can also be designed by "anticipating" pH changes. For example, in anionic ITP, the pH of the anode (cathode) can be titrated so that the solution pH is slightly above (below) the $pK_a$ of the weak base (acid). Similar strategies can be applied to cationic ITP systems.

Lastly, we note that atmospheric carbon dioxide dissolved in buffers can, in some cases, lead to unwanted side reactions[90] and byproducts (e.g., carbamate, carbonate), which may result in additional (undesirable and unknown) zones formed during ITP.[89] For example, dissolved carbon dioxide can react with buffer species containing primary amines to form carbamate which has a relatively high anionic mobility and focus just behind the LE.[89] To mitigate the effects of carbon dioxide and related species, barium hydroxide has been mixed with the TE prior to loading on chip to precipitate out dissolved carbon dioxide.[89]

*8.4.3 Operational regime map*

The design of buffers is particularly important for the case of relatively large fluidic channels used to process relatively large reagent volumes (and/or relatively large reagent ionic strengths). As one instructive design process, see the work of Marshall[178] and Marshall et al.[230] who considered the combined constraints of maximum permissible amount of electrical charge transferred (for stable pH buffering) provided by Eq. (138) together with system limitations imposed on current magnitude by Joule heating (as a function of LE concentration) (Eq. (126)) and bubble generation at the electrodes due to electrolysis. These constraints were combined to create an "operational regime map" useful in system design (**Fig. 22**). Contour lines represent achievable extraction times as functions of applied current and LE concentration (and separation capacity, units of Coulomb). Shaded areas in the plot are precluded by design constraints and result in unstable operation. Such operation maps can be used to guide design tradeoffs among assay time, volume processed, maximum allowable temperature achieved, and robust operation including repeatable electrolysis bubble release.

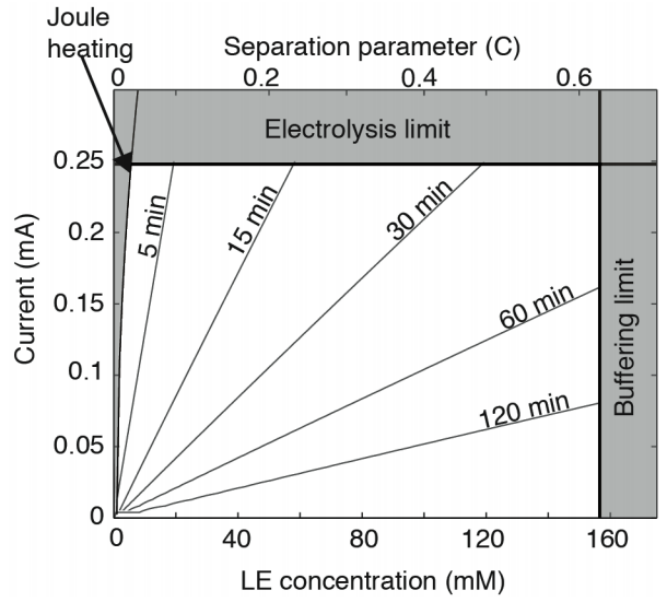

**Figure 22**. Example operational regime map for ITP purification experiments performed on a custom microchip designed by Marshall et al.[230]. Shown are contours of extraction time as a function of LE concentration and applied current. Shaded regions are precluded by design and operational constraints. Low LE concentration results in high Joule heating. At high currents, large bubbles form at the electrode surface and lead to poor electrical connection., Electrolysis can change the pH in the electrode reservoirs at high separation capacity. Figure is reproduced with permission from Marshall et al.[230]. Copyright 2014 Elsevier.

## 8.5 Other considerations for practical ITP experiments

There are, of course, many of other considerations associated with electrokinetic experiments and designing and operating microfluidic devices.[249] In this section, we briefly mention a few considerations particular to ITP experiments including operation with constant current or constant voltage, tradeoffs between sample volume and sensitivity, and channel materials.

ITP experiments are typically performed using either constant voltage or constant current operation. In channels of uniform cross section operated at constant current, system voltage typically increases linearly with time (since the total resistance of the channel increases as TE ions displace LE ions) while the ITP velocity is nearly constant. In this case, ITP dynamics are easier to analyze since, for example, finite injection samples will achieve a steady state condition (in the frame of reference of the moving ITP interfaces). Experimentally, constant ITP interfaces are easier to track in space and time. Typical values of applied current (e.g., using ionic strengths of order 100 mM) are a few microamps to tens of microamps for channels with cross-sectional areas of order 100 to 1000 $\mu m^2$, and order milliamps for cross sections of order 0.01 to 0.1 $mm^2$. In constant voltage operation, the current and ITP front velocity decrease with time (see supplementary information document of Bahga et al.[121]). Compared to constant current sources, constant voltage power supplies are generally more widely available, simpler, and offer higher maximum voltage and power (compared to constant current supplies). Typical values of applied voltages for ITP channels with order centimeter length ranges from order 0.1 to 1 kV. It is important to estimate the maximum voltage and current requirements of an experiment. One approach is to perform preliminary experiments where the channel(s) is filled with either pure TE buffer and/or pure LE buffer and perform simple current vs. voltage calibrations of the channel. Non-linear dependance between voltage and current in such experiments is indicative of Joule heating and, for large amounts of charge injected (compared to charge in buffered wells[86]), can also indicate insufficient reservoir buffering (c.f. **Section 8.4**).

Another consideration is the amount of sample volume that can be processed on chip. Electrokinetic microfluidic chips generally are designed to handle low sample volumes, typically on the order of few 100's of nanoliters to a microliter. Larger volumes (and, in particular, channels with larger cross-sectional areas) are more difficult because of the constraints of Joule heating. However, larger sample volumes are sometimes required as in the case of biological samples with trace concentration targets which can require processing of volumes of order 10 to 100 microliters of raw sample (or larger). A common approach in such cases is to employ channels with large aspect ratio cross sections, typically large width and small depth (along the vertical direction). Such channels are very compatible with microfabrication techniques and imaging and can offer larger volume while mitigating Joule heating. Example ITP chips to process relatively large volumes include those of Marshall et al.[230] and van Kooten et al.[222] who processed respectively 25 and 50 microliters. A recently developed commercial device (IONIC ITP system, Purigen Biosystems, Pleasanton, CA) developed an injection molded ITP chip to process 200 μL. There are, of course, systems which aim to both process large volumes and achieve high focusing and high plateau resolution. A classic approach to this is so-called column coupled ITP which employs a T-junction between a large cross-sectional area tube (for upstream volume processing) and a smaller cross-sectional area tube (for downstream high resolution).[250] Tapered-area channels are also used in microfluidics for similar purposes.[121,251] In addition to the aforementioned methods, EOF or pressure driven hydrodynamic counterflow can also be used to improve throughput and amount of processed sample in relatively small volumes (although sometimes at the cost of longer assay times), a topic we review in **Section 12**.

Lastly, we note that ITP microfluidic channels are typically fabricated from glass, fused-silica, plastics, or, less frequently, paper-based materials. Glass microfluidic chips are available from several vendors. Some examples include Caliper Life Sciences (A PerkinElmer company, MA, USA), Micronit (Netherlands), and Microfluidic ChipShop (Germany). Common polymers used PMMA,[30] polycarbonate,[252] polystyrene,[253] COC (cyclic olefin copolymer),[230] and elastomers, particularly PDMS[26,203]. Paper-based ITP systems are typically fabricated using nitrocellulose,[254,255] glass fiber,[256] or the proprietary Fusion 5 membrane materials[256,257]. Recently, 3D printed microchannels were fabricated for ITP application.[258] Few important considerations for choosing channel materials include low surface adsorption of the target molecules, ability to sustain a high voltage prior to breakdown (e.g., silicon channels are rarely used for ITP for this reason), amenability to surface functionalization (e.g., for surface-based ITP reaction assay), stable thermal properties, high thermal conductivity and diffusivity (e.g. as offered by glass), and compatibility with precision fabrication and microfabrication methods. We refer to Refs.[259–264] for comparisons and detailed discussions around various channel materials useful for on-chip electrophoresis applications including ITP.

## 9. Brief review of simulation tools

The theory presented in the sections above does not always allow for simple analytical calculations associated with the ITP process. Computer simulations have therefore played an important role in bridging the gap between theory and practice, and have greatly aided in the design and optimization of microfluidic ITP systems. Note however that ITP dynamics can be computationally challenging since it is a highly nonlinear process which involves multiple chemical species, chemical equilibrium reactions, sharp gradients in concentration, coupling with electric fields, and, at times, coupling with bulk flow. Significant improvements have been made in the modeling and simulation of electrokinetic processes over the last three to four decades. In this section, we briefly review simulations tools useful for ITP with an emphasis on those which are publicly available. As we shall see below, 1D dynamic simulators are at present the most widely used design tools for ITP. We note however that complex ITP phenomena (e.g., dispersion) and channel geometries (e.g., non-uniform cross section) may require simulations which are area-averaged, 2-D or 3-D for better accuracy.

### 9.1 One-dimensional ITP simulation tools

The earliest studies on numerical simulations of electrophoretic processes in general include the related works of Bier et al.[265], Saville and Palusinski[88], and Palusinski et al.[266]. Bier et al.[265] was the first study to implement any kind of an electrophoresis code applicable to ITP. In later (related) publications, Saville and Palusinski[88] and Palusinski et al.[266] presented generalized formulations electromigration-diffusion-advection processes including multiple weak electrolytes (including ampholytes with univalent ionization states) and provided a unified framework for simulating a variety of electrokinetic processes including ITP. Thereafter, several improvements in the physical modeling of electrokinetic processes have been suggested and implemented in numerical codes and simulation software packages, as mentioned below. Some of the key improvements include adaptive grid techniques to efficiently resolve high concentration gradients,[87,142] effects of ionic strength on species mobility and ionic activity[144,169], Taylor-Aris-type dispersion to account for sample dispersion in the presence of non-uniform EOF or pressure driven flow[87,267], axially varying channel cross-section,[268,269] and simulation of complex-formation equilibria[37,270] and non-equilibrium chemical reaction kinetics.[271]

At present, the commonly used dynamic electrophoresis simulation packages for modeling microfluidic ITP include SIMUL[144], GENTRANS[88,265,266,272], and SPRESSO[87,121,169]. These software are largely based on the theoretical formulation of Saville and Palusinski,[88] and have been periodically (and significantly) updated to include new computational capabilities and physical models such as those mentioned earlier. Bercovici et al.[87] in 2009 introduced a Taylor-Aris-type dispersion model to account for sample dispersion in presence of non-uniform EOF and pressure driven flow in the SPRESSO software. In similar vein, GENTRANS was later expanded by Caslavska et al.[267] to include Taylor-Aris-type dispersion effects. GENTRANS was also expanded by Breadmore et al.[37] to include calculations involving electrokinetic chiral separations with a neutral ligand. A similar extension of SIMUL 5 was presented by Hruška et al.[273] and Svobodová et al.[274], which led to SIMUL 5 Complex software. Dagan and Bercovici[271] extended SPRESSO to include non-equilibrium homogenous and heterogeneous reaction kinetics of chemical species. Ionic strength effects, which are typically modeled based on the Onsager-Fuoss model[173] and the extended Debye-Hückel theory[174] for ionic activity, were included originally in SIMUL 5[144] and later implemented in SPRESSO by Bahga et al.[169]. Recently, the software SIMUL 6[275] was launched as a successor of SIMUL 5. SIMUL 6 uses parallelization and multithread computation to significantly speed up computation time compared to SIMUL 5. In 2021, Avaro et al.[276] published an open source, web-based and cross-browser compatible ITP simulation tool called CAFES. CAFES enables rapid computations of the dynamic ion concentration and pH fields during ITP for a wide range of sample injection configurations. It has built-in database of buffer ions but also enables custom-designed buffer mixtures. CAFES has a simple graphical interface for user inputs and the simulations are directly performed on the user's computer (by their browser software), requiring absolutely no prior installation or compilation of the software.

The aforementioned 1-D simulators are strictly applicable only to straight channel, one-dimensional ITP. Hence, they predict only streamwise variation in concentration of species and cannot account for spanwise (and transverse) variations in concentration, for example, due to dispersion or to cross-sectional area changes. The latter requires two- or three-dimensional simulations which we discuss in **Section 9.2**. However, D, Hruška et al.[269] and Bahga et al.[121] used area-averaged 3-D species transport equations to approximate the effect of cross-sectional area variation in 1-D. They, respectively, expanded SIMUL and SPRESSO to include a quasi 1-D model to account for cross-sectional area changes in the channel. As an illustration, **Fig. 23** depicts the present capabilities of the SPRESSO software.

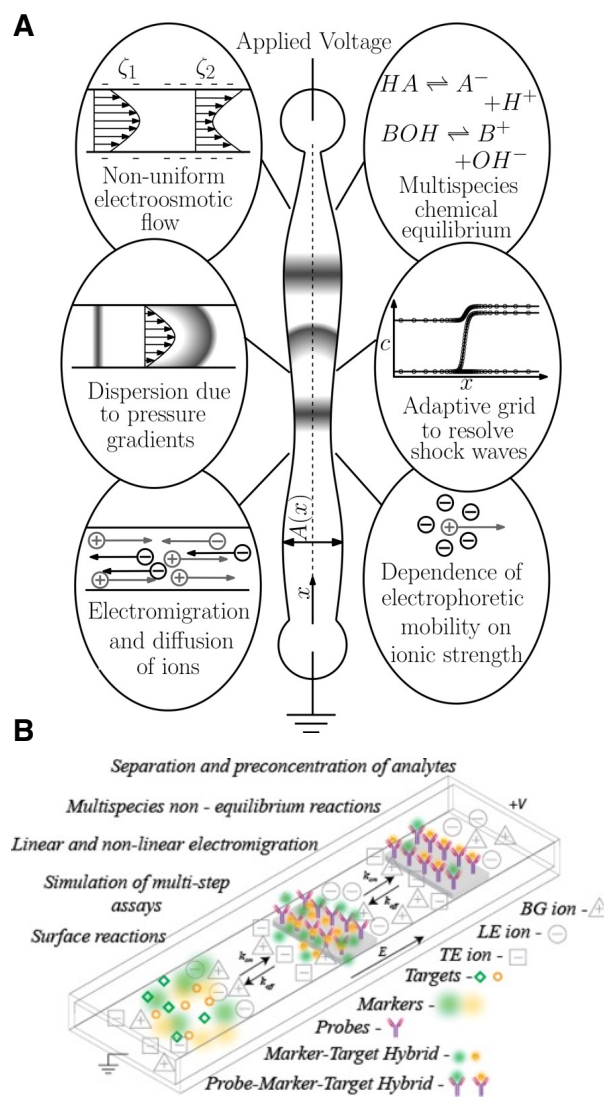

**Figure 23**. (A) Schematic highlighting various physical features of the SPRESSO simulation tool developed in Refs.[87,142,268]. Figure is reproduced with permission from Ref.[268]. Copyright 2012 John Wiley and Sons. (B) Illustration of the extended version of SPRESSO by Dagan and Bercovici[271] which includes a chemical reaction module to simulate electrophoresis-driven hybridization, binding, or chemical reaction processes. Figure is reproduced with permission from Ref.[271]. Copyright 2014 American Chemical Society.

SPRESSO is available freely to download from http://stanfordspresso.blogspot.com/. GENTRANS is available upon author request (from wolfgang.thormann@ikp.unibe.ch). CAFES simulations can be performed for free at: https://microfluidics.stanford.edu/cafes. We note that PeakMaster is another often used electrophoresis software, but it is only applicable for zone electrophoresis (and provides a rapid way of determining buffer equilibria and species mobility). The latter software is based on the linear theory of electromigration. Both SIMUL and PeakMaster software packages can be downloaded for free from https://web.natur.cuni.cz/gas/.

Further, we note that, although the underlying physical models are largely common among the simulation software and several other numerical studies of ITP, the numerical methods (in particular, the spatial discretization scheme) used are significantly different. The choice of the numerical method determines the tradeoffs among computational speed, accuracy, and spatial resolution. It is instructive to broadly categorize the numerical schemes (based on the spatial discretization) as dissipative or non-dissipative. Dissipative schemes use numerical diffusion to provide a stable, non-oscillatory solution at the expense of low accuracy (e.g., they predict much larger interface width than is physical). Examples of dissipative schemes include the upwind scheme,[277] flux-corrected transport scheme,[278] space-time conservation element solution element scheme,[279] and symmetric limited positive scheme.[268] In contrast, non-dissipative schemes offer higher accuracy but at the expense of higher computational time. Non-dissipative schemes typically require higher number of grid points to prevent oscillatory solutions. Examples of such schemes include second-order central difference[272] and sixth-order compact scheme[87]. Additionally, as mentioned above, numerical schemes which use adaptive grid refinement[87] (for e.g., SPRESSO), or those based on pseudo-spectral methods[280] have shown improvements in computational time and accuracy. The latter pseudo-spectral method was recently developed by Gupta and Bahga[280] for electrophoresis simulations including t-ITP and was released as a free software package called SPYCE which can be downloaded at http://web.iitd.ac.in/~bahga/SPYCE.html. We refer to Refs.[281–283] for studies that directly compare the performance and accuracy of several 1-D simulators across various electrophoretic processes including ITP. Lastly, we mention a free, open-source software tool called STEEP which enables detailed calculations of temperature effects of ionic species mobility and pH equilibria.[163]

**9.2 Multidimensional ITP simulation tools**

As mentioned above, ITP in microfluidic channels can exhibit strongly two- and three-dimensional concentration fields, for example, in the presence of streamwise variation in channel geometry or due to dispersion associated with pressure-driven flow and/or EOF. Additionally, complex sample injection strategies[284] and ITP processes very often involve microfluidic chips with branched channel networks,[36] and these require at least 2-D simulations to model the underlying electrokinetic transport. To study such effects, researchers have mostly relied on multiphysics software packages which include COMOSL Multiphysics[61,180,251,270,285] (COMSOL, Inc., Burlington, MA, USA) and CFD-ACE+[284,286] (CFD Research, Huntsville, AL, USA), and the freeware OpenFOAM.[287] For example, Garcia-Schwarz et al.[61] and Schönfeld et al.[139] used COMSOL to study 2-D fields associated with sample dispersion in ITP due to non-uniform EOF. Likewise, GanOr et al.[60] and Liu and Ivory[288] used COMSOL in 2-D to study the effects of counterflow in ITP. Paschkewitz et al.[289] used CFD-ACE to study ITP focusing in serpentine passages in a microfluidic chip.

In addition, custom numerical codes have also been developed to simulate and study ITP processes in 2-D. For example, Bhattacharyya et al.[62] developed a third-order QUICK scheme to study sample dispersion in the presence of counterflow in ITP. Likewise, Shim et al.[290] developed a finite-volume solver to study a five-component ITP system.

To our knowledge, the only study to date that reports 3-D simulation of ITP is the work of Bottenus et al.[251]. They used COMSOL to simulate 3-D ITP sample focusing and preconcentration in a cascade microchip which had a channel cross-sectional area that decreased 100-fold along the streamwise direction. It was later reported that this simulation took several weeks to complete.[19]

Though insightful, the main drawback for multidimensional tools to simulate ITP is the computational cost. At present, 2-D simulations take several hours to days, while 3-D simulations require several weeks or longer, depending on the computational power and memory. In contrast, 1-D simulations can take seconds to minutes to run and these can be performed on personal computers. For this reason, we believe that 1-D dynamic simulators will continue to play an important role, at least in the near future, for quick ITP assay designs including buffer chemistry, injection and initial configuration, operating electric field, detector placement, and overall optimization (including tradeoffs among resolution, sensitivity, and assay time).

## 10. Experimental tools and detection methods

In this section, we begin by providing scaling arguments for the resolution and sensitivity of separation processes which use plateau mode ITP. We then review various methods used for experimental visualization and/or analyte detection in microfluidic ITP. We place special emphasis on fluorescence, thermometric, and conductimetric detection approaches as these are the most compatible with microfluidic devices. Also, we focus on detection systems for analytical ITP, and do not review methods which couple ITP purification, focusing, and separation with downstream analyses such as zone electrophoresis, mass spectroscopy, or chromatography). See Refs.[22,24,243,291–294] for discussions on the latter topic.

### 10.1 Sensitivity and resolution for plateau zones

The width of plateau regions, $\Delta x_i$, is proportional to the amount of analyte $i$ collected (Eq. (95)). These widths, and the width of the interfaces between plateaus, $\delta$, therefore determine the minimum amount of analyte which can be detected via plateau identification. The step heights in the signal plateau regions, $\Delta s_i$, are related to differences between analyte effective mobilities and that of the LE. For example, an analyte mobility much lower than that of LE will cause a large step height.

In plateau mode, we can therefore define *sensitivity*, $S$, as a measure of whether an analyte which forms a local plateau region can be differentiated from both of its two adjoining plateaus. Well resolved peaks will have plateaus widths $\Delta x_i$ much larger than the interface width between zones $\delta$ (c.f. Figure 16). Hence, we recommend that $S$ can be formulated as

$$S = \frac{\Delta x_i}{2\delta} = \underbrace{\left[\frac{f(\mu_i)}{2}\left(\frac{\mu_L}{\mu_T} - 1\right)\right]}_{g(\mu_i)} \frac{F}{RT} \frac{c_{S_i}^W}{c_T^W} \frac{j}{\sigma^L} l \equiv g(\mu_i) \frac{F}{RT} \frac{c_{S_i}^W}{c_t^W} \frac{j}{\sigma^L} l. \tag{139}$$

Here $g(\mu_i)$ is a function of mobilities (between the square brackets of Eq. (139)) and, as an estimate, we have used the expression for the boundary width $\delta$ from Eq. (45). We see that, for a given applied current, decreasing leading ion conductivity increases electric field overall and decrease interface width, thereby increasing sensitivity. Of course, sources of dispersion (e.g., unwanted pressure gradients) would result in larger interface widths than our simple diffusion-only model used here (and hence reduce sensitivity).

Next, in plateau mode, we can define *resolution*, $R$, by our ability to detect one analyte mobility versus another. To do so, we should have a sufficient signal difference, $\Delta s$, relative to the LE ion (whose mobility is precisely known). This signal difference can be compared to background signal noise $\sigma_s$ (c.f. **Figure 24**) as follows

$$R = \frac{\Delta s}{\sigma_s}. \tag{140}$$

As an example, for monovalent strong electrolyte sample initially loaded in TE, the signal difference $\Delta s$ scales as

$$\Delta s \sim c_L^L - c_S^S = c_L^L\left(1 - \frac{c_S^S}{c_L^L}\right) = c_L^L\left[1 - \left(\frac{\mu_S}{\mu_L}\right)\left(\frac{\mu_C - \mu_L}{\mu_C - \mu_S}\right)\right]. \tag{141}$$

From Eq. (141), we see an increased LE concentration results in stronger signal difference. This is because high $c_L^L$ yields large absolute values of signal throughout all ITP zones, and the resolution $R$ scales as the difference (in units of the signal) between these absolute values. The noise of plateau signals, $\sigma_s$, is determined by several factors including noise in the detector (e.g., due to amplification and/or readout noise) and spatiotemporal variations in background signal.

Overall, for plateau mode ITP, the analysis presented above suggests decreasing $c_L^L$ (thereby decreasing LE conductivity $\sigma^L$) to increase sensitivity $S$ while increasing $c_L^L$ for increased resolution $R$. A simple approach to improve sensitivity is by using a tapered channels wherein the detection region has a smaller channel cross-sectional area, as has been demonstrated by several studies.[121,251,268,295] This is because a smaller cross-sectional area (like decreasing conductivity) increases the electric field for the same applied current. Bahga[121] presents analytical expressions for plateau length as a function of local cross-sectional area in an ITP channel.

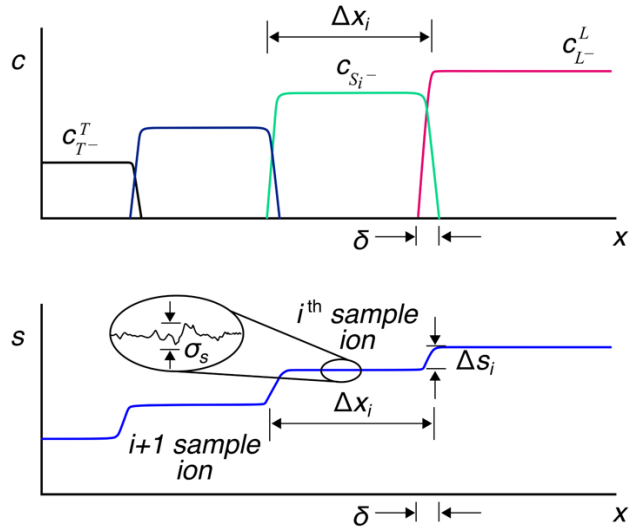

**Figure 24**. Schematic depicting typical (A) concentration versus $x$, and (B) generic detector signal $s$ versus $x$ (equivalently, time) for fully formed and separated plateau zones for an arbitrary number of analyte species $S_i$. Length of $i^{th}$ plateau zone is indicated by $\Delta x_i$ and the interface width between successive zones is indicated by $\delta$. $\sigma_s$ is a measure of the noise of the detector.

**10.2 Fluorescence detection**

Fluorescence-based detection is very attractive for on-chip microfluidic applications due to its ease of implementation, high sensitivity, and that it is a contactless approach. Microchannels typically have good optical access (e.g., channels made from glass or plastic) and sufficiently low background fluorescence (albeit with a short optical path length) for sensitive fluorescence detection. Fluorescence detection can be used to detect *directly* fluorescent analyte species or to detect *indirectly* analytes by using non-focusing tracers,[33] fluorescent mobility markers,[105] or fluorescent carrier ampholytes.[104] We will review these methods in more detail below. In principle, direct and indirect detection approaches are both applicable in detecting both peak and plateau mode analytes. However, in practice, direct detection is typically used for peak mode analyte detection, while indirect detection is used to identify separated analyte zones in plateau mode. Key components for fluorescence-based detection include a light source (e.g., LED, laser, mercury bulb), optical filters, microscope optics, and a detector (e.g., CCD and CMOS cameras, or pointwise detectors such as PMT and photodiode). Fluorescence detection systems are possible in various formats, including miniaturized and portable systems compatible with microfluidic systems.

*10.2.1 Direct detection*

In the direct detection approach, fluorescent molecules are directly focused in ITP and detected. The species of interest are either natively fluorescent (e.g., inherently fluorescent proteins and molecules), or fluorescently labeled (e.g., SYBR-based labeling of nucleic acids, fluorescently tagged proteins) prior to or during ITP. Some of the earliest demonstrations of direct fluorescence-based detection in ITP were in the works of Reijenga et al.[296], Kaniansky et al.[297], Schmitz et al.[298], and Jarofke[153]. Reijenga et al.[296] combined simultaneous fluorescence detection and UV absorption in ITP to directly detect quinone in soft drinks and bovine serum albumin. Kaniansky et al.[297] used ITP and fluorescence measurements to detect as low as 250 femtomoles of amino acids labelled with 2,4-dinitrophenyl. Likewise, Schmitz et al.[298] pre-stained lipoproteins with a fluorescent lipophilic dye called 7-nitrobenz-2-oxa-1,3-diazole (NBD)-ceramide and detected them using ITP from human serum samples. Jarofke[153] used ITP to detect up to 3 ng/mL of fluorescent histamines.

Fluorescence detection methods gained significant popularity in late l990s with the rapid growth in microfabrication techniques and significant improvements in the sensitivity of optical detectors, and it has been widely employed in ITP.[245,299–303] Direct detection can be used to visualize and/or track plateau zones as well as the interface region between zones (for e.g.., using a fluorescent molecule which focuses in peak mode between two consecutive plateau zones). This is useful, for example, for extraction of trace analytes such as nucleic acids from raw biological samples (see **Fig. 25C**),[32,79,93] or before interfacing with downstream zone electrophoresis (ZE) to improve resolution.[30] It can also be used for highly sensitive quantification of trace amounts of analytes and estimate the degree of preconcentration in ITP.[110,304] We discuss the physics around manipulation and detection of trace analytes in peak mode ITP in **Section 5** and mention applications and achievable LODs in **Section 11**. Direct detection of the ITP peak, especially using spatial detectors such as CCD and CMOS cameras, also provide access to detailed two- or three-dimensional information of the shape of the focused species. This has led to new fundamental physical understanding of the ITP processes and has been useful in developing methods to minimize sample dispersion for in-line quantification.[61,62,222,230] For example, see an optimized chip design and associated ITP visualization in **Fig. 25A** from the study of van Kooten et al.[222]. As another example, Khurana and Santiago[97] and Garcia-Schwarz et al.[34] (see **Fig. 25B**) quantified the scaling and relations among peak mode width, sample concentration, and current using fluorescence imaging of AF488 dye in peak mode ITP. In the same vein, Garcia-Schwarz et al.[61] and Bhattacharyya et al.[62] measured 2D depth-integrated peak shapes in ITP using fluorescent species to study mechanisms underlying sample dispersion in ITP. Direct fluorescence readouts have also been useful in the design of several ITP-enhanced chemical reaction assays which require monitoring, separation, and control of co-focused species.[39,93,111,203] ITP-aided reactions are reviewed in detail in Section 8. We also note here that for milli-scale ITP devices, fluorescence visualization may be performed directly without the need of a microscope (but of course, with the presence of an appropriate excitation light source and filters).[209,230]

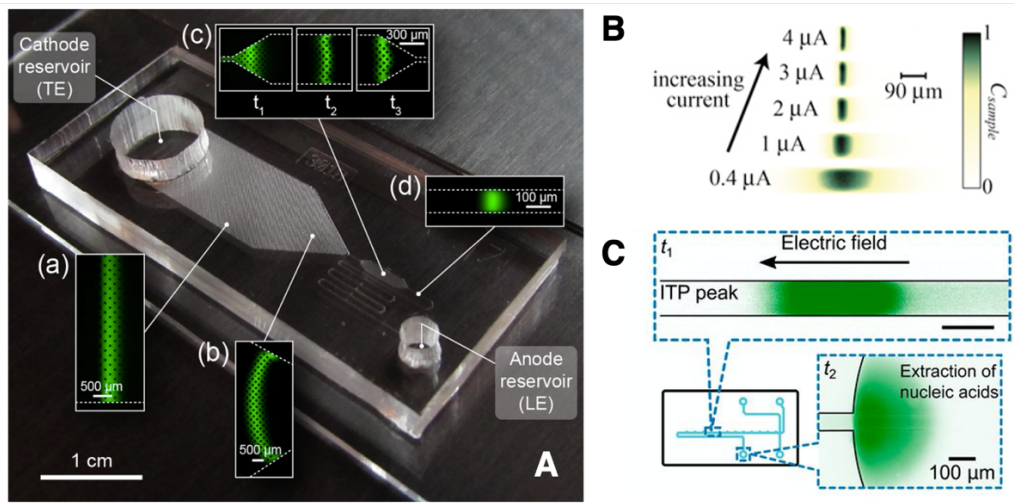

**Figure 25**. Direct detection and visualization of peak mode ITP in various microfluidic systems. (A) Fluorescence images of peak mode ITP focusing and preconcentration of dye SYTO9 at various axial positions in a custom large volume PDMS/glass microfluidic chip. Figure is reproduced from with permission from Ref.[222]. CC BY 4.0. (B) Peak mode ITP focusing of fluorescent dye AF488 versus varying applied currents between 0.4 and 4 microamps. Experiments were performed in a commercially available borosilicate glass microfluidic chip (NS12A, Caliper Life Sciences, MA). Figure is reproduced with permission from Ref.[34]. Copyright 2012 JoVE. (C) Visualization of ITP peak during extraction, purification, and preconcentration of nucleic acids from a raw nasopharyngeal swab sample. Nucleic acids were stained using the fluorescent intercalating dye SYBR Green I. Experiments were performed in an NS12A Caliper chip (same as (B)). Figure is reproduced with permission from Ref.[93]. Copyright 2020 National Academy of Sciences.

An obvious limitation of direct fluorescence detection in ITP is the availability of inherently fluorescent analytes or difficulties with labeling analytes. Only a limited number of species offer native fluorescence in convenient wavelengths or contain functional groups such as thiols, free amines, or hydroxyls which enable derivatization.[105] Simultaneous detection of multiple species (in the same zone) can also be hampered by methods to discriminate among excitation and emission wavelengths simultaneously. This can be addressed to a degree using multichannel (i.e., different wavelengths) detectors, but multiplexing is often limited to three to five different channels. Another consideration is the dye photostability and/or pH insensitivity of the fluorophore to enable concentration quantification and assay reproducibility (e.g., unlike fluorescein, AF488 has excellent photostability and its fluorescence is insensitive to pH for pH value above ~4)[97,305]. Interestingly, pH dependent dyes have also been used to indirectly detect and identify species in ITP.[1,306,307] Lastly, note that the lower limit of detection of all fluorescent detection methods is limited by the detector noise and background auto fluorescence of the chip and reagents.

It is worth noting that most microfluidic ITP studies involving trace analyte detection have almost exclusively used fluorescence-based detection. This is due its high sensitivity, ease of implementation, and compatibility with standard microfluidic setups (c.f. **Sections 5** and **11**). Several strategies can be combined and used to optimize the sensitivity of detection in peak mode ITP.[97,110] For example, the concentration of the focused analyte is proportional to the leading ion concentration, thus a higher LE concentration can enable increased peak concentration. Likewise, a low TE concentration locally increases the electric field in the reservoir/initial sample zone and thus favors higher sample influx rates. In addition, semi-infinite sample injection allows for continuous injection of analyte into the peak and can be more beneficial than a finite injection mode for trace analytes. In a similar vein, counterflow ITP and increases in channel length can also used to increase focusing time and the amount of sample focused in peak mode ITP. Reduction in channel cross-sectional area also allows for higher current density (by mitigating effects of Joule heating) and can also be used to improve sensitivity.[121,188,308]

### *10.2.2 Indirect detection*

In the indirect fluorescence detection (IFD) approach, non-fluorescent analytes (with minimal or no *a prior* knowledge of the physicochemical properties) are identified indirectly by adding non-reacting fluorescent tracer species.[33,105,296,309,310] In IFD approaches, the fluorescently labeled species are typically present in trace quantities in the LE and/or TE buffers, and so contribute negligibly to local conductivity. This is unlike indirect detection methods using UV absorption where the UV-sensitive species are often on the same order of the buffering ions and can contribute significantly to the local conductivity.[307,311–314]

We here can classify IFD methods into two broad classes. The first IFD approach, called the non-focusing tracer (NFT) technique,[33] utilizes fluorescent species tracers that does not focus in ITP (but electromigrate) and whose fluorescence intensity varies and adjusts spatially depending on the physicochemical properties of the separated zones. This is schematically depicted in **Figure 26**. As discussed in Chambers and Santiago,[33] detection using the NFT approach can be achieved using fluorescent (non-focusing) co-ions or counterions. The NFT ion can be either an *overspeeder* or an *underspeeder*. These can electromigrate in the direction of the ITP wave but do not focus because they have mobility magnitudes which are higher than the LE or lower than the TE ion, respectively. When a fluorescent counterion NFT species is used, the ion is called a *counterspeeder*. The counterspeeder migrates from the LE to the TE traversing all zones. Chambers and Santiago[33] also presented the theory around detection using the NFT technique. By conserving ion fluxes, they derived an expression for the concentration ratio of NFT ion $i$ in the in zone A (analyte $A$) versus zone B (analyte $B$) in terms of the effective mobilities of the species as

$$\frac{c_{i,A}}{c_{i,B}} = \frac{\mu_A^A}{\mu_B^B}\left(\frac{\mu_i^B - \mu_B^B}{\mu_i^A - \mu_A^A}\right) = \frac{I_A}{I_B}. \tag{142}$$

The second equality in Eq. (142) assumes that the fluorescence intensity is proportional to the concentration of the NFT species (which is valid in the absence of self-quenching[296,315]). The mobilities in the expression are effective mobilities. Despite the striking similarity of Eqs. (142) and (79) (the latter equation represents adjusted concentrations of trailing zone analytes), it is important to note that the concentrations in Eq. (142) are necessarily trace quantities, unlike the concentrations in Eq. (79) which are on the order of the buffering ions. Also, Eq. (142) holds true for both univalent and multivalent species. For a fully ionized, high-mobility counterspeeder, the intensity ratio approaches the effective mobility ratio, i.e., $I_A/I_B \approx \mu_A^A/\mu_B^B$. The sensitivity of counterion NFT is nearly uniform across all analyte mobilities. On the other hand, the most sensitive regime for co-ionic NFT is where $\mu_i^A \approx \mu_A^A$; i.e., where the mobility of the analyte is nearly equal to the NFT. The NFT-based approach, like thermometric detection, is a fairly universally applicable method for plateau mode ITP. Shown in **Fig. 26C** is an example from Chambers and Santiago[33] which shows NFT-based detection of HEPES and MOPS using the counterspeeder species rhodamine 6G (R6G). Garcia-Schwarz et al.[34] later demonstrated the separation and NFT-based indirect detection of two amino acids (arginine and lysine) in cationic ITP using R6G as an underspeeder. In a similar vein, Smejkal et al.[316] demonstrated separation and detection of eight carboxylic acids using R6G as the counterspeeder.

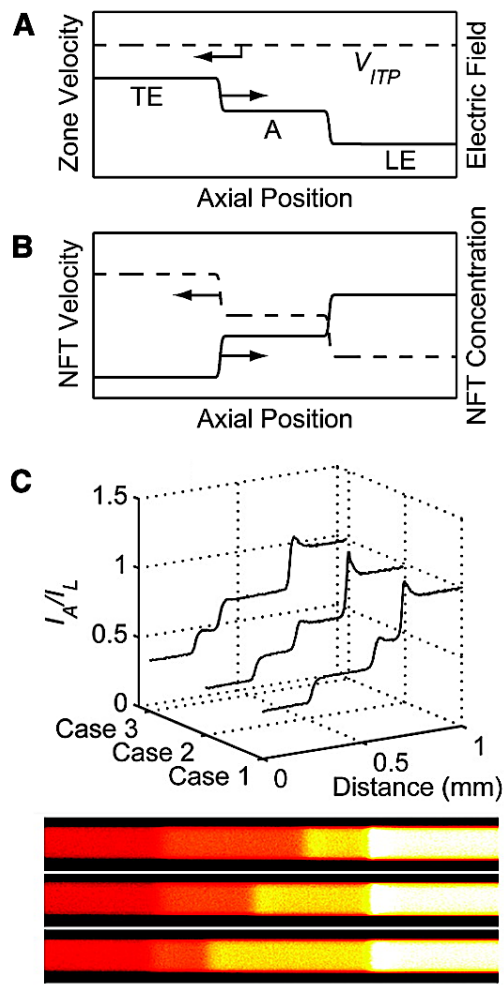

**Figure 26**. Qualitative representation of the indirect fluorescence detection method using a non-focusing tracer (NFT) ion. (A) Depiction of plateau mode ITP with a single analyte $A$ under steady state conditions. (B) Adjustment of the NFT concentration and velocity profiles within each zone. (C) Experimental detection of HEPES and MOPS in plateau mode ITP using a counterionic fluorescent NFT rhodamine 6G. Top panel shows the ratios of zone intensities (each zone intensity normalized by the LE value), and bottom panel shows corresponding experimental images. Figure is adapted with permission from Ref.[33]. Copyright 2009 American Chemical Society.

A major advantage of the NFT approach is that only a single fluorophore is required to detect multiple analytes. A drawback of the NFT technique is that self-quenching can become an issue at sufficiently high NFT concentrations (e.g., within the LE zone) and when there can be undesirable chemical interaction of the NFT with analytes (e.g., via complex formation).[317] Other challenges may include adsorption of the dye to the channel walls (e.g., R6G is cationic and can be adsorbed to negatively charged glass channel walls).[33]

A second approach to IFD uses of the use of fluorescent mobility marker species.[104,105,310,318,319] These are fluorescent spacer ions which focuses between plateaus (including LE and TE) and thereby help determine the effective mobility of non-fluorescent analytes in plateau zones (see also Refs. [89,318,320–322]). Khurana and Santiago[105] demonstrated this approach using a set of carefully selected fluorescent mobility markers (e.g., Oregon Green carboxylic acid, fluorescein, and Bodipy) for the indirect detection of amino acids and achieved detection sensitivity of ~10 μM. Their assay was limited by the possible number of fluorescent markers with appropriate mobilities required for separation and also had the disadvantage of requiring a careful a priori choice of markers depending for specific target analytes.

An analyst can use many fluorescent mobility markers to relax the requirement of a priori knowledge of analyte mobility and to improve both resolution and dynamic range of possibly detectable species. A good example of this is the fluorescent carrier ampholyte assay (FCA) of Bercovici et al.[104,310] and Kaigala et al.[319] Carrier ampholytes are amphoteric small molecules, typically synthetic polypeptides with varying isoelectric points, and can aid separations in ITP.[323–326] The latter mixtures of hundreds or thousands of carrier ampholytes are typically used to establish a background pH gradient in isoelectric focusing experiments.[327] In the FCA assay, this mixture was added to the ITP buffers and at least many of them accumulate between LE and TE. Analyte ions (which are not fluorescent) present in sufficiently high concentrations displace these labeled ampholytes and create "gaps" in an otherwise contiguous (albeit non-uniform) fluorescence signal. **Figure 27** shows the concept of the FCA assay. Integration of the fluorescent signal (e.g., from left to right) with and without analytes help identify the fraction of ampholytes focused to the left or right of an analyte "gap", and hence quantify the analyte effective mobility relative to the mobility distribution of the FCAs. The gap width quantifies the amount of analyte present (see Eq. (143)).[1,194] As with the NFT approach, this approach can be used only for analytes present in sufficient concentrations such that they are at least near plateau mode. Bercovici et al.[310] demonstrated this technique for the detection and quantification of four model analytes consisting of 20 μM MES, 40 μM MOPS, 30 μM ACES, and 50 μM HEPES; see **Figure 27C**. Bercovici et al.[104] also applied the technique to the label-free detection of 2-nitrophenol (2NP) and 2,4,6-trichlorophenol (TCP) toxins in tap water on a hand-hand portable ITP device (compatible with point-of-care), and achieved detection limits of ~ 1 μM.

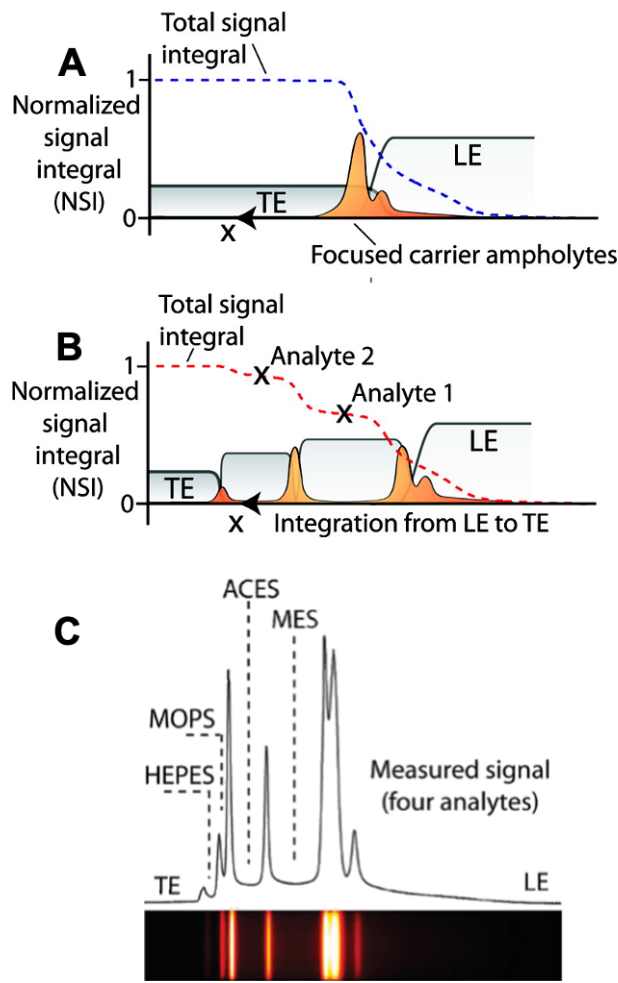

**Figure 27**. Schematic and experimental demonstration of indirect detection of ITP separation using the fluorescent carrier ampholyte (FCA) assay. (A) All the labeled CAs focus at the LE-TE interface in the absence of sample, resulting in a continuous fluorescence signal. (B) Analytes focused under ITP displace subsets of the labeled CAs, resulting in gaps in the measured fluorescence signal. Plateau regions in the normalized signal integral (NSI) indicate the presence of specific focused analytes. The NSI value can provide quantitative information about the effective mobility of the focused analytes. (C) Experimental demonstration of the FCA assay for the indirect detection of four analytes consisting of 20 μM MES, 40 μM MOPS, 30 μM ACES, and 50 μM HEPES. Shown is the heatmap of the raw signal (below) and the corresponding intensity profile (top). Figure is adapted with permission from Ref.[310]. Copyright 2010 American Chemical Society.

## 10.3 Thermometric detection

In ITP, the conductivity of zones typically decreases from the highest value in the LE to the lowest value in the adjusted TE. This variation in conductivity affects both the volumetric Joule heating rate and the associated temperature rise within each electromigrating zone. Temperature within each zone typically achieves a quasi-steady state equilibrium value which is determined by the balance of heat conduction fluxes (away from the channel axis and through the liquid and across channel walls) and volumetric Joule heating rate (see more details in **Section 8**).[15,114,162,227] For a given, local current density $j$, the local temperature rise $\Delta T$ within a zone scales inversely with the zone conductivity $\sigma$ as $\Delta T \propto j^2/\sigma$. Thus, the temperature increases (in roughly a stepwise fashion) from the LE to the TE zone (and across analyte zones), as shown by the experimental data in **Fig. 28**. This variation in temperature over zones is the basis for thermometric detection in ITP.

The temperature (and its gradients) can be measured using point-wise detectors such as thermocouples[1,17,18,157] and thermistors[2,328,329] which are typically bonded to the exterior channel walls, or by point-wise measurements of the differential absorbance of narrow-band diode lasers,[330] or by contactless thermal imaging, for example, using an infrared (IR) sensor or camera.[162,178] Though early works used thermistors immersed in liquids,[130] such an approach is not recommended since the sensor can itself disturb and interfere with the ITP process. Current thermometric detection approaches do not involve direct contact with the liquid in the channel. Such sensors rely on the assumption that a rapid thermal equilibrium is attained between the solution within the channel and the channel wall over time scales much smaller than the duration for the electromigrating zone to pass by the detector. We also note that there exist temperature-dependent fluorescent dyes (e.g., Rhodamine B) which could, in principle, be used as an approach for thermometric detection in ITP. This approach has been used for other microfluidic applications,[331] but to our knowledge remains unexplored for ITP.

The thermometric detection method is a universal approach[1] (i.e., applicable to all species) because the measured temperature signal is a function of the conductivity (and, consequently the constituent ion mobilities) of the separated zones, and not inherently species specific. For this reason, this method cannot be used for peak mode ITP since the trace sample ions in peak mode contribute negligibly to the local conductivity. Note further that this method (and in general, all universal detection methods including thermometric and conductimetric) cannot distinguish pure zones (consisting of single focused species) from so-called shared ITP zones (zones which contain a stable mixture of two or more species in a plateau).[193] We will limit our discussion here to the common case of detection of pure zones.

To well resolve zone boundaries using thermometric detection, the length scale of zones of varying temperature $l_T$ should be larger than the interface width $\delta$ between zones (i.e., a zone truly in plateau mode, see Section 7). $l_T$ should also be larger than the characteristic integration length (along the channel axis) of the temperature sensor. Sharp changes in the measured temperature can be used to quantitatively estimate the length of the plateau zones. This measured zone length can, in turn, be related to the species concentration in the plateau, and from these data one can infer the original analyte concentration. For finite sample injection width $l_{inj}$ of analyte $X$ which is focused into an ITP plateau zone of length $l_X$ (measured) with steady state concentration $c_X^X$ (measured, or predicted from model/simulation), the initial injected analyte concentration $c_{X,inj}$ can be obtained from species conservation as[1,194]

$$c_{X,inj} = \gamma \, c_X^X \frac{l_X}{l_{inj}}, \tag{143}$$

where $\gamma = S_{det}/S_{inj}$ is the area ratio between the detection and injection regions.[33,121] Note that this above relation is also applicable to fluorescence (c.f. **Section 10.2**) and conductivity-based (c.f. **Section 10.4**) detection methods.

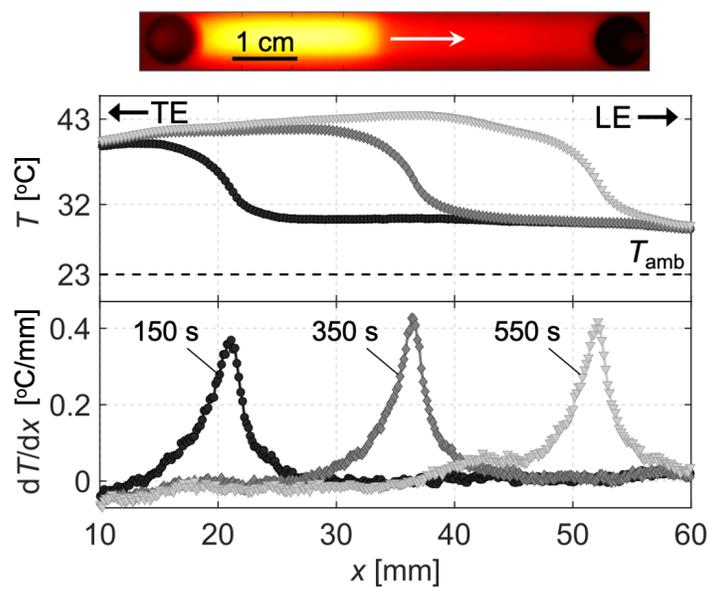

**Figure 28**. (Top) Heatmap shows experimentally measured spatial temperature profile during ITP. Measurements were obtained using contactless, infrared thermal imaging. (Middle) Spanwise-averaged temperature and (Bottom) temperature gradient versus channel length shown at three intermediate times during ITP. Figure is adapted with permission from Ref.[162]. Copyright 2020 Elsevier.

A limitation of pointwise temperature sensors such as thermocouples and thermistors is that they usually require more complex device fabrication and/or setup to achieve sufficient thermal contact to the wetted channel wall. Direct thermal imaging is much more versatile and provides spatiotemporal information within the entire field of view. One limitation of the latter is that IR cameras tend to have lower resolution than visible light cameras (one recent optical setup provided a resolution of 10 pixels per mm)[162].

Generally speaking, the spatial resolution of thermometric methods is poorer than methods based on UV absorbance or fluorescence.[1] Typical minimum detectable zone lengths using the thermometric detection are around 2 to 5 mm.[1,107,227,328] Resolution of concentration value (via accurate measure of plateau zone length) can be improved by increasing time of separation, which results in longer plateau lengths (c.f. **Section 6** and **10.1**). Lowering LE concentration can also increase resolution by proportionally decreasing the concentrations of trailing zones (this also increases Joule heating for the same applied current density) The latter approach is, in turn, constrained by the minimum desired buffering capacity of the system at low ionic strengths and other problems associated with significant Joule heating (c.f. **Section 8**). Another limitation of thermometric approaches is that the temperature rise can depend on the thermophysical properties of the channel wall. Thus, temperature detection methods require careful calibration for each different channel design and material if quantitative information is desired. Lastly, sensitive thermometric detection also requires that plateau zone conductivities be sufficiently different to result in measurable temperature differences The latter requirement may be a design tradeoff with the need to maintain sufficiently low temperatures. High absolute temperatures can modify or damage analyte species (e.g., in bioassay applications); and high temperatures and temperature gradients can result in poor reproducibility or even unstable flow fields. A potential way to mitigate temperature effects is to ensure good thermostatic temperature control along the entire chip, and particularly near the detection region.

## 10.4 Conductivity detection

As discussed in **Sections 2** to **4**, the conductivity of plateau zones in ITP varies in accordance with charge neutrality, species effective mobilities, and current continuity. Specifically, the conductivity typically decreases monotonically from the LE to the TE across zones, and the ratio of conductivity between two generic zones A (species *A*) and B (species *B*) can be expressed in terms of the effective mobility and concentration of the species as

$$\frac{\sigma^A}{\sigma^B} = \frac{\mu_C^A c_C^A - \mu_A^A c_A^A}{\mu_C^B c_C^B - \mu_A^B c_A^B}, \qquad (144)$$

where subscript $C$ refers to the counterion, and the concentrations are the "adjusted" values governed by the Jovin and Alberty relations (c.f. **Section 4**). This variation is the basis for conductivity-based zone detection methods for ITP.[332–336] Note the ratio in Eq. (144) also helps govern the temperature fields for thermometric detection (as discussed in **Section 10.3**). Conductivity detection is a fairly universal method, and can be used to detect ions that focus and segregate into plateau zones in ITP.[1] As do other techniques, conductivity methods provide information on zone length which is useful for quantification of amount (or initial concentration) of analytes, as given by Eq. (143) (in addition to direct measurements of zone conductivity). As with other methods, conductivity-based methods cannot be used to detect peak mode analytes since such ions contribute negligibly to local conductivity.

Early applications of this method measured directly the conductance of the solution using sensing electrodes in direct contact with the solution within the channel, and a high frequency alternating current/voltage operation to avoid Faradaic reactions.[1,168,337,338] This approach is currently disfavored because direct electrode contact with the solution can lead to undesirable bubble formation, polarization and unwanted passivation of the sensing electrodes, and signal interference from the high separation voltages in the main channel.[339,340] Moreover, the direct contact approach which was originally developed for capillaries with large inner dimensions, sensing electrodes were typically introduced in the channel through laser drilled holes and this required complicated fabrication procedures. Direct contact is also less convenient and incompatible with small channel dimensions typical of microfluidic applications. To address these limitations, a new class of contactless conductivity detection methods was developed for capillary ITP systems in 1980 by Gaš et al.[341] and later improved by Vacík et al.[342]. Gaš et al.[341] demonstrated good reproducibility and better resolving power in ITP separations using the contactless conductivity detection compared to thermometric detection. It is also worth noting that the studies of Gaš et al.[341] and Vacík et al.[342] were among the earliest reported demonstrations of the contactless conductivity technique across all electrophoresis separation methods. Contactless conductivity methods were then significantly improved in 1998 by Fracassi da Silva et al.[343] and Zemann et al.[103], mainly geared towards capillary electrophoresis applications. The early 2000s saw several studies which developed novel microfabrication techniques for integrating contactless conductivity on chip for ITP and several other on-chip electrophoresis applications.[250,344–350] A notable example is the study of Graß et al.[351] from 2001, which was the first to describe the fabrication and testing of a fully integrated contactless conductivity detector on a poly(methylmethacrylate) (PMMA) chip for ITP separations. **Fig. 29C** shows an example isotachopherogram for separation and detection of oxalate and acetate based on conductivity measurements from the latter study.

The contactless conductivity method, commonly now known as capacitively coupled contactless conductivity detection (C⁴D), typically uses high frequency alternating current measurements between two (or four) closely spaced electrodes placed directly on or around the external surface of the microchannel. Hence is no direct electrode material-to-solution contact. This configuration can be achieved by either direct microfabrication of electrodes on chip (see **Fig. 29A**) or by attaching an externally connected C⁴D cell in close proximity to the channel within a chip.[103,343] **Fig. 29B** shows a simple equivalent circuit model for C⁴D measurement cell consisting of two electrodes (viz., excitation and sensing) separated longitudinally along the channel.[347] The main circuit components include the solution resistance $R_s$, wall capacitance $C_w$ (since the wall acts as a dielectric layer in series), and a stray capacitance $C_0$ due to the direct capacitive coupling between the electrodes. See Refs.[103,335] for more detailed models which include effects due to the electric double layer capacitance and resistance. We refer to Refs.[352–357] for some excellent reviews on recent developments in instrumentation, fabrication, and design of contactless conductimetric detection methods for on-chip electrophoresis applications.

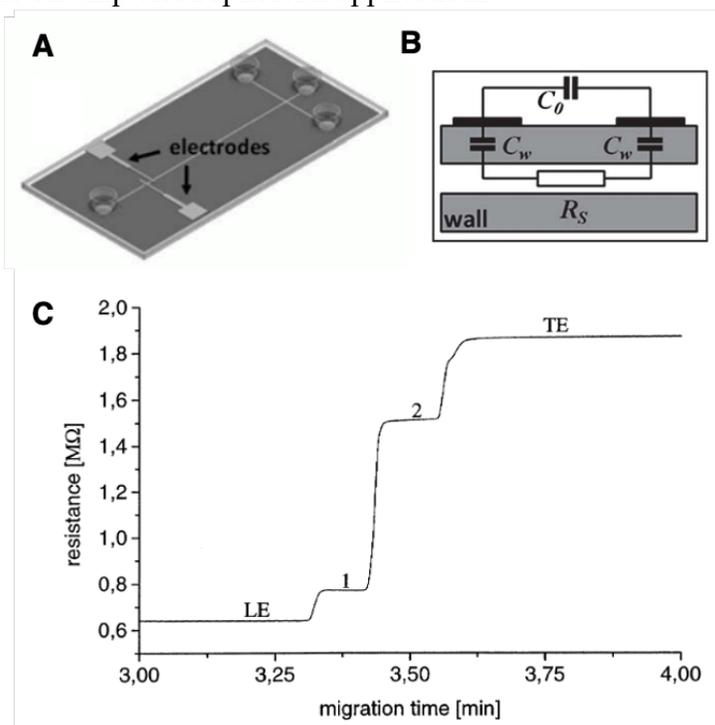

**Figure 29**. Depiction of capacitively coupled contactless conductivity detection (C⁴D) detection for ITP separations. (A) Integration of planar electrodes for C⁴D on a microchip. Figure adapted with permission from Ref.[345]. Copyright 2008 John Wiley and Sons. (B) Equivalent circuit model for a typical C⁴D microcell, consisting of solution resistance $R_s$, wall capacitance $C_w$, and a stray capacitance $C_0$. Figure is reproduced with permission from Ref.[347]. Copyright 2012 The Royal Society of Chemistry. (C) Detection of ITP separation of oxalate and acetate in a custom PMMA microfluidic chip using C⁴D. Figure is reproduced with permission from Ref.[351]. Copyright 2001 Elsevier.

C⁴D modules are now available commercially and are commonly used as detectors for on-chip microfluidic ITP applications. For example, Cong et al.[358] and Bottenus et al.[359] used commercial C$^4$D units to monitor ITP separation of lanthanides on custom-built PMMA and fused-silica microfluidic devices, respectively. As another example, Koczka et al.[183] combined UV detection and C$^4$D and performed ITP analyses on a variety of samples such as red wine, fizzy drink, and juice. The study of Koczka et al.[183] also importantly highlighted a major limitation associated with the resolution power of C$^4$D systems. The resolving power of C$^4$D systems is limited by the separation distance between the sensing electrodes (which are typically around 1 to 2 mm). For example, a separation distance of 1 mm requires that the ITP plateau zone width be greater than about 1 mm to achieve good resolution. Another major consideration for C$^4$D is the requirement of a thin channel wall (on the side where the electrodes are placed), typically limited to a few 100 microns. A thinner wall enables better capacitive coupling with the solution in the channel, and results in higher sensitivity. Typical sensitivity values using C$^4$D are in the sub-micromolar range.[183] It is also important to note that conductivity methods are currently limited to pointwise measurements in ITP, i.e., signal versus time can be measured only at a single or at most few spatial locations along the channel. This is in contrast to fluorescence imaging or thermal imaging methods which provide detailed spatiotemporal information.

### 10.5 Other methods

We here briefly mention several other methods including uncommonly used methods applicable to on-chip ITP detection. A few of these methods have special historical significance as early approaches to ITP detection.

UV-based detection was widely used in early studies of ITP (both off-chip and on-chip). This method utilizes at least one species focused in (near plateau mode) ITP which exhibits significant UV absorbance.[1,102,360] For example, a UV-detectable counterion can be used to detect unlabeled species in plateau mode ITP as it traverses from the LE to TE and adjusts its concentration (and associated UV absorbance) across various zones.[296,361] Another implementation involves the use of UV-detectable spacer ions which form intermediate zones between non-UV-detectable (UV-detectable) analyte zones, thus enabling indirect (direct) analyte detection.[321,322] Of course, a straight-forward application for UV-based detection involves direct detection of analytes which significantly absorb light in the UV wavelengths.[129,362] Importantly, UV methods require concentrations of the UV-detectable species to be on the same order as that of the buffer (unlike fluorescence methods which are applicable down to trace concentrations). Also, UV-absorbance methods are often not very sensitive for microfluidic applications due to the small optical path lengths in the microchannels. Lastly, like the UV-based method, chromophores[183,363,364] and radioactive[365–367] species can also be used for detection in ITP (of course, using the appropriate detectors), although such methods usually have relatively low sensitivity.

Another class of detection methods are based on electrochemical measurements (amperometric,[368,369] potentiometric,[370–373] and potential gradient[374–376]). Recall that the electric field and conductivity vary across ITP zones and, respectively, increases and decreases monotonically from the LE to the TE. This effect forms the basis for electrochemical detection methods in ITP. These methods involve current/voltage measurements made locally in-line using direct contact of source and sensing electrodes with the solution. In such methods, it is important that detection-associated measurements do not significantly affect the driving current/voltage supplied through the main channel for ITP separation. As mentioned in **Section 10.4**, fabricating such systems with direct electrode contact with the solution can involve complicated procedures

(rarely compatible with microfluidics) and can often lead to undesirable effects at the electrode surface (e.g., bubble formation, Faradaic reactions, passivation).

Lastly, we note that it is generally challenging to visualize or detect non-fluorescent trace analytes in peak mode ITP. In this regard, Karsenty et al. developed a method based on current monitoring within a channel with multiple constrictions to detect the peak mode sample location. In a different approach, Eid et al.[377] used a fluorescent tracking dye (AF647) compatible with their downstream assay to detect the ITP peak which contained nucleic acids. Interestingly, non-fluorescent dyes such as tartrazine and cresol red can also been used in visualization of ITP experiments.[364]

## 11. Applications of microfluidic ITP

This section summarizes various references that employ microfluidic ITP for practical applications. Applications reviewed here include using ITP for analyte purification, preconcentration, separation, reactions, and sensitive detection. We will place a special emphasis on bioassay systems involving ITP. The discussions are not meant to be comprehensive but to present examples aimed at indicating the variety of systems which have been explored.

### 11.1 Purifying and concentrating trace analytes for sensitive detection

ITP is a powerful purification and preconcentration tool which can be used for a variety of on-chip applications. We will here mostly focus on peak mode ITP applications which involve direct detection of trace analytes. ITP can also be used as a preconcentration step prior to separation using zone electrophoresis to improve the latter's resolution and sensitivity; a topic we discussed in more detail in **Section 10.1**. Some of the early demonstrations of peak mode (called by some early researchers the spike mode) were performed in the late 1970s in capillary ITP systems in the work of Svoboda and Vacík[360] and Everaerts et al.[378], where peaks in UV isotachopherograms were used for direct quantitative evaluation of trace analytes involving organic weak acids. Picomolar levels of trace analytes were detected using this approach.

The years between late 1990s and early 2000s saw the first implementations of ITP in microfluidic chips.[25–30] In several of these studies, ITP was primarily used as a preconcentration method in tandem with downstream separation using zone electrophoresis. Nevertheless, we briefly review below some of the ITP aspects of these studies since these studies were among the earliest to demonstrate and quantify the degree of preconcentration achievable using peak mode ITP. Wainright et al.[30] reported a ~500-fold increase in the concentration of their fluorescently labeled eTAG reporter molecules in peak mode transient ITP performed in a PMMA microchannel. Their limit of detection was in the sub-picomolar range of reporter molecules. Similar fold increases were reported by Jeong et al.[29] for ITP experiments involving fluorescein as sample performed in a PDMS chip in around 2 min. In 2006, Jung et al.[110] reported order million-fold preconcentration of AF488 sample in a few minutes using peak mode ITP performed on a commercially available glass microchannel. They detected initial sample concentrations as low as 100 fM. Later, Jung et al.[304] implemented high-numerical aperture light collection and a PMT to improve sensitivity and reported detection of as small as 100 aM initial concentration of AF488. Sample preconcentration was also, more recently, demonstrated in a nitrocellulose paper-based microfluidic device by Moghadam et al.[254] The latter group reported 900-fold increase in AF488 sample concentration in peak mode ITP in ~2 min (see **Fig. 30**). **Fig. 30** also shows that the amount of sample focused in peak mode ITP in their device increases linearly with time (see data for $x/L_i < 0.6$) for constant current operation (as discussed in **Section 5.2**). Deviation from linearity for $x/L_i > 0.6$ was attributed to dispersion due to non-uniform EOF.

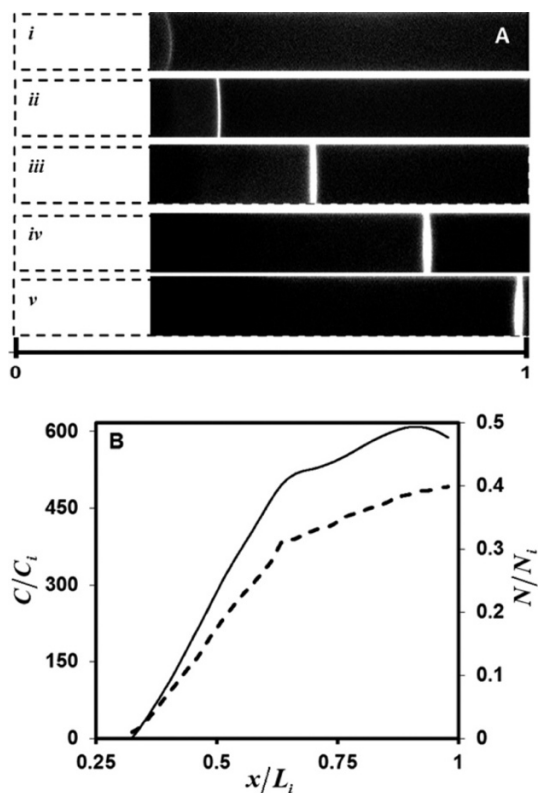

**Figure 30**. (A) Experimental images of ITP preconcentration and focusing of AF 488 sample in peak mode ITP implemented on a nitrocellulose-based paper device. Snapshots taken at (i) 1, (ii) 30, (iii) 60, (iv) 90, and (v) 140 s after applying the electric field. (B) Sample concentration normalized by the initial sample concentration ($C/C_i$), and accumulated moles of sample normalized by the initial moles of sample in the TE reservoir ($N/N_i$) versus normalized axial length. Refer to Moghadam et al. for details on experimental setup and device design. Figure is reproduced with permission from Ref.[254]. Copyright 2014 American Chemical Society.

Peak mode ITP has also been used for other microfluidic applications and assays involving biomolecules such as nucleic acids and proteins. In such applications, ITP is typically used to simultaneously purify and preconcentrate target molecules from background contaminants and ions present in the raw biological samples, and followed by either direct on-chip detection or for downstream analysis (e.g., PCR). For example, Persat et al.[32] purified and preconcentrated nucleic acids from whole blood using ITP and demonstrated compatibility of the extracted volume with downstream PCR. Bercovici et al.[39] reported order 1,000 to 10,000-fold preconcentration of nucleic acids in ITP. Similarly, Bottenus et al.[188] and Bottenus et al.[251] demonstrated order 10,000-fold preconcentration of proteins in ITP prior to on-chip detection. Refer to Rogacs et al.[31] for a comprehensive review applications of microfluidic ITP for nucleic acid purification. We review such bioassay ITP systems in more detail in **Section 11.3**. Preconcentration provided by peak mode ITP can also be applied to initiate and accelerate chemical reactions involving molecules that focus in ITP; a topic we discuss in **Section 11.4**. Lastly, refer to **Sections 10.2.1**, **5,** and **Supplementary Information S2** for discussions around methods to increase trace analyte focusing rates in ITP and improve sensitivity.

**11.2 Applications of ITP for separations**

Applications of ITP in microfluidics for separations can broadly be categorized in two types as follows. The first type segregates (and therefore separates) co-ionic species into plateaus and these are detected while the species are still in plateau mode ITP. The second type is to use ITP as an initial preconcentration process and then end/disrupt ITP to effect a capillary electrophoresis type separation. In this section, we shall concentrate on the first type. We summarize the second type in **Section 12**.

Separation and detection while still in ITP mode requires that all analytes of interest necessarily form plateau zones. These zones can be detected either directly (e.g., UV, fluorescence) or indirectly (e.g., conductivity, fluorescent markers such as NFT, temperature); c.f. **Section 10**. Plateau mode ITP was among the earliest forms of ITP separations, starting at least as far back as the 1970s, and it has been extensively used in capillary systems.[1,168,239] This approach has also been used implemented in microfluidic ITP separation processes. For example, Cong et al.[358] and later Bottenus et al.[359] (c.f. **Figure 31**) demonstrated separation of lanthanides in microfluidic devices. Similarly, Cui et al.[36] separated and detected three fluorescent proteins directly using ITP. Indirect detection of plateau mode analytes can also be achieved by using focusing spacer ions with known mobilities (which are used as "mobility markers").[104,105,321,322] Such spacer ions can either be trace or abundant ions, which respectively focus in either peak or plateau mode ITP between the target analytes of interest. In the latter approach, "gaps" in signal between the separated markers correspond to the existence of a plateau mode analyte with the appropriate mobility range. Such indirect methods and applications are reviewed in more detail in **Section 10**. Alternately, peak mode ions can also be separated and detected using background spacer ions (sometimes, also in combination with a sieving matrix). Such approaches have been used for the separation of unreacted reactants from products in ITP-aided reactions[94,200] (c.f. Section 11.4) and for size-based selection of target analytes (e.g., RNA).[209]

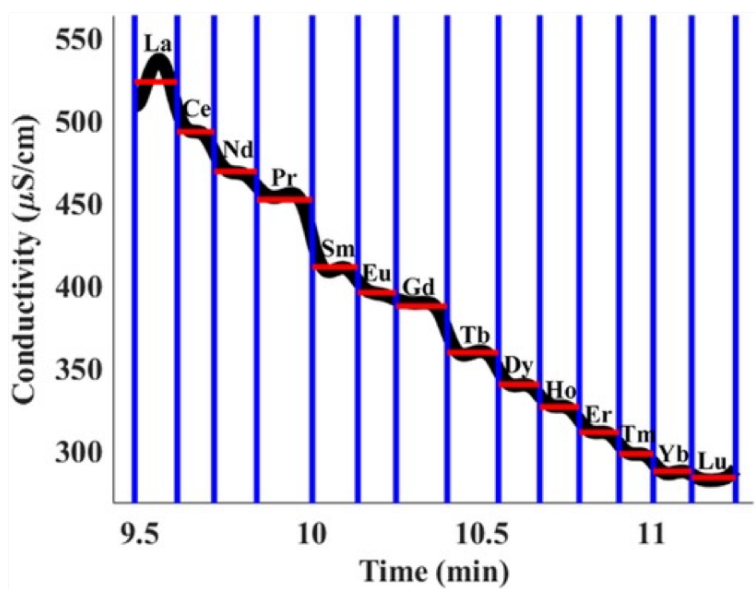

**Figure 31**. Experimental demonstration of the separation of trivalent lanthanides using plateau mode ITP. Shown are measured conductivity signals versus time. The distance between the vertical lines represents the characteristic temporal band length. The mean conductivity signals for each lanthanide are indicated by horizontal lines. Figure is reproduced with permission from Ref.[359]. Copyright 2019 John Wiley and Sons.

## 11.3 Bioassay systems leveraging ITP

### 11.3.1 DNA and RNA purification and on-chip assays

ITP has been extensively used for the development of several DNA- and RNA-based assays over the last decade.[32,78,79,93,111,202,207,209,379] ITP NA assays are currently the most common applications of microfluidic ITP in biology and biotechnology. Microfluidic ITP applications for NA analyses can be broadly categorized as either those which use ITP for NA purification and preconcentration (i.e., sample prep) prior to downstream analyses (e.g., PCR), or those that combine ITP-based preconcentration and/or separation followed by direct on-chip target NA detection. NAs are highly soluble, have a high electrophoretic mobility, and a relatively low pKa of ~3 and so are ideally suited for manipulation and focusing using anionic ITP. These ITP systems are typically buffered near pH 7 to 9 to ensure that background contaminants (e.g., proteins, downstream inhibitors) from raw biological samples such as blood, urine, and plasma do not focus and interfere with NA-based ITP assays. In all NA-based ITP applications, NAs are typically detected in peak mode ITP (e.g., using fluorescence) since they are usually present in trace quantities (e.g., order nM or lower concentration) compared to the buffering ions (e.g., order mM concentration). Further, the free solution electrophoretic mobilities of DNA and RNA are nearly independent of length or sequence.[380] Hence, all DNA and RNA typically focus into a single, concentrated peak in ITP. For fluorescence detection, nucleic acids are typically either prelabeled synthetically (e.g., using fluorescently tagged oligos) prior to the assay or are labeled on-chip using intercalating fluorescent dyes (e.g., SYBR). We refer to Rogacs et al.[31] for a comprehensive review on the design, development, and applications of ITP systems for nucleic acid purification and sample prep. We also refer to three review articles[38,65,381] which discuss in detail ITP applications for NA-based assays.

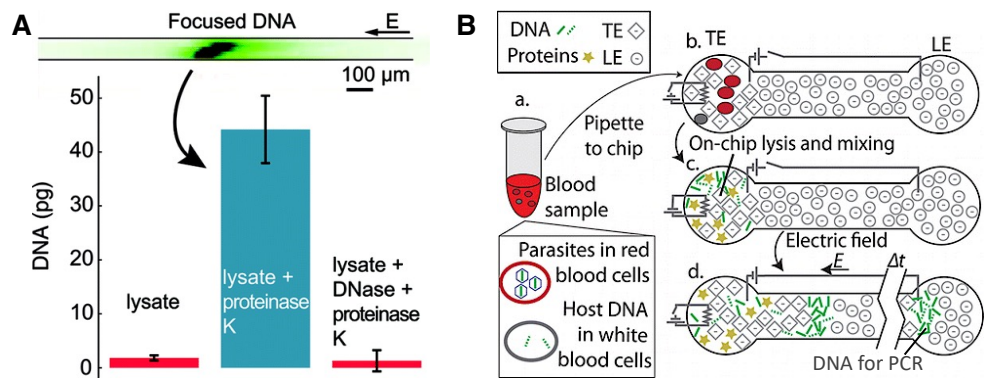

**Figure 32**. (A) Purification of genomic DNA from whole blood using ITP. Shown are experimental visualization of the ITP peak containing label DNA (top), and quantification of the amount of focused DNA (bottom). Figure is reproduced with permission from Ref.[32]. Copyright 2009 American Chemical Society. (B) Schematic of an ITP assay for cell lysis and nucleic acid extraction implemented on a printed circuit board-based device. Purified DNA is extracted into the LE reservoir and is used for off-chip PCR analysis for malaria detection. Figure is reproduced with permission from Ref.[382]. Copyright 2012 American Chemical Society.

The studies of Kondratova et al.[383] and Kondratova et al.[79] in 2005 were the earliest to report extraction of DNA using ITP from biological samples such as plasma and serum. They demonstrated their method using fairly standard laboratory equipment (not microfluidics) and agarose gels and as a downstream step following a series of purification steps. The study of Schoch et al.[109] published in 2009 was likely the first to demonstrate extraction of nucleic acids using ITP in microfluidic channels. They used ITP and a sieving matrix to selectively extract and preconcentrate small RNAs from cell lysate (refer also to Han et al.[209] for another recent example of simultaneous RNA extraction and size selection using ITP). Subsequently, Persat et al.[32] demonstrated ITP extraction of DNA from whole blood (**Fig. 32A**) and showed the compatibility of the extract with PCR. Thereafter, several ITP systems were developed which combined ITP extraction of nucleic acids with downstream nucleic acid amplification. For example, Marshall et al.[379] used ITP to extract DNA from malaria-containing erythrocytes mixtures and then detected and quantified the extracted pathogenic DNA using off-chip PCR. A similar assay was later integrated on a printed circuit board device (**Fig. 32B**).[382] As another example, Borysiak et al.[384] used ITP to extract DNA and integrated it with on-chip LAMP amplification to detect *E. Coli* from whole milk. In a similar vein, Eid and Santiago[377] combined ITP extraction and downstream recombinase polymerase amplification (RPA) to detect inactivated *Listeria monocytogenes* from whole blood.

Similar methods have more recently been implemented on paper-based devices.[236,257,385] A notable example is the work of Bender et al.[236] who used a paper-based device for simultaneous ITP extraction of DNA and RPA-based target amplification and demonstrated detection of HIV-1 DNA from whole blood within clinically relevant viral abundances. As another example, Sullivan et al.[385] recently developed a paper-based microfluidic ITP device for nucleic acid extraction from whole blood. Notably, their device integrates whole blood fractionation, plasma protein digestion, and ITP extraction of nucleic acids on a single device. Such portable paper-based devices are likely compatible with point-of-care applications. Research efforts toward rapid, point-of-care ITP applications are currently being pursued with the main goals of reducing intermediate manual steps, on-chip integration of sample lysis and pre-extraction protocols, improving sensitivity and assay robustness, and minimizing bulky instrumentation.

These studies have demonstrated ITP as an attractive method for NA purification because of its selectivity to NAs without the use of membrane or toxic chemicals typical of conventional approaches. Besides, ITP provides high yields of NA regardless of input sample amount, and does not incur losses for small size NAs unlike conventional solid phase or liquid phase extractions methods.[31] Nucleic acid extraction efficiencies of up to 80% have been demonstrated in microfluidic ITP for DNA sample input in the range of 250 pg - 250 ng.[230] A similar high efficiency using ITP has been also demonstrated for input sample amounts as small as total RNA from a single cell.[386] ITP-based NA purification has also been shown to be superior (by over two orders of magnitude) in extraction efficiency compared to commercially available kits for small size nucleic acids (e.g., <60 nt).[387]

We note that most ITP-based NA assays described above require sample lysis (e.g., using Proteinase K, heat, surfactant) prior to loading on chip. It is important that the lysis procedure be compatible with ITP. We refer to Ref.[31] for a detailed discussion on this topic including different procedures recommended for RNA vs DNA extraction. Another limitation of ITP is the common requirement to dilute the raw sample (typically between 1:10 and 1:100 either in TE or LE) prior to loading on-chip. This is typically required because raw biological samples typically contain overly high salt concentration (~order 100 mM strong salts). The ITP processing of such mixtures can lead to strong variations in pH or challenges associated with separation capacity.

Most of the studies mentioned above used ITP as a sample preparation process prior to downstream target NA detection. However, ITP systems have also been developed to perform direct on-chip target NA detection. A common strategy used in such application involve ITP-accelerated hybridization reactions between probe and target NA, which is typically followed by on-chip separation of excess, unreacted species from reacted molecules (e.g., in homogenous reaction assays; c.f. **Section 11.4**) to improve sensitivity.[94,207] We refer to the discussion in **Section 11.4** on ITP-based reactions for a detailed review of such ITP systems (see also the review of Eid and Santiago[38]).

Another mode of application of direct on-chip detection and quantification of NAs in ITP has been by combining ITP with downstream separations including various forms of on-chip capillary electrophoresis,[24,30] for example, for the analysis of DNA fragments such as PCR-amplicons. For example, Liu et al.[388] used transient ITP to preconcentrate PCR amplicons and coupled capillary gel electrophoresis to separate, quantify, and analyze DNA fragments between ~70 to ~1400 bp. Similarly, Bahga et al.[211] used bidirectional ITP to first preconcentrate a 1 kbp dsDNA ladder and then separate fragments using bidirectional ITP-induced zone electrophoresis all in a single straight channel in solution. Few other recent examples of direct on-chip target NA detection are the works of Bender et al.[236] mentioned earlier (simultaneous ITP and RPA) and the ITP-CRISPR approach of Ramachandran et al.[93] (see **Section 11.4** for more details).

*11.3.2 Protein-based assays*

Like nucleic acids, proteins can also be focused, preconcentrated, and separated in ITP.[66,204,373,389] There have, however, been generally a fewer number of applications involving on-chip (microfluidic) ITP for proteins analyses compared to NA-based systems. This is likely due to the challenges associated with working with proteins including their sometimes-limited solubility and the fact that they exhibit a wide range of pH- and buffer-dependent electrophoretic mobilities due to varying isoelectric points (pI) and size. Such variability makes it difficult to develop generally applicable ITP assay designs and chemistries. Despite these challenges, protein-based systems have been developed in both cationic[77,177,256] and anionic[36,93,200,390] ITP formats, depending on the charge of target proteins. Application of ITP involving proteins can also be broadly categorized as either those which extract, preconcentrate and/or separate target proteins (e.g., for direct on-chip detection and quantification), or, as systems which use ITP to aid chemical reactions involving proteins (e.g., for immunoassays). We will here only briefly review a few such applications as illustrations. We refer to a recent article by Han and Khnouf[66] for a detailed review focused on ITP-based protein systems and applications.

Cui et al.[36] used anionic ITP to focus, preconcentrate, and separate three fluorescent proteins, namely, green fluorescent protein (GFP), Allophycocyanin, and *r*-phycoerythrin, into distinct zones. Jacroux et al.[77] used a series of two cationic ITP systems to separate fluorescently tagged cardiac troponin I (CTnI) from two other proteins, *r*-phycoerythrin and albumin, with similar isoelectric points. This study highlighted that a judicious choice of buffer and separation conditions are required for protein separations in ITP. Other systems for the detection of CTnI were later developed as presented in Refs.[76,188,256]. Another interesting example is the work of Qu et al.[177] who developed an ITP system to extract proteins and nucleic acids simultaneously from raw biological samples using bidirectional ITP (c.f. **Section 12**). In this system, proteins (e.g., albumin) focused in cationic ITP while nucleic acids focused in anionic ITP. This method is shown in **Fig. 33**. Briefly, raw sample is diluted in a mixture containing TE+ and TE- ions and is sandwiched between two zones (located on opposite sides) where one of the zones contains LE+ and the other LE- (**Fig. 33**). Here + and – indicate cationic and anionic components. On application of an electric field, the cationic and anionic ITP shock waves originate from the same TE/sample reservoir and migrate away in opposite directions. We note that this implementation of bidirectional ITP of Qu et al.[177] is very different from the common case of bidirectional ITP where the cationic and anionic ITP shocks interact[212] (c.f. **Section 12**).

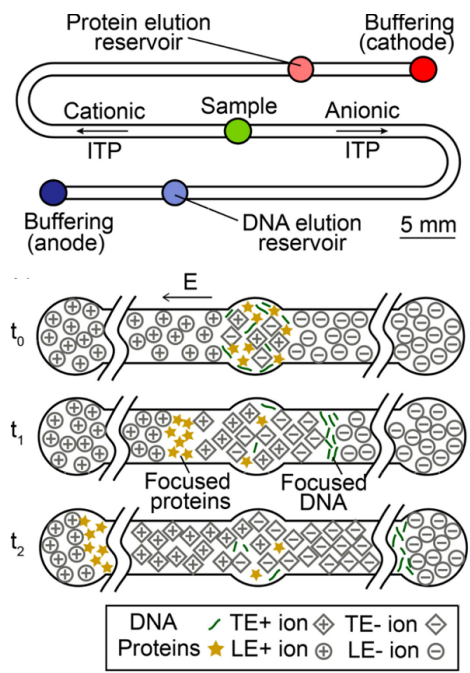

**Figure 33**. Schematic of an ITP chip (top) and assay (bottom) for simultaneous extraction of proteins and nucleic acids from raw biological samples using simultaneous, non-interacting cationic and anionic ITP, respectively. The cationic and anionic ITP shock waves start from a common sample reservoir, and they respectively electromigrate on opposite directions to the cathode and anode, respectively. Figure is reproduced with permission from Ref.[177]. Copyright 2014 American Chemical Society.

Proteins analyses have also been performed by combining microfluidic transient ITP with zone/gel electrophoresis and mass spectrometry. Such systems have been developed, for example, for the analyses of functional proteins in infant milk[391], urinary proteins,[392] and several other trace protein analyses.[393,394] Immunoassays that use simultaneous ITP preconcentration and reaction acceleration have also been developed in various formats including bead-based assay[202,389], paper-based lateral flow devices[255,256], chips with surface functionalization of antibody[389], and devices with integrated biosensors which contain binding aptamers[390] (c.f. **Fig. 34A**). We review in detail several such homogeneous and heterogeneous reaction-based ITP assays for proteins in **Section 11.4**. Lastly, we note that several techniques can be used to "modify" the mobilities of proteins to enable their focusing in ITP. Examples include complexing the protein of interest with specialized aptamers[200] (c.f. **Fig. 34B**), nucleic acids[93], or antibodies[256], prior to ITP. Co-focusing of proteins and nucleic acids in ITP can also aid NA-based assays (for e.g., RPA and CRISPR)[93,236] which rely on enzymatic action of proteins on DNA and RNA.

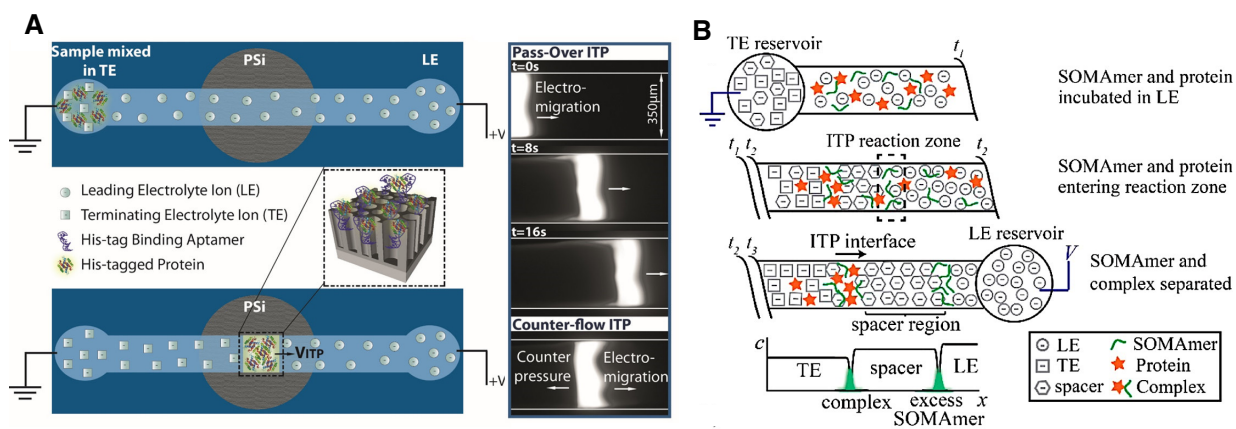

**Figure 34**. (A) Schematic illustration of the porous silicon biosensor for ITP-based protein detection. Target proteins are focused in ITP and they bind to the aptamers immobilized on a PSiO$_2$ biosensor. Shown on the right are raw fluorescence images of target protein focusing in peak mode ITP under pass-over and counterflow operational modes. Figure is reproduced with permission from Ref.[390]. Copyright 2017 American Chemical Society. (B) Schematic of the ITP–spacer assay of Eid et al.[200]. Initially at $t_1$, SOMAmer and target protein are mixed with the LE buffer, and spacer ions are mixed with the TE buffer. Then, at $t_2$, SOMAmers and target proteins bind to form low-mobility complexes which electromigrate but are oversped by spacer ions. At $t_3$, unreacted SOMAmer molecules focus between the LE and spacer, whereas SOMAmer–target complexes focus between the spacer and TE. Figure is reproduced with permission from Ref.[200]. Copyright 2015 American Chemical Society.

*11.3.3 Single-cell analyses*

      Single cell analyses have gained significant interest over the last few years due since they help understand heterogeneity in cell populations and uncover gene level interactions between subcellular components. To this end, various microfluidic ITP systems have been developed to study RNA and DNA at the level of single cells, including subcellular components. In 2014, Shintaku et al.[63] first developed a microfluidic ITP system that enables selective lysing of single living cells, followed by ITP extraction and quantification of cytoplasmic RNA and its physical separation from the DNA-containing nucleus. In this method, first, a brief bi-phasic electric pulse was used to selective lyse the outer cell membrane while keeping the nuclear membrane intact. ITP was then used to selectively focus cytoplasmic RNA (sans the intact nucleus). They used on-chip fluorescence measurements to quantify the heterogeneity in the amounts of DNA and RNA among single mouse lymphocyte cells from the same. Soon thereafter, Kuriyama et al.[386,395] extended the method of Shintaku et al.[63] and developed an integrated microfluidic ITP system for the fractionation and recovery of the nucleus and cytoplasmic RNA from single cells (**Fig. 35**). Electric fields (for lysis and ITP) were used to automatically extract cytoplasmic RNA and nucleus

from the same cell at two different channel reservoirs. They used downstream RT-qPCR and qPCR to quantify cytoplasmic RNA and genomic DNA, respectively. Abdelmoez et al.[396] later found that the distinct focusing kinetics exhibited during ITP-based NA extraction from single cells can be used to distinguish soluble RNA molecules from mitochondrial RNA.

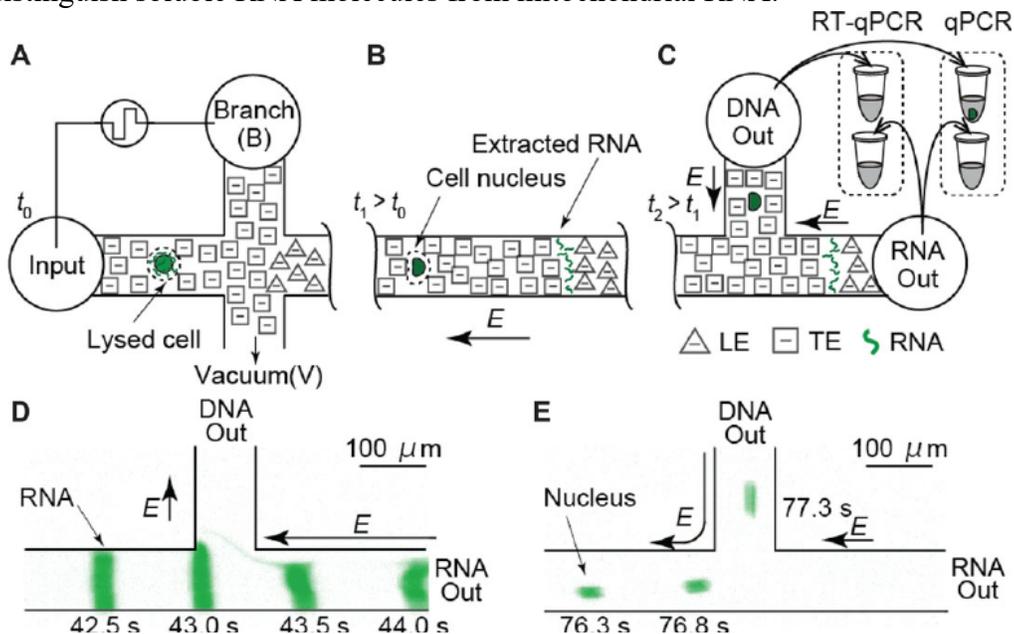

**Figure 35**. Schematic illustration of the single-cell microfluidic assay of Kuriyama et al.[386]. (A) A short b-phasic electric pulse is used to selectively lyse the outer membrane of the cell. (B) ITP is used to focus and preconcentrate nucleic acids from the cytoplasm while the nucleus does not focus in ITP (but electromigrates in the same direction as the ITP peak). (C) Automated electric field control is used to collect the cytoplasmic nucleic acids and nucleus into separate reservoirs. (D) and (E) show experimental visualizations of the fractionation process of (C). Figure is reproduced with permission from Ref.[386]. Copyright 2015 John Wiley and Sons.

In 2018, Abdelmoez et al.[397] significantly expanded the aforementioned ITP-based single cell approaches and studied correlations of gene expression between nuclear and cytoplasmic RNA (using off chip RNA sequencing). Their chip included an automated protocol to capture single cells using an integrated hydrodynamic trap. This study was also the first to integrate ITP-based systems with downstream genomic sequencing technologies. The ITP-integrated sequencing approach presented in Abdelmoez et al.[397] was termed SINC-Seq.[397] A very recent work by Oguchi et al.[398] integrated SINC-Seq with nanopore sequencing and studied mRNA isoform diversity within the nucleus and cytoplasm. As more applications continue to emerge, it is important to keep in mind that current ITP systems for single cell analyses are limited in throughput (typically limited to one cell per chip/channel) compared to several other high throughput single cell systems, for example, those based on droplets.[399,400]

**11.4 Applications of ITP for accelerating chemical reactions**

*11.4.1 Applications of homogeneous reactions using ITP*

In 2008, Kawabata et al.[401] first demonstrated the use of on-chip ITP to control chemical reactions. They used ITP to effect an immunoassay reaction to measure $\alpha$-fetoprotein (AFP), a liver cancer marker in blood. ITP was used to focus and preconcentrate a DNA-coupled anti-AFP antibody prior to its reaction with AFP (which did not focus in ITP) in an adjacent zone on chip. The coupling of DNA with the antibody increased the mobility of the DNA-antibody complex and this enabled it to focus in ITP. Kawabata et al.[401] used on-chip CGE (capillary gel electrophoresis) to separate reaction products and quantify AFP using LIF. They demonstrated a 140-fold increase in signal due to the ITP format and a detection sensitivity of 5 pM of AFP in 136 s. Kawabata et al.[401] called their assay the Electrokinetic Analyte Transport Assay (EATA) and it involved a homogeneous reaction where only one of the two reactants focused and preconcentrated in ITP. Such reactions were later modeled by Eid et al.[200]. Later, Park et al.[402] presented a method to improve the sensitivity and reproducibility of the EATA assay, and Kagebayashi et al.[403] developed a fully automated microfluidic platform that integrated the EATA assay.

Nucleic acids have approximately size-independent and high magnitude mobilities and are thus attractive for ITP applications. Thus, most of the existing applications of ITP to accelerate biochemical reactions have also involved some form of nucleic acid hybridization reactions including both DNA and RNA. Such reactions involve the reaction between single stranded nucleic acids and their complementary nucleic acid strand. Goet et al.[404] first demonstrated that ITP can be utilized to control and bring together two samples and focus them in ITP to initiate chemical reactions. They suggested and qualitatively showed this approach for DNA hybridization reactions. Persat and Santiago[111] first applied ITP to initiate, control, and perform hybridization reactions between nucleic acids which are co-focused in ITP. In their study, Persat and Santiago[111] showed experiments quantifying hybridization reactions between microRNA (miRNA) and molecular beacons (functionalized with a fluorophore-quencher pair), both co-focused in peak-mode ITP (**Fig. 36**). Using this approach, they demonstrated miRNA detection with 10 pM sensitivity and showed that the ITP reaction was also specific. Subsequently, Bercovici et al.[405] presented an assay for the detection of bacterial urinary tract infections using ITP-enhanced, homogeneous hybridization reactions between bacterial 16S rRNA and complementary molecular beacons. This study integrated sample purification from urine samples and demonstrated a sensitivity of 100 pM of pathogenic target RNA. Later, Bercovici et al.[39] developed a comprehensive model for ITP-enhanced second order homogenous reactions (c.f. **Section 7.1**) and experimentally demonstrated that ITP can accelerate nucleic acid hybridization reactions by ~14,000 fold. The latter was the first study to quantitatively demonstrate the degree of reaction acceleration attributable specifically to the effects of ITP.

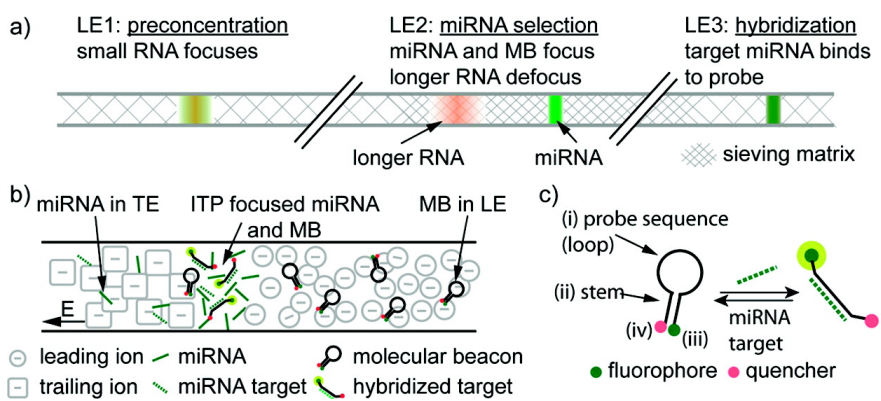

**Figure 36**. Schematic depiction the ITP hybridization assay of Persat et al.[111]. (A) Three-stage ITP strategy used for purification and hybridization. LE1 allows for strong preconcentration of small RNA. LE2 contains a higher polymer concentration to selectively focus miRNA. LE3 allows for specific hybridization. (b) Schematic of molecular beacon (MB)-based hybridization reaction in ITP. MBs are loaded in LE and miRNA in TE. They co-focus, preconcentrate, and react in ITP. (C) Reaction of target miRNA with probe MBs results in sequence-specific increase in fluorescence of the ITP peak. Figure is reproduced with permission from Ref.[111]. Copyright 2011 American Chemical Society.

The aforementioned studies of nucleic acid hybridization utilized molecular beacon reporters to detect a fluorescence signal upon reaction. In their native, unreacted state, these reporters have a hairpin structure with quenched fluorescence and this structure disrupted upon hybridization, increasing fluorescence. An important limitation of such strategies is due to finite fluorescence signal of the quenched state which limits the dynamic range of the signal increase (typically less than two orders of magnitude).

A few studies have addressed the aforementioned limitation of quenched reporters by effecting physical separations of reacted vs. unreacted species after ITP focusing. Bahga et al.[406] integrated ITP-enhanced DNA hybridization and capillary zone electrophoresis in solution using a novel bidirectional ITP scheme. In their approach, anionic ITP was used to speed up the DNA hybridization reaction of target with a complementary molecular beacon. After sufficient reaction time, the anionic ITP front interacted with a counter migrating cationic ITP front which disrupted the ITP peak and initiated zone electrophoresis which separated the hybrid products from unreacted beacons. They demonstrated 5 pM detection sensitivity and assay time of 3 min, and showed the approach can be used to detect multiple-length ssDNA targets. A limitation of the bidirectional ITP approach is the complexity of buffers design to effect efficient focusing and separation. See Bahga et al.[211] for detailed principles of bidirectional ITP and procedures to choose appropriate buffers. Later, Garcia-Schwarz and Santiago[206,207] developed an alternate strategy wherein the unreacted molecular beacons were captured (and removed from a moving ITP zone) after reaction using photopatterned, DNA-functionalized hydrogels. In their multi-stage assay, Garcia-Schwarz and Santiago[206,207] used ITP to first enable rapid reaction between miRNA and molecular beacons in solution, and this was followed by an on-chip separation stage. In the separation stage, the ITP peak migrated into a region of hydrogel functionalized with probe molecules to capture unreacted beacons. Note that latter separation stage used a heterogeneous reaction involving immobilized capture probes in the hydrogel, and we will discuss such heterogeneous reactions in the next subsection. The ITP peak which migrated beyond the functionalized hydrogel region consisted only of the hybrid product (i.e., devoid of unreacted molecular beacons) and was used for detection. Their assay achieved ~1 pM detection limit and exhibited 4 orders of magnitude dynamic range in detection. Eid et al.[94] developed another multi-stage strategy to separate reaction products from unreacted molecules on chip by using ionic spacers and a sieving matrix (**Fig. 37**). In their assay, the spacer ion had a mobility in between the LE and TE and formed a plateau between the LE and TE zones in ITP. The reactants focused in peak-mode ITP in solution between the spacer ion and LE zone during the first stage of their assay. After reaction, the ITP peak (and spacer zone) migrated into a region with a sieving matrix (Eid et al.[94] used 1.8% hydroxyethyl cellulose, HEC; polyvinylpyrrolidone, although PVP is another choice[111]) wherein the spacer ion migrated faster than the reaction products but slower than the unreacted probes, thus effecting separation. After sufficient separation time (~40 s), the reaction products and unreacted species were refocused among the two ITP interfaces, which was followed by detection (**Fig. 37**). Using this approach, Eid et al.[94] obtained a 220 fM ssDNA target detection sensitivity in 10 min and ~3.5 orders of magnitude dynamic range. The sieving matrix approach of Eid et al.[94] required a well-tailored choice of sieving matrix and spacer ions depending on the application, and this can limit the its applicability.

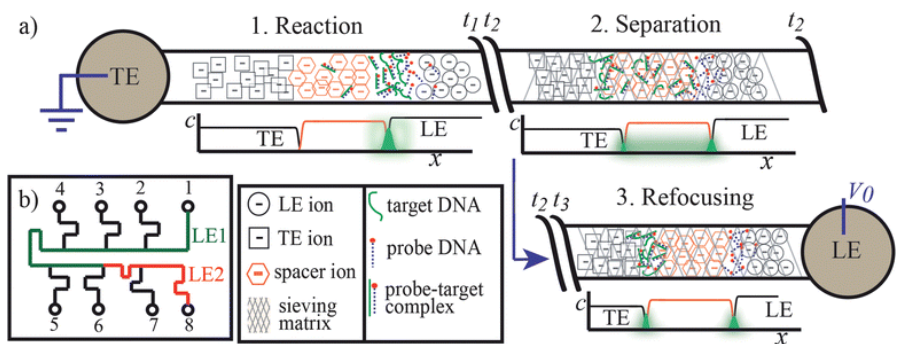

**Figure 37**. Schematic of the ITP-spacer assay of Eid et al.[94]. Their assay included three on-chip stages: (1) reaction between probe and target DNA in peak mode ITP in free-solution. In this step, spacer ions electromigrate slower than DNA. (2) The ITP peak electromigrates into a 1.8% HEC sieving matrix region where the spacer molecules overspeed the target molecules and probe–target hybrids. (3) The reaction products are fully separated from the unreacted probe and the molecules are refocused among the two ITP interfaces. Figure is reproduced with permission from Ref.[94]. Copyright 2013 The Royal Society of Chemistry.

A more recent, single-step approach to eliminating free probes in ITP-aided reaction assays involved the use of synthetic nucleic acid analogues with weakly charged backbones such as peptide nucleic acid (PNA) or Morpholino as reaction probes. Assays which rely on PNA and Morpholino probes make use of the significantly low electrophoretic mobility of probes in solution compared to the higher mobility of the probe-target hybrid. Typically, in such assays, a significant excess of probes is mixed with target ssDNA in the TE reservoir to enable rapid target-probe binding prior to application of electric field. Furthermore, a TE ion is chosen so as to focus only probes which hybridize to the target focus. Ostromohov et al.[407] utilized PNA probes and ITP-based "focus upon hybridization" approach and demonstrated ssDNA detection sensitivity of 100 fM with a dynamic range of 5 orders of magnitude. See Ostromohov et al.[407] for a model and theory associated with the approach. Morpholino probes are more soluble than PNA,[408] and this reduces adsorbance to surfaces and enables design of longer probe sequences (for example, Ostromohov et al.[407] used 14 nt PNA probes whereas Rosenfeld and Bercovici[98] used 25 nt Morpholino probes). Zeidman-Kalman et al.[201] demonstrated a similar assay based on Morpholino probes and used it to develop a method to measure the dissociation kinetics of hybridization. Later, Rosenfeld and Bercovici[98] implemented an assay based on Morpholino probes in a microfluidic paper-based analytical device (μPAD) to detect ssDNA with a sensitivity of 5 pM in 10 min. The "focus upon hybridization" approach has advantages over excess probe elimination using gel separation or sieving matrix as it is a single-step assay and uses simple buffer chemistry.

Unlike DNA and RNA, proteins are large molecules with low mobilities and highly variable ionic charge in solution (due to a wide range of isoelectric points). Electrophoretic mobility depends on local pH, ionic strength, and temperature. For this reason, designing ITP-aided reactions involving protein focusing is typically more difficult than for nucleic acids. Eid et al.[200] presented an ITP assay where they used modified aptamers called SOMAmers (slow off-rate modified aptamer) to bind to typically non-focusing C-reactive protein (CRP) and enable its focusing and preconcentration in ITP. In this assay, only SOMAmers focused in ITP, while native CRP did not focus in ITP but remained as a background ion in LE. As the focused ITP peak swept past the LE, CRP reacted with the preconcentrated SOMAmers and was recruited into the ITP peak. Similar to Eid et al.[94], Eid et al.[200] used a multi-stage assay with an ionic spacer and sieving matrix to separate unreacted SOMAmers from the SOMAmer-protein complex prior to detection to improve sensitivity. More recently, Ramachandran et al.[93] demonstrated ITP to control and enhance a reaction involving clustered regularly interspaced short palindromic repeats (CRISPR)-associated enzyme, Cas12 (**Fig. 38**). They showed that the complex comprised of the enzyme and target-specific RNA (guide RNA, gRNA) co-focused in ITP along with free nucleic acids. In their assay, the CRISPR-gRNA first recognized the target nucleic acid via sequence complementarity, and then the CRISPR enzyme became activated and cleaved reporter molecules which released fluorescence. The co-focusing of the CRISPR-gRNA complex together with target and reporter nucleic acids accelerated enzymatic activity and enabled them to develop a rapid, 35-min assay to detect the RNA of SARS-CoV-2 (the virus that causes COVID-19). The ITP-controlled, CRISPR enzyme reaction process in this assay was completed in 5 min. Unlike the assay of Eid et al.[200], the CRISPR assay of Ramachandran et al.[93] involved ITP preconcentration and focusing of all reactants including protein enzymes and nucleic acids in ITP.

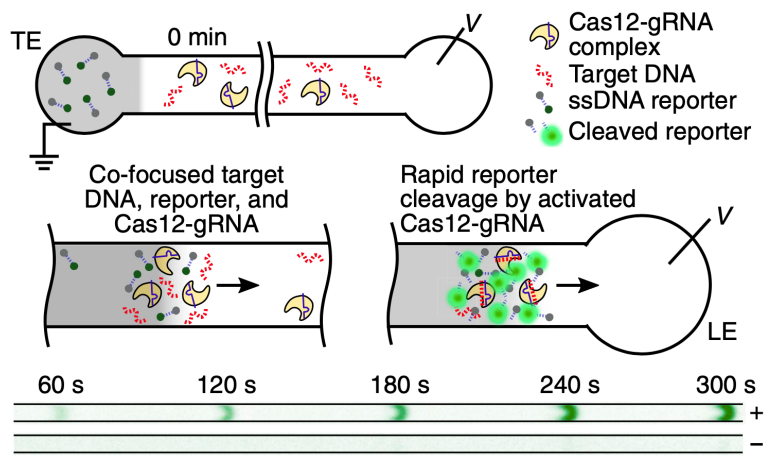

**Figure 38**. ITP-CRISPR enzymatic assay of Ramachandran et al.[93] for nucleic acid detection. ITP co-focuses CRISPR-Cas12-gRNA complex along with the target and reporter nucleic acids. Recognition of the target DNA by the enzyme complex activates the enzyme. Upon activation, the enzyme cleaves ssDNA reporter molecules, which results in an increased fluorescence signal of the ITP peak. Scale bar: 0.5 mm. Figure is reproduced with permission from Ref.[93]. Copyright 2020 National Academy of Sciences.

*11.4.2 Applications of heterogeneous reactions using ITP*

In 2014, Karsenty et al.[202] first demonstrated control and enhancement of surface reactions using ITP. They studied nucleic acid hybridization aided by ITP wherein the target molecules focused in ITP and the probes were immobilized on a surface of the microchannel. The reactive surface was formed within a microchannel by fabricating a trench on one of the channel walls and prefilling the trench with probe-functionalized paramagnetic beads and immobilizing them. The nucleic acid probes were biotinylated and bound to streptavidin coated magnetic beads prior to loading on chip. As the ITP peak passed over the reaction surfaces, fluorescently labeled target molecules from the ITP peak reacted with surface probes and the probe-target hybrids were immobilized on the beads, enabling detection. A limitation of this simple pass over approach (c.f. Section 7.2) of Karsenty et al.[202] was the short incubation time of target molecules in ITP with surface probes for reaction; a value limited by finite residence time of the ITP zone. Despite this, Karsenty et al.[202] demonstrated ~100-fold improvements in signal and LOD for a 3 min assay compared to standard flow-through assay.

Also in 2014, Han et al.[203] demonstrated that ITP can be used to improve the speed and sensitivity of traditional microarray-based assays involving nucleic acid hybridization (**Fig. 39**). Notably, Han et al.[203] was the first study to demonstrate multiplexed DNA hybridization reactions using ITP. The study involved reactions between 20 probe and target pairs across 60 microarray spots, with the probes immobilized on the spots and the targets focused in ITP. They developed an operation scheme where a high electric field was applied to rapidly focus nucleic acids, and this was followed by turning off the electric field to allow the ITP peak to diffuse and homogenize within a narrow channel cross section. ITP focusing and electromigration was then reinitiated but at a slower migration velocity over the microarray spots for efficient hybridization. They demonstrated a 30-fold decrease in assay time compared to conventional 15 h microarray assays and simultaneously improved sensitivity by an order of magnitude without sacrificing specificity.

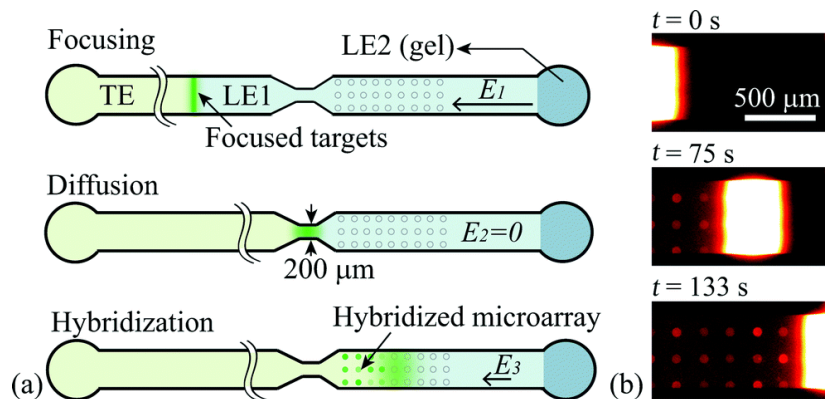

**Figure 39**. Schematic of the multiplexed-microarray heterogeneous reaction system of Han et al.[203]. (A) ITP is used to focus target DNA in peak mode ITP. When the ITP peak reaches the constriction, the electric field is turned off briefly to allow the ITP peak to homogenize prior to downstream hybridization. A low electric field is used for the latter hybridization step to maintain a homogenous ITP peak across the channel width. (B) Experimental images showing increase in fluorescence of the microarray spots when the immobilized probes hybridize with fluorescent DNA target molecules. Figure is reproduced with permission from Ref.[203]. Copyright 2014 The Royal Society of Chemistry.

In 2014, Khnouf et al.[389] demonstrated a proof-of-concept method for ITP-based enhancement of sensitivity of immunoassays involving heterogeneous reactions. They presented two methods involving the reaction of a target model protein (BSA) with surface immobilized antibody probes. In the first method, they used antibody-coated magnetic beads initially loaded in the LE buffer and then immobilized on chip using an external magnet to create a reactive surface, prior to ITP. In the second approach, they directly bound capture antibodies on the surface a PMMA microchannel using chemical treatment. In both methods, electrokinetic injection was used to introduce a fixed amount of the target protein. The target was focused and preconcentrated in ITP in solution and then reacted with the probes when the ITP zone passed over the reactive surface region. The ITP-based immunoassay achieved a 100-fold preconcentration of the protein and an LOD in the picomolar range.

In 2017, Paratore et al.[204] significantly expanded on the surface-reaction method of Karsenty et al.[202] and designed an ITP-based surface immunoassay and demonstrated detection of enhanced green fluorescent protein. Paratore et al.[204] used paramagnetic streptavidin beads which were functionalized with biotinylated capture antibodies off-chip (prior to ITP) and were then immobilized within a trench in the microchannel. Notably, this study explored the use of counterflow ITP (CF mode; c.f. Section 8.2.1) and the stop-and-diffuse mode (SD mode; c.f. Section 8.2.1) to enhance reaction efficiency. The SD mode is advantageous for low diffusivity target molecules and high viscosity of the medium. The SD mode provided a 10-fold better LOD of 300 fM compared to the simple PO mode (LOD of 3 pM), and is a useful approach when the location of the ITP peak is tractable in real-time (e.g., using current/voltage monitoring or optical methods). The CF mode performed better than the SD mode and lowered the LOD to 220 fM, but it requires more complicated experimental setup involving accurate flow control. For this reason, CF mode is recommended only for applications requiring extremely low levels of detection. Under optimal assay conditions, Paratore et al.[204] demonstrated a 1300-fold improvement in LOD compared to a standard immunoassay in 6 min. The study further noted the sensitivity of the ITP-based immunoassay is strongly influenced by the pH and salt concentration, factors which require careful optimization depending on the protein target.

We next review applications involving an ITP-focused reactant and an immobilized reactant in a gel or porous column. Garcia-Schwarz et al.[206] first demonstrated volumetric heterogeneous reactions in ITP. In their multi-stage assay for miRNA detection, Garcia-Schwarz et al.[206] first performed homogeneous ITP reaction between target miRNA and (excess) complementary molecular beacons. In the next stage of their assay, excess, unreacted beacons were removed from ITP prior to detection by using an in-line capture region which consisted of a hydrogel functionalized with capture probes complementary to the reporters. This approach decreased background signal from molecular beacons in ITP and improved assay sensitivity. Later, Garcia-Schwarz et al.[207] adopted this approach and developed a highly specific ITP assay with single-nucleotide specificity for miRNA detection. In particular, they demonstrated enrichment and specific detection of let-7a miRNA from a mixture of eight miRNAs of the let-7 miRNA family which differed by single to few nucleotides. Their 15 min assay was 1000-fold higher in sensitivity compared to Northern Blotting and 10-fold faster compared to RT-PCR.

Another application of ITP to enhance volumetric heterogeneous reaction is the work of Shkolnikov and Santiago.[205,409] They developed a method which coupled in line ITP and affinity chromatography (AC), and applied it to the detection and separation of sequence-specific nucleic acids. The affinity substrate was poly(glycidyl methacrylate-co-ethylene dimethacrylate) porous polymer monolith (GMA-EDMA PPM), and it allowed for easy immobilization of affinity probes. Moreover, the affinity column had negligible nonspecific binding and, unlike gels, the affinity column was non-sieving. Shkolnikov and Santiago[205,409] used ITP to first preconcentrate all nucleic acids from the sample in free solution. The ITP peak then electromigrated to the affinity-based column region to capture target DNA molecules. They demonstrated capture of target DNA from 10,000-fold higher abundant background DNA, starting from 200 µL of sample in under 1 mm column length and <10 min. They also developed and experimentally validated a model to predict the spatiotemporal behavior of the coupled ITP and AC process, including figures of merit such as capture efficiency and capture length.

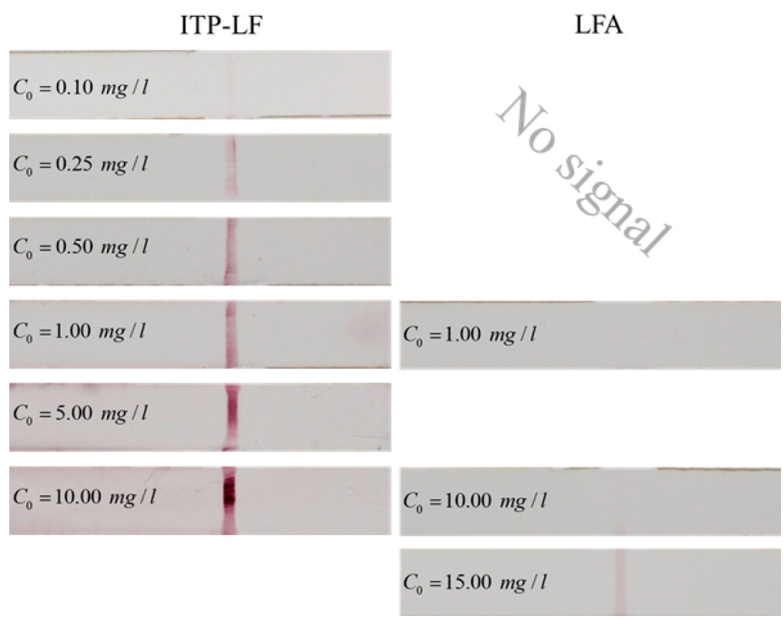

**Figure 40**. Improvement in the sensitivity of lateral flow assays using ITP. Qualitative estimation of the LOD of ITP-enhanced versus conventional lateral flow assay (LFA). Here, goat anti-mouse IgG labeled with 40 nm colloidal gold was used as the target, and the assay time was 5 min. Qualitatively, no signal was observed in the conventional LFA for target concentrations below 10 mg/L, whereas, the ITP-enhanced assay detected as low as 0.1 mg/L target. Figure is reproduced with permission from Ref. [255]. Copyright 2015 American Chemical Society.

ITP has also been implemented on paper-based microfluidic devices to improve the sensitivity of lateral flow assays (LFA). Moghadam et al.[254] demonstrated ITP sample extraction on a nitrocellulose-based paper microfluidic device and showed ITP can preconcentrate sample by 900 fold and achieve about 60% sample recovery efficiency. Later, Moghadam et al.[255] integrated ITP extraction in a later flow immunoassay format and demonstrated both fluorescent (using AF488 dye) and colorimetric (using gold nanoparticles) readouts. IgG secondary antibodies were used as the target and capture reagents. ITP was used to focus target analyte into a thin band and transport them via electromigration to the LFA capture line. The preconcentration of the targe prior to the reaction with the capture probes on the LFA capture line dramatically increased reaction rate associated with the slow reaction kinetics of surface binding in LFA (**Fig. 40**). They demonstrate a 400-fold improvement in the LOD for a 90 s assay time compare to conventional LFA. They estimate the gain factor in LOD in terms of the fraction of bound probe in ITP-LFA versus conventional LFA formats scales as $pk_{off}t_{ITP}$. Here, $p$ is the effective focusing ratio, $k_{off}$ is the reaction off-rate, and $t_{ITP}$ is the ITP residence time. Importantly, they find the LOD gain factor due to ITP is independent of the dissociation constant, $K_D$, and the gain can be tuned by controlling applied current which, in turn, governs the focusing ratio and ITP residence time. In a similar approach, Guo et al.[410] devised a novel approach to alter the electrophoretic mobility of proteins by immunobinding the target proteins to primary and secondary antibodies, and showed BSA (negatively charged protein) focused in cationic ITP in a fiberglass paper-based microfluidic device. Guo et al.[256] expanded on the work of Guo et al.[410] and developed an ITP-LFA assay for the simultaneous detection of two cardiac makers (cTnT and cTnI) from human serum. Similar to Moghadam et al.,[255] Guo et al.[256] utilized ITP (specifically cationic ITP) to preconcentrate target cardiac protein markers prior to the LFA capture line, and enhance the reaction of target with monoclonal antibody (mAb) probes on the LFA capture line.

*11.4.3 ITP reactions involving molecules and particles or cells*

We here discuss ITP reaction assays which involve particles such as cells and functionalized micron-size beads which either focus in peak-mode ITP or are involved in reaction with focused small molecules. For small particles, particle number density multiplied by the number of reactive sites per particle can be treated as an effective volumetric concentration affected by ITP focusing. For sufficiently small particle-to-particle distances, we can ignore diffusion to and from particle surfaces and treat the particle-bound reactant in a manner very similar to a dissolved species (this results in a pseudo homogenous ITP reaction). If the particle diameter is on the order of or significantly larger than the axial length of ITP zone, we can treat the particle as a finite immobilized surface.

Goet et al.[224] first demonstrated that micron size beads can focus in ITP and presented an analysis of focusing conditions. Later, Shintaku et al.[225] developed a multiplexed, bead-based assay to speed up nucleic acid hybridization reactions (**Fig. 41A**). In their work, Shintaku et al.[225] used commercially available ~6.5 µm magnetic beads (Luminex, Austin, TX) functionalized with 10 different ssDNA probe sequences complementary to various targets in the sample. Upon application of electric field, both beads and target DNA focused and preconcentrated in peak-mode ITP, enabling acceleration of reaction kinetics. They applied moderate-to-high voltages to induce instabilities in the ITP peak in order to promote mixing of the beads with the target. The assay was operated in a regime where the probes (via focused beads) were in excess and the rate of product formation was limited by the influx of target into the ITP peak. They demonstrated a 60-fold reduction in reaction time due to bead-based ITP reaction to achieve the same level of sensitivity and specificity as a well-stirred standard reaction.

Schwartz et al.[247] first demonstrated that ITP can be used for in-line labeling of bacterial cells (**Fig. 41C**). They used a stationary cationic ITP peak (electromigration countered by pressure driven flow) to preconcentrate and focus fluorescently labeled antimicrobial peptides (AMP). Bacterial cells were continuously flowed through the LE, but they did not focus in ITP. As the cells flowed through the ITP peak, they rapidly reacted with the AMPs and were labeled prior to inline downstream quantification. In this case, to achieve efficient labeling, the reaction time scale must be short compared to the advection time scale of the cells across the ITP peak. This provides a physical limit on the maximum electric field and throughput of the system. The work of Schwartz et al.[247] used AMPs which nonspecifically bound to the negatively charged outer membrane of bacteria, and thus provided the ability to selectivity detect bacteria. However, the method was not able to distinguish between different bacteria. To overcome this limitation, Phung et al.[411] developed a two-stage ITP assay for in line fluorescence in situ hybridization (FISH) labeling of bacterial cells (**Fig. 41B**). Specificity in FISH was obtained by designing appropriate FISH probes for target bacterial cells in the sample. The FISH probes and bacterial cells simultaneously focused into a small volume and reacted in ITP, and a counter-pressure-driven flow was used to increase hybridization time. Similar to Eid et al.[94], unhybridized probes were then separated from cells in line prior to detection using a sieving matrix and spacer ion. Phung et al.[411] compared their ITP-based FISH to off-line standard FISH, and reported a decrease in total assay time from 2.5 h to 30 min.

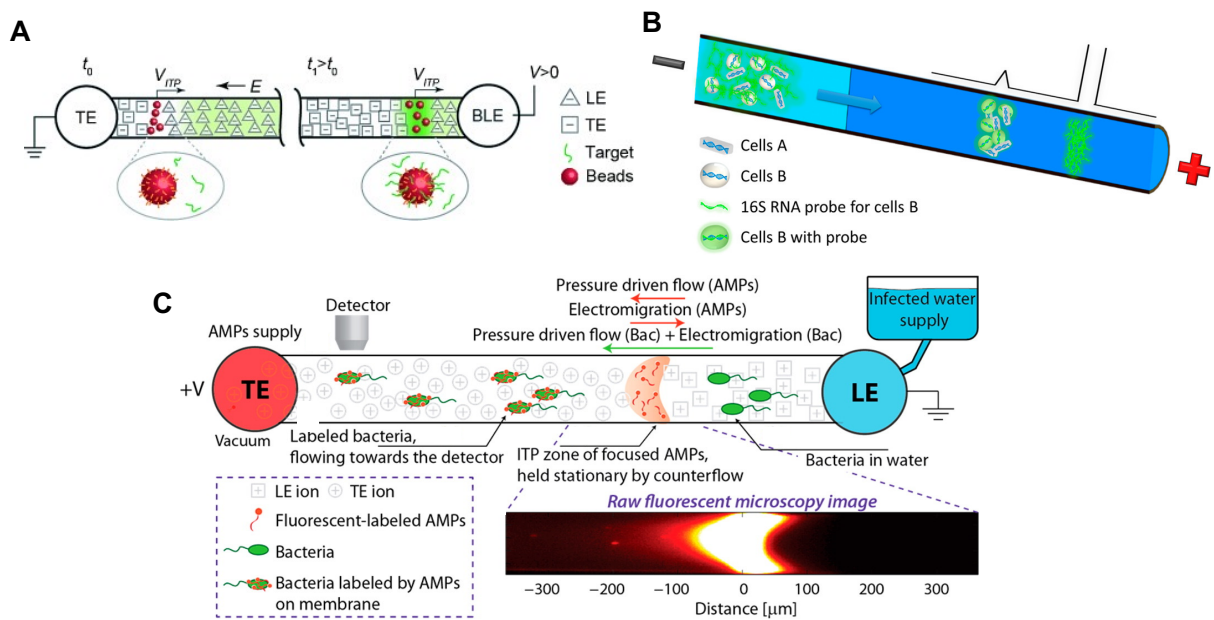

**Figure 41**. Three example ITP systems involving reactions among molecules and particles and/or cells. (A) Bead-based hybridization assay of Shintaku et al.[225]. Beads which are functionalized with probe DNA co-focus with target nucleic acids in ITP. This co-focusing and preconcentration accelerates reaction between probe and target. Figure is reproduced with permission from Ref.[225]. Copyright 2014 John Wiley and Sons. (B) Assay of Phung et al.[411] for ITP-based fluorescence in situ hybridization assay of intact bacterial cells. Intact cells co-focus and preconcentrate along with 16s RNA probes in ITP. Figure is reproduced with permission from Ref.[411]. Copyright 2017 American Chemical Society. (C) Microfluidic assay of Schwartz et al.[247] for continuous bacteria detection. The ITP peak which contains focused antimicrobial peptides is held stationary using counterflow, while bacterial cells are flowed through the channel via pressure-driven flow. As the cells flow though the ITP peak, they are labeled and detected downstream. Figure is reproduced with permission from Ref.[247]. Copyright 2014 American Chemical Society.

## 12. Miscellaneous configurations of microfluidic ITP

This section briefly reviews miscellaneous configurations used for ITP processes and analyses. Our intent is not to provide an extensive review but to present illustrative examples of (and some references for) the wide variety of configurations possible.

### 12.1 ITP preconcentration followed by electrophoretic separation

Capillary electrophoresis (CE) typically provides the best separation performance of any electrokinetic separation method.[43] This is due to CE's capabilities for analyte mobility resolution and peak capacity (the number of separated species in one experiment), among other figures of merit. In plateau mode, ITP offers some separation capacity but, by comparison to CE, ITP's clear advantage is superior sensitivity due to its strong capability for, and highly selective, analyte focusing (i.e., increase of concentration). In peak mode, ITP also offers the ability to achieve highly focused and spatially resolved peaks containing multiple species. An excellent way to combine the separation power CE and the focusing capabilities of ITP is to couple ITP preconcentration with subsequent capillary electrophoresis. Bahga and Santiago[24] cover in detail methods, configurations, chemistries, advantages, and disadvantages of such coupling. In this section, we review just a couple of these many coupling strategies.

The most common and easiest to implement method to couple ITP and CE (and even mass spectrometry) is transient ITP (t-ITP). [72] **Fig. 42** shows an example t-ITP chip layout using a column-coupled configuration (essentially, a T-junction) for coupling ITP with downstream zone electrophoresis.[24] In this configuration, the sample is mixed with TE and placed in the west reservoir, while the rest of channels including the north and east reservoirs are filled with LE. In the first step, electric field is applied from east to west (i.e., from the LE to the TE reservoirs) to focus and preconcentration target analytes in (peak- or plateau-mode or a combination thereof) ITP. The second step is initiated once the ITP peak migrates past the T-junction, here, electric field is applied from the north to the east reservoirs (both of which are LE reservoirs). The latter step introduces LE ions behind the TE ions. LE ions overspeed TE ions, eventually exposing the analytes locally only to LE ions—hence disrupting ITP and commencing capillary zone electrophoresis type separation. Various other chip designs (including simple straight channels) and loading configurations can be used for t-ITP.[24] Recently, changes in temperature have been used to effect t-ITP using thermally responsive gels which can be used to strongly vary micromolar analyte mobility.[412] t-ITP is likely most effective when preconcentration is achieved in peak mode. By far, the biggest disadvantage of t-ITP is the high degree of electromigration dispersion imparted on the analyte ions as LE ions are used to displace TE ions (or vice versa). This dispersion limits the resolution achievable by CE as it increases the effective injection sample width at the start of CE. We refer to the Ref.[24] for a detailed review on coupling t-ITP with zone electrophoresis and Refs.[29,95,108,110,304,306,324,413–415] for various applications of t-ITP in microfluidics.

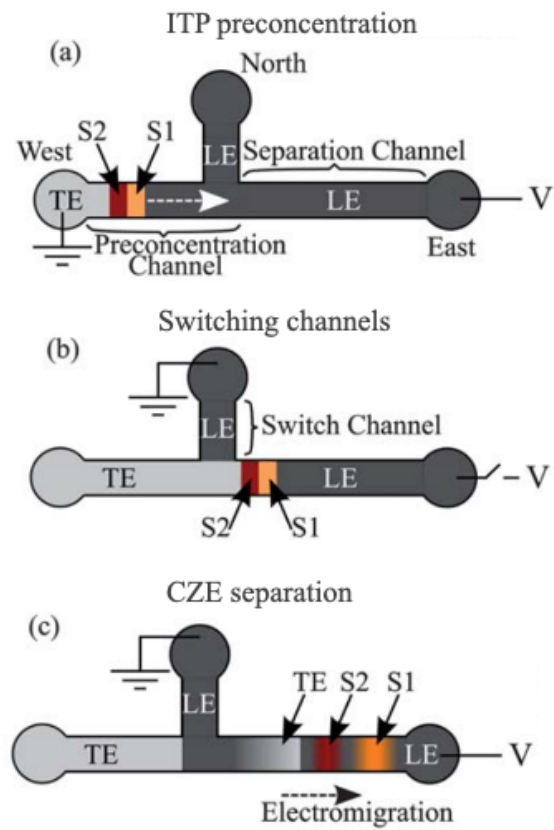

**Figure 42**. Schematic of an example column-coupling configuration for t-ITP/ZE implemented in a microfluidic chip. The west reservoir is loaded with TE and sample, and the remaining channel (including the main channel and the north and east reservoirs) is loaded with LE (A) ITP preconcentration of sample is initiated by applying an electric field between the East and West reservoirs. (B) As the ITP zone passes the branched section, the applied electric field is switched and is now applied between the east and north reservoirs. (C) Disruption of ITP and initiation of CZE separation. Figure is reproduced with permission from Ref.[24]. Copyright 2013 The Royal Society of Chemistry.

Bidirectional ITP offers a method of preserving and leveraging the initial high spatial resolution and sample concentration of a peak-mode ITP process while also coupling CE separation. In bidirectional ITP, counter-migrating cationic and anionic ITP processes are established in the same channel. As shown in **Fig. 43**, the cationic and anionic ITP shock waves typically originate in each of two end-channel reservoirs and migrate toward each other. This is achieved using two sets of LE and TE ions, namely, cations LE+ and TE+ for cationic ITP, and anions LE- and TE- for anionic ITP. LE+ and TE- ions are loaded in the cathodic reservoir, TE+ and LE- are loaded in the anodic reservoir, and the main channel is filled with a mixture containing LE+ and LE-. On application of electric field, the cationic shock wave migrates from the anodic reservoir towards the cathode, and the anionic shock wave migrates from the cathodic reservoir towards the anode. At an intermediate time, these two countermigrating shock waves interact with each other and disrupt the ITP process, resulting in the formation of a new (background) zone composed of TE+ and TE-wherein analyte ions under CE type separation. This configuration was used by Bahga et al.[211] and Bahga et al.[406] to preconcentrate and subsequently separate DNA molecules, including a 1kbp DNA ladder and hybridized DNA products from unhybridized ssDNA. We refer to Refs.[416–423] for detailed discussion around theory, simulations, and applications of bidirectional ITP. A disadvantage of the bidirectional ITP method is that designing assay chemistries can be complex (requiring two sets of carefully chosen and compatible LE and TE), and this can limit the possible range of analyte mobilities that can be separated.

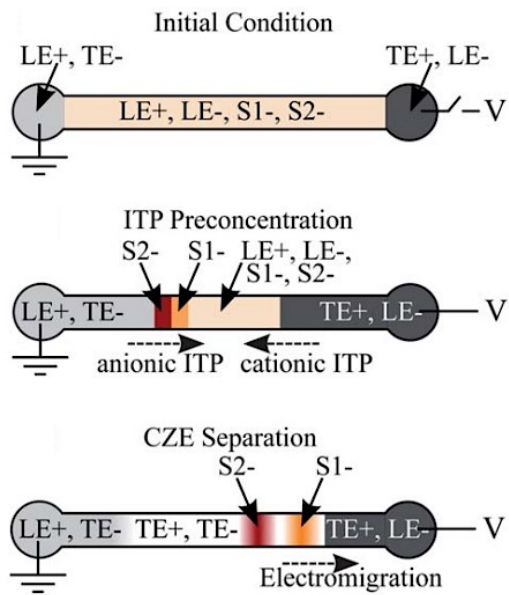

**Figure 43**. Schematic depicting ITP/ZE achieved using bidirectional ITP in a straight channel. Sample ions focus and preconcentrate in anionic ITP prior to the interaction of the anionic ITP shock wave with a countermigrating cationic ITP shock wave. Upon interaction of these shock waves, ITP is disrupted and electrophoretic separation is initiated via the formation of a new "background" zone composed of TE+/TE- ions. Figure is reproduced with permission from Ref.[24]. Copyright 2013 The Royal Society of Chemistry.

## 12.2 Cascade ITP

Recall from **Section 6** that the concentration of plateau zones in ITP scale proportionally with the LE ion concentration. Also, plateau mode concentration is determined by LE ion concentration, LE ion mobility, and plateau species mobility. Cascade ITP leverages these features using a sequence of LE zones with varying LE ion concentrations to improve detection sensitivity. In one approach, analytes are initially focused behind one LE zone with high concentration, and this enables strong sample focusing and high separation capacity. Subsequently, the focused zone electromigrates into a downstream region formerly occupied by an LE of lower ionic strength. In the latter zone, the local electric field is higher, and this can be used to raise the concentration of a peak-mode analyte distribution. The approach was first introduced by Boček et al.[423] for capillary ITP systems and used it to demonstrate complete separation of ions which typically formed mixed zones in ITP. Cascade ITP was later introduced in microfluidic ITP by Kaniansky et al.[250] using a column-coupled configuration to load electrolytes. More recently, Bahga et al.[212] adapted bidirectional ITP to achieve a cascade of LE concentrations and used it to demonstrated cascade

ITP in a single, straight channel chip. See also Guo et al.[256] for an example of cascade ITP implemented on a paper-based microfluidic device.

## 12.3 Counterflow and gradient elution ITP

One approach to increasing the processing capacity of an ITP system while minimizing sample length is to apply flow of the bulk liquid in a direction opposed to the direction of ITP. The flow can be achieved using an externally applied pressure gradient[424] or using the inherent electroosmotic mobility of the channels.[425] In either case, the effect is enable a longer duration of ITP in a relatively short channel. **Fig. 44** shows an example implementation of counterflow ITP from the work of Phung et al.[426]. In step (i), the sample is diluted in TE and loaded in a semi-infinite injection mode. In step (ii), constant pressure-driven flow is applied to approximately balance the electrophoretic motion of the ITP peak. Between step (ii) and (iii), sample ions continuously focus at the stationary ITP interface. Finally, in (iv), pressure driven flow is stopped and the peak is allowed to electromigrate to a detector placed downstream. Phung et al.[426] used this setup to detect bacterial cells in ITP and showed that counterflow improved LOD by a factor 4 compared to the case with no flow. See Refs.[60,180,328,379,427,428] for various other studies and applications that using counterflow in ITP.

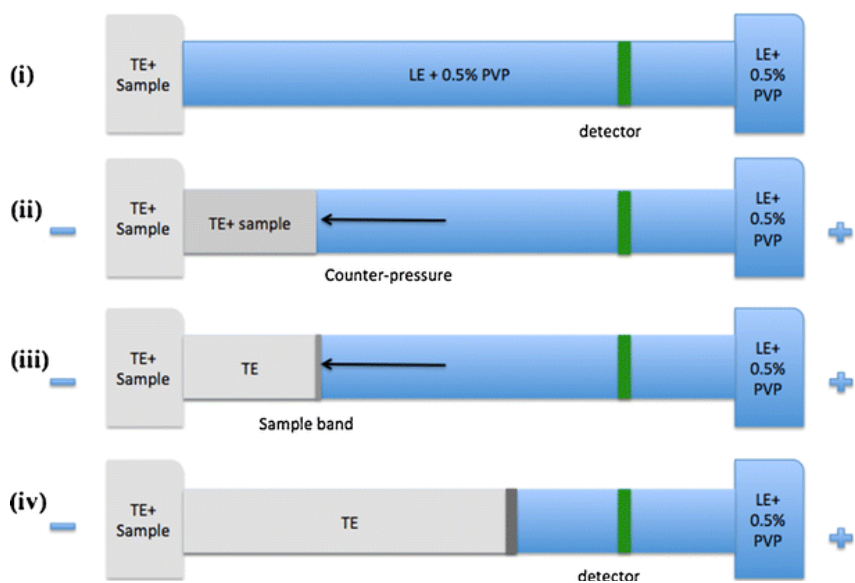

**Figure 44**. Schematic of counterflow ITP. In step (i), sample is mixed with TE and loaded via semi-infinite injection. In Step (ii) Electric field is applied and ITP is initiated. A counterflow is also applied using a hydrodynamic pressure head to balance electromigration. In (iii), the stationary ITP zone accumulates sample over time. In step (iv), counterflow is turned off, and the ITP peak migrates downstream for detection. The shape of the LE-to-TE interface is idealized as the pressure-driven flow tends to disperse this interface. Figure is reproduced with permission from Phung et al.[426]. Copyright 2015 Springer Nature.

Another implementation of counterflow in ITP is so-called gradient elution ITP, first shown by Shackman and Ross[429]. In a first step, the effect of the applied electric field and a (dominant) counter-flow (EOF and/or pressure driven flow) lead to a configuration where LE anions fill the entire length of a free-standing capillary connected to a reservoir filled with TE buffer and anionic analyte species, as shown in **Fig. 45**. In this configuration, analyte ions focus at the LE/TE interface near the channel entrance, as indicated in the figure. After sufficient analyte focusing, the counterflow is controlled to permit analyte ions to enter in order indicative of analyte effective mobility. We know of no quantitative analysis of this process, but the operation is reportedly robust and has been applied to ITP separation and controlled elution of amino acids and DNA.[430–434] Disadvantages of the technique include a requirement for precise pressure-driven flow control and repeatable and controllable EOF channel mobility.

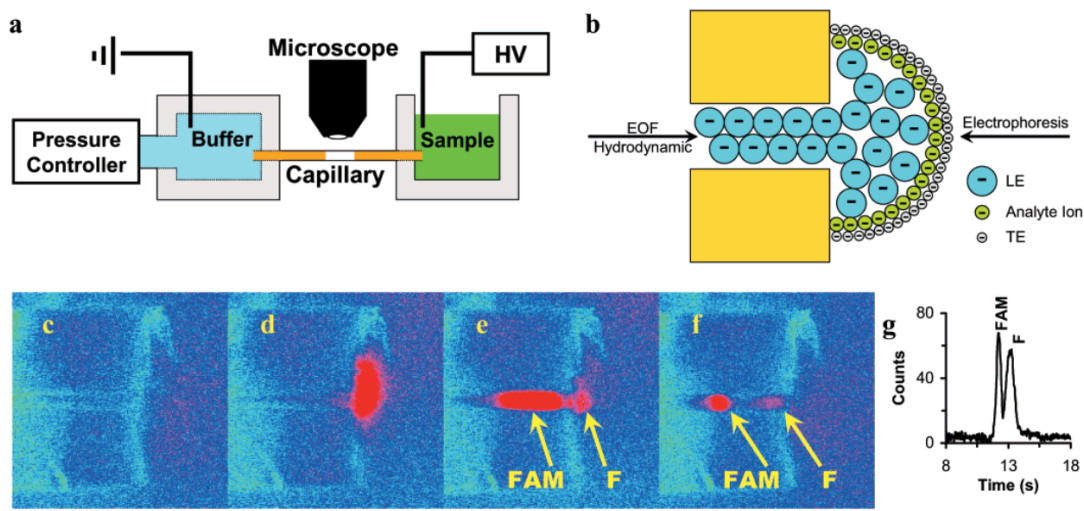

**Figure 45**. Gradient elution ITP. (A) Schematic of experimental setup used by Shackman and Ross[429]. (B) Combined EOF and hydrodynamic flow is used to focus and accumulate analyte ions between the LE and TE interface at the entrance of the capillary/channel. After this initial focusing, counterflow is regulated to control the release of analyte species in to the main channel, based on the species mobility values. (c)-(f) Experimental images showing the controlled injection of carboxyfluorescein (FAM) and fluorescein (F) in to the channel for (c) 1800, (d) 850, (e) 730, and (f) 590 Pa of applied pressure. (G) Electropherogram showing separation of F and FAM using gradient elution ITP. Figure is reproduced with permission from Ref.[429]. Copyright 2007 American Chemical Society.

## 12.4 Free-flow ITP

Free-flow ITP is a two-dimensional process that couples bulk flow and ITP electromigration in perpendicular directions. The system configuration is similar to that of so-called free-flow electrophoresis.[245,435,436] **Fig. 46** shows experimental visualizations of free-flow ITP from Janasek et al.[245]. In the figure, pressure-driven flow is established along the axial direction of channel (downward) with relatively shallow depth (into the image) and relatively large spanwise width (horizontal). Along the left and right spanwise (vertical) edges of the chamber are trenches of TE and LE buffers integrated with linear electrodes which establish a horizontal electric field directed from right to left. Sample ions introduced at the (top) entrance undergo ITP-type focusing as they flow downward. Free-flow ITP allows for continuous focusing and spatial separation (e.g., if focused to the point of plateau-mode ITP). Refer to Turgeon et al.[246] and Novo and Janasek[437] for detailed reviews around the theory and applications of microfluidic free-flow ITP.

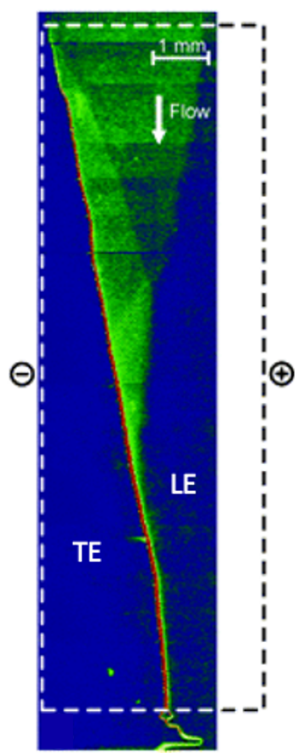

**Figure 46**. Experimental demonstration of free-flow ITP by Janasek et al.[245]. Shown is a heatmap of fluorescence intensity in $x$ (horizontal) versus $t$ (vertical), where red and blue respectively represent the highest and lowest fluorescence intensities. Here, 5 μM of fluorescein is preconcentrated and focused in free-flow ITP under an applied electric field of 525 V/cm. Flow is directed from the top to bottom at a velocity of 10 μL/min. Figure is adapted with permission from Ref.[245]. Copyright 2006 American Chemical Society.

## 13. Outlook for ITP in microfluidics

Microfluidic ITP has been pursued with increasing interest and shown significant developments over the last two decades. New applications for the use of ITP in microfluidics continue to emerge, especially in fields of biological science including molecular biology and diagnostics, chemical analysis, and food and plant sciences. These applications include the use of ITP both as an analytical tool and as an intermediate, preparatory tool for downstream analyses. We here offer comments around several areas of microfluidic ITP that would benefit from improvements and improved understanding.

Although microfluidic ITP technologies were initially developed (and are easily suited) to handle small sample volumes (typically, a few microliters), many emerging, particularly in the field of molecular diagnostics and trace analyte identification, require processing of large sample volumes—as much as several 100s of microliters to one or more milliliters to achieve sufficient sensitivity. To the best of our knowledge, the commercial microfluidic ITP system of Purigen Biosystems, Inc. (IONIC ITP system) currently processes the highest sample input volumes, around 200 μL in under an hour. Key impediments to the scale up of ITP channels are Joule heating and the buoyancy effects associated with larger volume systems. The requirement to mitigate Joule heating effects can significantly influence chip and system design and limit minimum assay time. For example, Joule heating can drive the material, fabrication method, and geometry of system. Buoyancy effects in ITP systems with relatively large cross section can arise due to the difference in mass density of various ITP zones (e.g., adjusted TE and LE) and this can create an important orientation dependence.

Microfluidic ITP technologies can also strongly benefit from the development of more (or fully) integrated and automated systems and workflows. This is of particular interest for molecular diagnostic systems which seek to provide automated and rapid raw-sample-to-answer solutions. One example is nucleic acid detection and/or identification technologies, including the integration of on-chip sample lysis, ITP-based purification, focusing, mixing, and chemical reactions (such as enzymatic amplification) followed by sensitive on-chip detection (e.g., via fluorescence). Typical challenges associated with such integrated assay systems using ITP include developing optimized buffer chemistries compatible with the various phases of the assay, optimal channel designs for ease of use and for interfacing various steps, and operation schemes which, in either open- or closed-loop, control and regulate the entire assay. Further, reproducible sample and buffer loading configurations (e.g., using automatic pressure valves) within the reservoirs and channels are important to ensure robustness of assays.

Another area for improvement is to better leverage the preconcentration ability of ITP to improve the performance (particularly the sensitivity but also the resolution) of downstream separation systems. As we have discussed such efforts have been ongoing for decades, but it is the opinion of the current authors that ITP is nevertheless underutilized as a preconcentration technique prior to separation and identification methods such as capillary electrophoresis, chromatography (including electrochromatography), and mass spectroscopy. In our experience, ITP can often be integrated as a preconcentration method using existing hardware by careful design of the chemistries and the initial configuration of sample versus buffers. Another related and emerging research area is the development and integration of hybrid matrix systems for ITP, for example, those involving liquid- and gel-based matrices, including functionalized gels and matrices for specific heterogenous assay and processing steps (including separations). Such systems should enable new modalities and functionalities.

Lastly, we recommend that easy-to-use modeling and simulation tools be developed (or existing ones expanded) to enable simple assay design and optimization of buffer chemistries and geometries. Current simulation tools are limited to testing ideas based on ad-hoc user inputs for chemistry and geometry, and preclude an easy approach to optimization. We suggest research into development of "inverse problem" approaches and simulation tools for ITP wherein the user provides target performance requirements and constraints (e.g., estimates of analyte mobilities and p$K_a$'s) as input to the simulation and the simulation outputs suggested possible design solutions ranked by performance metrics.

## Author Information

### Corresponding Author

*E-mail: juan.santiago@stanford.edu

### ORCID

Ashwin Ramachandran: 0000-0002-2512-8401
Juan G. Santiago: 0000-0001-8652-5411

**Notes**

The authors declare no competing financial interest.

**Biographies**

**Ashwin Ramachandran** received his B. Tech in Aerospace Engineering from the Indian Institute of Technology Bombay in 2015, and MS and PhD in Aeronautics & Astronautics and PhD minor in Bioengineering from Stanford University in 2021. His PhD research focused on the modeling and development of electrokinetic techniques for biochemical analyses and water desalination. He is currently a Postdoctoral scholar at Princeton University in the Departments of Mechanical and Aerospace Engineering and Molecular Biology.

**Juan G. Santiago** received his PhD in Mechanical Engineering from the University of Illinois at Urbana-Champaign in 1995. He holds the Charles Lee Powell Foundation endowed professorship at Stanford University and is Professor and Vice Chair of the Mechanical Engineering Department. His research includes the development of microsystems for on-chip chemical and biochemical analysis, methods for DNA quantification and analyses, and capacitive deionization. He is a Fellow of the American Physical Society, a Fellow of the American Society of Mechanical Engineering, and a Fellow of the American Institute for Medical and Biological Engineering. He serves and has served on the editorial board of several journals and has co-founded several companies in microfluidics. He is the founding Editor-in-Chief of the Cambridge University Press journal *Flow*.

# Acknowledgements

A.R. acknowledges support from the Bio-X Bowes Fellowship of Stanford University. J.G.S. and A.R. gratefully acknowledge financial support from Ford Motor Company and Stanford Bio-X Interdisciplinary Initiatives Committee (IIP).

**TOC Graphic**

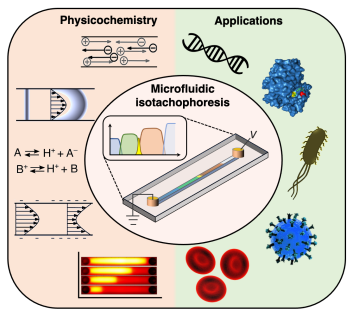